# Molecular tuning of DNA framework-programmed silicification by cationic silica cluster attachment

Xinxin Jing[1,2,8], Haozhi Wang[1,8], Jianxiang Huang[3,4,8], Yingying Liu[1,8], Zimu Li[1], Jielin Chen[1], Yiqun Xu[1], Lingyun Li[1], Yunxiao Lin[1], Damiano Buratto[3,4], Qinglin Xia[5], Muchen Pan[5], Yue Wang[5], Mingqiang Li[1], Ruhong Zhou[3,4,*], Xiaoguo Liu[1,*], Stephen Mann[6,7,*] and Chunhai Fan[1,*]

1. School of Chemistry and Chemical Engineering, New Cornerstone Science Laboratory, Frontiers Science Center for Transformative Molecules, Zhangjiang Institute for Advanced Study and National Center for Translational Medicine, Shanghai Jiao Tong University, Shanghai 200240, China.
2. Institute of Molecular Medicine, Shanghai Key Laboratory for Nucleic Acid Chemistry and Nanomedicine, Renji Hospital, School of Medicine, Shanghai Jiao Tong University, Shanghai, China.
3. Institute of Quantitative Biology, College of Life Sciences, and Department of Physics, Zhejiang University, Hangzhou, China.
4. Shanghai Institute for Advanced Study, Zhejiang University, Shanghai, China.
5. Division of Physical Biology, CAS Key Laboratory of Interfacial Physics and Technology, Shanghai Institute of Applied Physics, Chinese Academy of Sciences, Shanghai 201800, China.
6. School of Materials Science and Engineering, Shanghai Jiao Tong University, Shanghai 200240, China.
7. Max Planck-Bristol Centre for Minimal Biology, School of Chemistry, University of Bristol, Bristol BS8 1TS, United Kingdom.
8. These authors contribute equally, Xinxin Jing, Haozhi Wang, Jianxiang Huang and Yingying Liu.
*e-mails: rz24@columbia.edu, liuxiaoguo@sjtu.edu.cn, s.mann@bristol.ac.uk, and fanchunhai@sjtu.edu.cn

## Abstract

The organizational complexity of biominerals has long fascinated scientists seeking to understand biological programming and implement new developments in biomimetic materials chemistry. Nonclassical crystallization pathways have been observed and analyzed in typical crystalline biominerals, involving the controlled attachment and reconfiguration of nanoparticles and clusters on organic templates. However, the understanding of templated amorphous silica mineralization remains limited, hindering the rational design of complex silica-based materials. Here, we present a systematic study on the stabilization of self-capping cationic silica cluster (CSC) and their assembly dynamics using DNA nanostructures as programmable attachment templates. By tuning the composition and structure of CSC, we demonstrate high-fidelity silicification at single-cluster resolution, revealing a process of "adaptive templating" involving cooperative adjustments of both the DNA framework and cluster morphology. Our results provide a unified model of silicification by cluster attachment and pave the way towards the molecular tuning of pre- and post-nucleation stages of sol-gel reactions. Overall, our findings provide new insights for the design of silica-based materials with controlled organization and functionality, bridging the gap between biomineralization principles and the rational design of biomimetic material.

# Introduction

Nature has created a wide range of biominerals with evolutionary lineage, hierarchical structuration and adaptive functionality[1-7]. The organizational complexity of biominerals across multiple length scales has long intrigued researchers interested in the biological programming of inorganic materials and the potential abstraction of these concepts in biomimetic materials chemistry[8,9]. Recent studies on crystalline biominerals such as bones[10], teeth[11] and shells[12] have advocated a nonclassical crystallization pathway involving the controlled attachment of non-crystalline nanoparticles and clusters to preformed organic templates followed by *in situ* reconfiguration to initiate crystal nucleation and directional growth on the organic matrix[4]. This novel crystallization mechanism has also been demonstrated in a range of biomimetic systems including anatase ($TiO_2$)[13], calcium carbonate ($CaCO_3$)[14-17], hydroxyapatite ($Ca_{10}(PO_4)_6(OH)_2$)[18,19], lead selenide (PbSe)[20,21] and magnetite ($Fe_3O_4$)[22,23].

Amongst the wide diversity of biominerals, silica is the most abundant amorphous phase. It is naturally deposited in diatoms[24], sea sponges[25], radiolaria[26] and plants[27] via highly controlled and programmed processes involving the *in vivo* condensation of silicic acid in the presence of specific proteins to produce exquisite complex microarchitectures with unusual mechanical properties. For example, the axial filament at the center of the silica spicule of the demosponge *T. aurantium*[5,28] consists of an integrated hybrid structure consisting of specific enzymes (silicateins) packed in a hexagonal superstructure with amorphous silica occupying the interstitial spaces among the protein units. Much effort has been undertaken to synthesize various types of microscopically ordered amorphous silica by integrating classical sol-gel processes with supramolecular templates based on surfactant, protein or DNA self-assembly[29-32], or through the use of positive charged proteins and polyamines to control silica nucleation and growth[33]. Nevertheless, the general lack of a unified theoretical model hampers the microscopic understanding of templated silicification, thereby constraining the rational design of new silica-based materials with complex microscopic organization and functionality.

Silica sol-gel reactions are typically undertaken via a successive process involving the hydrolysis of silicon alkoxides to produce silanol monomers, linear and cyclic oligomers, and pre-nucleation clusters ~1.0 nm in size, which give rise to primary silica nanoparticles that associate into orderless gels or sols depending on the extent of crosslinking and condensation, respectively (**Extended Data Fig. 1a**)[34,35]. Although this polymerization mechanism is distinctly different from the classical models of crystallization, in principle the transitioning from primary nanoparticles to 3D networks or colloidal dispersions resembles particle attachment mechanisms of non-classical crystallization in which disordered pre-nucleation clusters serve as reconfigurable building units for the formation of crystalline phases[36]. However, elucidation of the pre- and post-nucleation stages of silicification is difficult because of the transient nature of siloxane oligomers and their rapid dehydration and progressive polymerization into large structures.

To address this challenge, we tuned the composition, structure and arrested growth of silica pre-nucleation clusters by controlling the co-condensation of N-trimethoxysilylpropyl-N,N,N-trimethylammonium chloride (TMAPS) and tetraethyl orthosilicate (TEOS) to generate monodisperse suspensions of ultrastable self-capping cationic silica cluster (CSC) (**Extended Data Fig. 1b**) which are characterized using a combination of cryogenic transmission electron microscopy (cryo-EM), synchrotron-based small angle X-ray scattering (SAXS), dynamic light scattering (DLS), high-speed atomic force microscopy (HS-AFM), matrix-assisted laser desorption ionization mass spectrometry (MALDI MS), solid-state magic angle spinning 13C and 29Si nuclear magnetic resonance (MAS NMR), ion trap mass spectrometer (IT MS) and Monte Carlo/molecule dynamics (MC/MD) hybrid simulations. Our results imply a general mechanism for the widespread presence of template-directed amorphous mineralization in natural and synthetic systems and provide a step to programmable biomimetic silica-based hybrid materials.

## Results

### Synthesis and characterization of cationic silica clusters (CSC)

CSC were discovered at pH 8 by co-condensation of mixed aqueous solutions of TMAPS and TEOS. By tuning the TMAPS/TEOS ratio in TE-MgCl$_2$ buffer, we obtained a series of size-controlled CSC after 6-day maturation of the mixture (**Fig. 1a, Supplementary Fig. 1 to 3, Supplementary Table 1**), which indicated that the size of the CSC increased as the TMAPS/TEOS ratio was decreased. Synchrotron SAXS measurements showed highly uniform diameters of gyration ($D_g$) for these CSC, from ~6.8 nm ~4.7 nm ~3.9 nm to ~1.3 nm (**Fig. 1b and 1c**). We used the $D_g$ values to designate the different samples obtained under different molar ratios, eg. 3.9 nm-CSC. In each case, the aqueous CSC dispersions displayed a ζ-potential typically between +30 to +40 mV and were stable for at least two months. Binary phase diagrams of hydrodynamic size plotted for different TMAPS/TEOS molar ratios indicated that the sizes of the CSC were primarily dependent on TMAPS concentration (**Fig. 1d, Supplementary Table 2**). Cluster sizes smaller than 1.0 nm were determined for TMAPS concentrations exceeding 61.6 mM, suggesting that high concentrations of TMAPS inhibited the polymerization process. Corresponding phase maps of conductivity (**Fig. 1e**) and ζ-potential (**Supplementary Fig. 4, Supplementary Table 3**) showed analogous increases in the positive surface charge along with a stepwise increase in the conductivity with increasing TMAPS concentration. A progressive decrease in pH from an initial value of 8.0 occurred as the TMAPS or TEOS concentrations decreased or increased, respectively, due to increased consumption of hydroxide ions during the sol-gel process (**Supplementary Fig. 5**). Corresponding cryo-EM images indicated that the generation of homogeneous CSC at pH 8 was replaced with a heterogeneous mixture of silica pre-nucleation clusters and CSC at an initial pH of 6 due to the rate of silicon alkoxide hydrolysis being higher than the condensation rate under acidic conditions (**Supplementary Fig. 6**). Taken together, the results were consistent with the presence of positively charged trimethylammonium side chains at the surface of the co-condensed silica clusters that acted as self-capping agents for constrained polymerization and stabilization of the CSC.

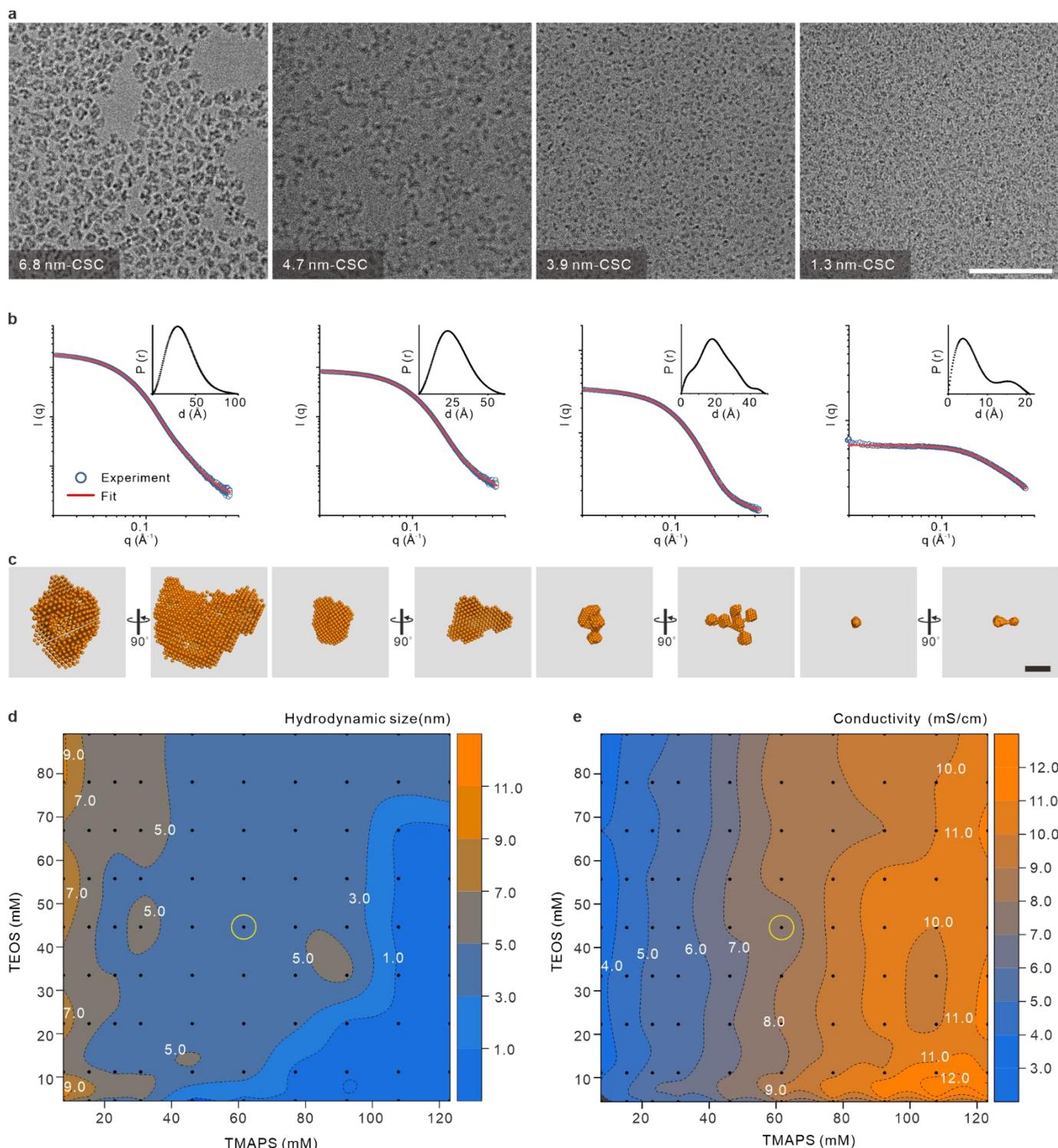

**Fig. 1 | Morphology, size and conductivity of typical CSC. a.** Cryo-EM images for typical CSC with calculated diameters of gyration ($D_g$) of 6.8 nm, 4.7 nm, 3.9 nm and 1.3 nm. Scale bar, 50 nm. **b.** SAXS 2D profiles and Pair Distance Distribution Function (PDDF, P(r)) plots of these CSC. In each case, the SAXS profiles revealed monodisperse populations of CSC. The PDDF plots of 6.8 nm- and 4.7 nm-CSC showed characteristic deflected Gaussian Distribution features, indicating ellipsoidal structures. Interestingly, the PDDF plots of 3.9 nm-CSC showed multiple shoulder peaks at ~5 Å, ~30 Å and ~45 Å, indicating a branched necklace or random coil-like structure. The PDDF plots of 1.3 nm-CSC showed bimodal distributions with different peak intensities, indicating a dumbbell shape with different sphere sizes. **c.** *Ab initio* calculations of the molecular envelopes from the SAXS data revealed that 3.9 nm-CSC were composed of seven subunits with diameters of ~1 nm. Scale bar, 2 nm. **d** and **e**, Hydrodynamic size and conductivity binary phase diagrams of CSC (N = 90). The yellow circles highlight the TMAPS/TEOS ratio for 3.9 nm-CSC.

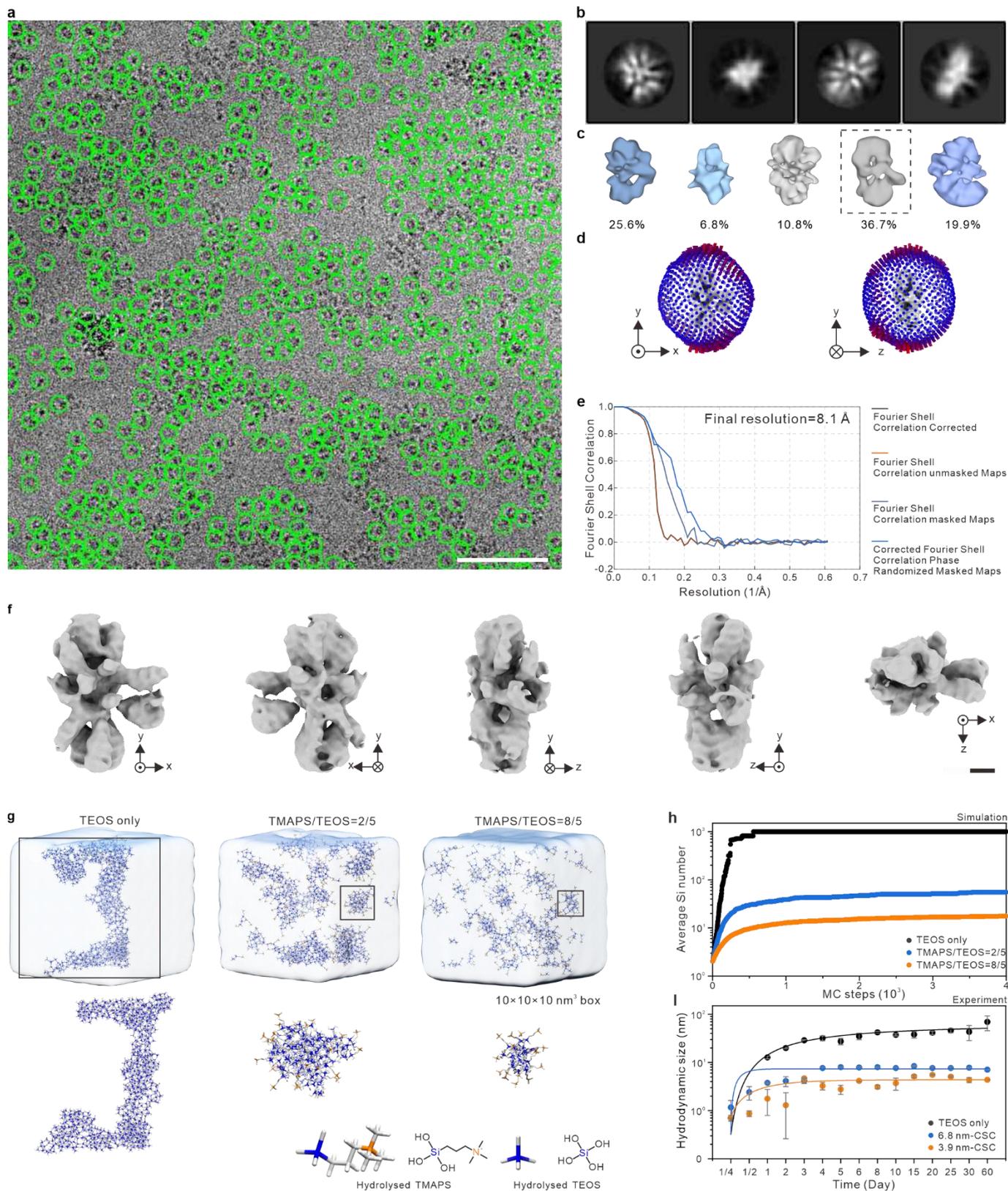

**Fig. 2 | Single-particle analysis and MC/MD simulations of 3.9 nm-CSC. a**, Representative image. The green circles highlight the auto-picked CSC particles. Scale bar, 50 nm. **b**, Representative 2D class averages. **c**, Representative 3D class averages. **d**, Histogram representing the orientational distribution of the particles from the 36.7% class. **e**, Fourier shell correlation plots. **f**, Five different views of the electron density maps of the particles from the 36.7% class. Scale bar, 1 nm. **g,** MC/MD simulations of the initial stages in silica sol-gel condensation. The presence of TMAPS interrupts the continuous polymerization of hydrolyzed TEOS to produce ultrastable CSC (middle and right) with sizes determined by the TMAPS/TEOS molar ratio. Top images; total simulation systems using a $10*10*10$ nm$^3$ water box. Bottom images; representative polymerized silica clusters for three different TMAPS/TEOS ratios; Si atoms (blue), N atoms (orange), O atoms (gray) and H atoms (white). **h**, Calculated average Si atom numbers of single clusters formed by condensation of TEOS alone or TMAPS/TEOS at ratios of 0.4 or 1.6 for increasing MC steps. **d**, Corresponding experiments using TEOS alone or TMAPS/TEOS ratios of 0.4 and 1.6 showed time-dependent changes in cluster hydrodynamic size. In the TEOS only system, the silica particles increased continuously to 100 nm after 60 days while the CSC reached their maximum sizes after 4 days and remained stable for at least 60 days when prepared from mixtures of TMAPS and TEOS.

High-resolution cryo-EM images of samples prepared at a TMAPS/TEOS ratio of 1.40 showed highly monodisperse CSC with a size distribution of 3.3 ± 0.7 nm (**Fig. 2a**). The reconstructed electron density 3D model of discrete 3.9 nm-CSC revealed non-spherical and loosely packed features, with a maximum 3D spatial resolution of 0.96 nm (**Fig. 2b-f**). This was consistent with DLS and synchrotron SAXS measurements that gave a hydrodynamic diameter and calculated $D_g$ of 3.6 ± 0.1 nm and ~3.9 nm, respectively.

To reveal the structural features of the 3.9 nm-CSC, we performed *ab initio* calculations of the molecular envelopes derived from the SAXS data, which revealed a hierarchically arrangement of seven loosely packed primary units of ~1.2 nm (**Fig. 1c**). The *Kratky* plots from the SAXS 2D profiles of CSC prepared at lower TMAPS/TEOS molar ratios indicated that the clusters increased in density as their size increased (**Extended Data Fig. 2**). We identified distinctive curves including linear divergence (1.3 nm-CSC, TMAPS/TEOS = 2.80), partial parabolic convergence (3.9 nm-CSC, TMAPS/TEOS = 1.40) and parabolic convergence (4.7 nm- and 6.8 nm-CSC, TMAPS/TEOS = 0.70 and 0.35, respectively)[37], which reflected the structural evolution of the CSC from linear to random-coil to spherical, according to the *Porod* rule and *Power* law[38] ($I(q) \propto q^{-n}$). These structural changes imply that the 3.9 nm-CSC consist of a flexible chain of corner-shared organosiloxane/siloxane tetrahedral that can be readily deformed or compressed.

The ability to tune the molecular structure of the CSC via changes in the TMAPS/TEOS molar ratio was demonstrated by hybrid MC/MD simulations of the early stages of silicification. For this, we simulated the initial co-condensation behavior in mixed solutions of TMAPS and TEOS placed in a weakly alkaline buffer (pH = 8.0 ± 0.1) within a water box of 10 nm$^3$ (**Fig. 2g**). The simulations confirmed an effective self-capping effect of TMAPS side chains in which the positively charged -$N^+(CH_3)_3$ groups preferentially covered the surface of the silica clusters. Potential mean force (PMF) analysis performed on two different CSC showed that the outward projection of -$N^+(CH_3)_3$ groups provided both repulsive coulombic forces and steric hindrance that inhibited inter-cluster aggregation and successive polycondensation (**Extended Data Fig. 3**). By contrast, in the absence of TMAPS, simulations of the silica clusters resulted in rapid growth into particles that attained the size limitation of the water box. The MC/MD simulations also demonstrated that the size of the CSC was inversely dependent on the TMPAS/TEOS molar ratio (**Fig. 2h and 2i**). For example, the calculated average number of Si atoms in two typical CSC were 53 ± 3 (TMAPS/TEOS = 2/5) and 18 ± 1 (TMAPS/TEOS = 8/5).

SAXS volume of correlation ($V_c$)[38] and MALDI MS were used to determine the molecular weights ($M_w$) of the CSC prepared at different TMAPS/TEOS molar ratios. The $V_c$ Mw values of the different CSC were 1,500, 8,300, 18,100 and 46,000 Da, which correlated with the corresponding MALDI MS m/z values of 1,221.1, 8,262.2, 18,001.5 and 52,011.3 (**Fig. 3a and 3b**).

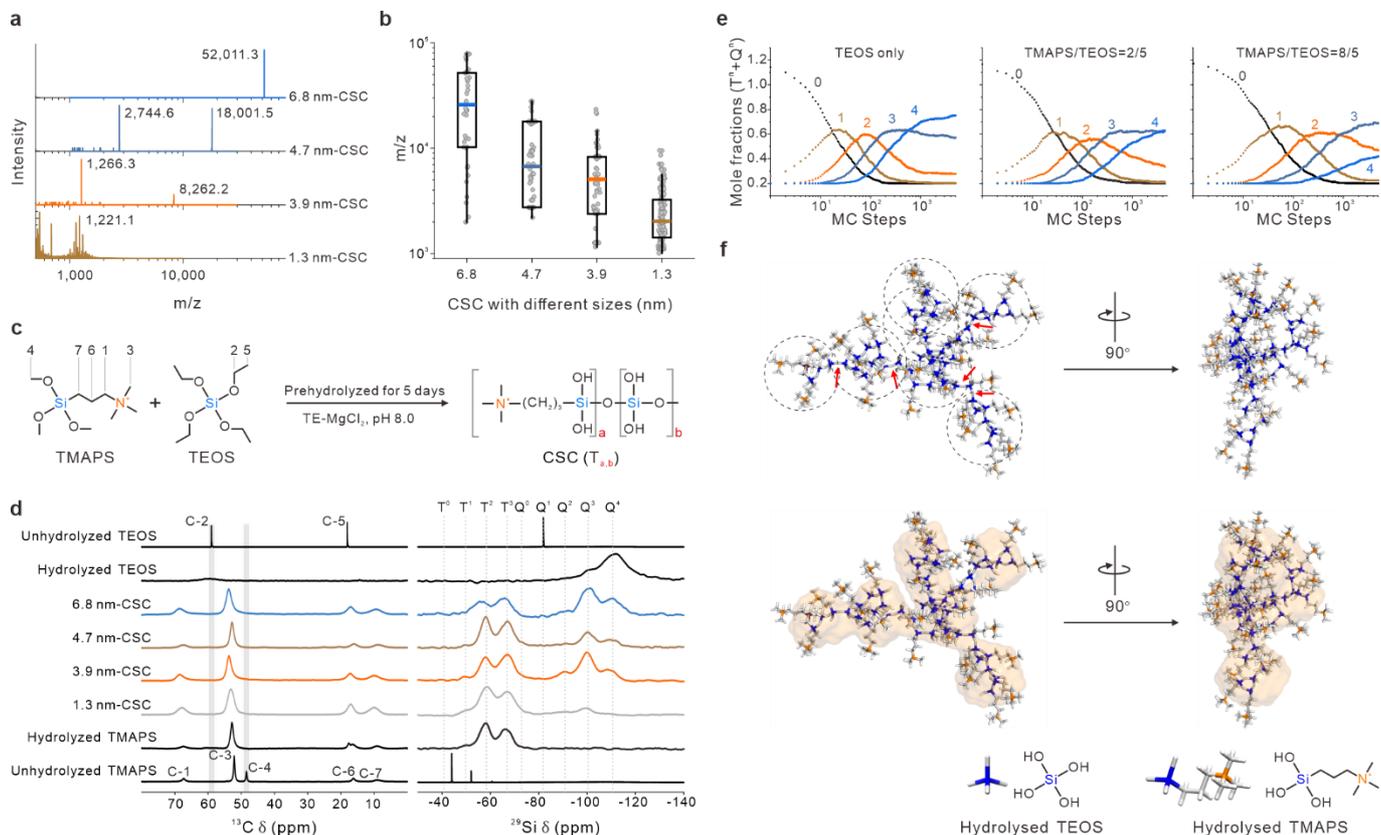

**Fig. 3 | Molecular weight, NMR analysis and structural modelling of CSC. a, b,** Mass spectra of CSC with different TMAPS: TEOS molar ratios (**a**) and scatter plot of m/z values for typical CSC (**b**); from left, $N$ = 40, 120, 90 and 173. **c,** Labelling of C atoms in TMAPS and TEOS used to assign $^{13}$C NMR spectra chemical shifts. $T_{a,b}$ labels represent the number of TMAPS and TEOS units present in the CSC. **d,** $^{13}$C (left) and $^{29}$Si MAS (right) NMR spectra of various freeze-dried CSC and single-component samples of unhydrolyzed and hydrolyzed TMAPS or TEOS. Chemical shifts labelled with $T^n$ or $Q^n$ (n = 0, 1, 2, 3 or 4) represent Si atoms either from TMAPS or TEOS with different crosslinking degrees, respectively. **e,** Simulated evolutions of $T^n + Q^n$ distributions for three typical reaction conditions undertaken with increasing TMAPS/TEOS molar ratios. **f,** Representative atomic structure models of a 3.9 nm-CSC ($T_{32,29}$) based on siloxane-linked tetrahedra. Top image; front and side views showing 7 subunits (dashed circles) and Si-O-Si linkages between adjacent subunits (red arrows). Bottom image displays the overlapping of the atomic structure and coarse grained SAXS model of the 3.9 nm-CSC.

Solid-state $^{13}$C and $^{29}$Si MAS NMR spectroscopies were used to determine the molecular structure of the CSC produced by co-condensation (**Fig. 3c**). In general, the smallest CSC (1.3 nm-CSC) had a calculated molecular form of $T_{5,3}$, where *a* and *b* represent the number of TMAPS and TEOS units in the general form $T_{a,b}$. This was equivalent to the protonated and chlorinated cluster, $C_{30}H_{86}Cl_5N_5O_{19}Si_8^+$ produced by the loss of 8 water molecules via dehydration during formation of siloxane linkages. In contrast, the 3.9 nm-CSC showed two characteristic m/z peaks of 1,266.3 and 8,262.2, which corresponded to $T_{5,5}$ and $T_{32,29}$ structures with the loss of 12 and 78 water molecules, respectively. Specifically, $^{13}$C MAS NMR spectra of freeze-dried CSC were consistent with complete hydrolysis of the TEOS and TMAPS alkoxide groups, as evidenced by the disappearance of both C2 (59 ppm) and C4 (48 ppm) peaks that were present in the spectra of pure TEOS and TMAPS molecules (**Fig. 3d,** left panel). Notably, four $^{13}$C peaks from TMAPS (C1, 67 ppm; C3, 53 ppm; C6, 17 ppm; C7, 9 ppm) were retained in all the samples without significant changes in intensity ratios, suggesting that hydrolyzed TMAPS groups were incorporated into the CSC. The presence of significant proportions of $T^2$ and $T^3$ peaks in the $^{29}$Si MAS NMR spectra confirmed co-polymerization of TMAPS and TEOS in the CSC as well as localization of TMAP side chains at the cluster surface (**Fig. 3d**, right panel, **Supplementary Table 4**), where $T^n$ (n = 0, 1, 2 or 3) represents Si atoms in the TMAPS molecules with different degrees of crosslinking. Moreover, the $^{29}$Si MAS NMR spectra indicated a relatively low degree of crosslinking in the CSC, with an apparent increase in peak intensities from $Q^4$ to $Q^3$ to $Q^2$ as the size of CSC decreased, where $Q^n$ (n = 0, 1, 2, 3 or 4)[39] represent Si atoms in TEOS molecules with different degrees of crosslinking. This agreed well with the calculated distribution of $Q^n$ from MC/MD simulations (**Fig. 3e**). Specifically, the $Q^4/Q^3/Q^2$ percentages for the 3.9 nm-CSC and 1.3 nm-CSC were 12.0/80.7/7.2 and 0.0/70.9/29.1, respectively, suggesting the presence of linear or cyclic structures in both cases[40], in agreement with the cryo-EM images of the corresponding CSC (**Supplementary Fig. 7**).

Taken together, the SAXS *Kratky* plots, NMR spectra and MALDI MS data indicate a progressive process of condensation and densification as the CSC increase in size to produce ultrastable clusters. The growth of pre-nucleation silica clusters with molecular masses of ~1.2 kDa is constrained by segregation of TMAPS side chains at the cluster/water interface such that polycondensation becomes self-limited. This was consistent with the atomic structure model for the 3.9 nm-CSC ($T_{32,29}$), which showed a branched chain of corner-shared siloxane tetrahedra comprising seven subunits with TMAP moieties distributed preferentially along the periphery (**Fig. 3f**).

Importantly, analyses on the early stages of silicification revealed the presence of metastable intermediate silica building blocks, which resembled those proposed in the particle attachment mechanism of nonclassical bio-crystallization[4]. As previous studies on the hexactinellid sponge *Euplectella sp*[41] and diatom shells[42] indicate that discrete silica nanoparticles are closely associated with the microscopic biomineralized architectures, our results provide further support that the production and assembly of stabilized CSC could play a key role in biosilicification.

**CSC-directed silicification of programmable DNA frameworks**

Having substantiated the existence of ultrastable CSC, we were motivated to investigate the dynamic interactions between the silica clusters and biomacromolecules, e.g., double-stranded DNA (dsDNA) frameworks using HS-AFM (**Fig. 4 and Extended Data Fig. 4**). An aqueous suspension of 3.9 nm-CSC at a concentration of 0.8 mg/L was injected into the liquid HS-AFM sample pool containing two-dimensional (2D) square-latticed triangular DNA origami frameworks. HS-AFM images revealed cluster-by-cluster attachment feature of CSC onto the DNA framework. We observed that freely diffusing CSC located within ~10 nm of the DNA framework were attracted onto the template surface typically within 5-15 seconds (**Fig. 4a**). The attachment appeared to be irreversible with no apparent disassociation back into the external solution over 180 s of consecutive imaging (**Fig. 4a, Supplementary Video 1**). Continuous imaging of the samples over *ca.* 400 s indicated that the CSC were randomly attached onto the DNA framework, suggesting that appropriate inter-DNA helix distances were required for ordered assembly. To test this, we employed two-layer honeycomb-latticed

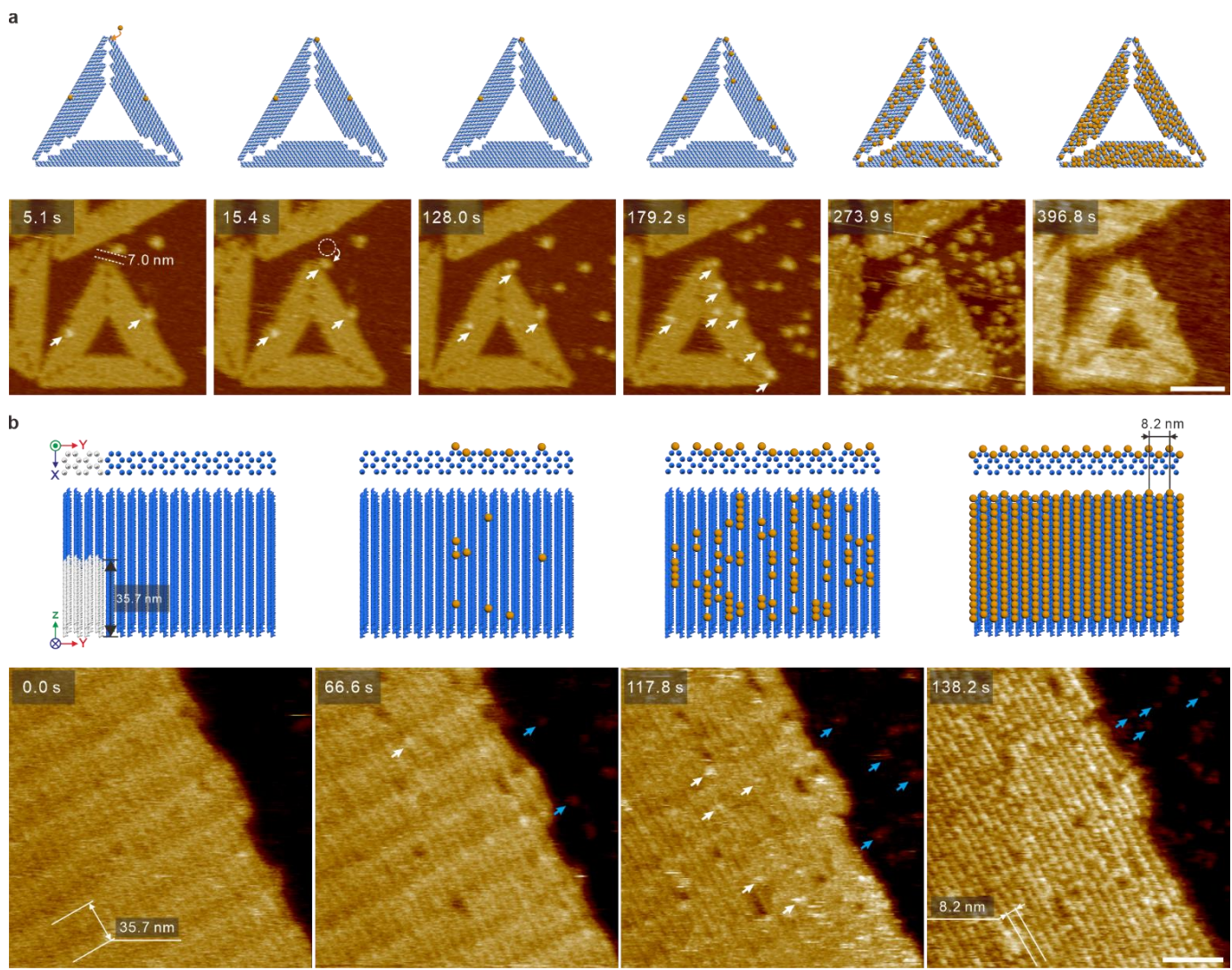

**Fig. 4 | In-situ HS-AFM imaging of DNA framework-programed silicification by CSC attachment. a,** Schematics and HS-AFM images of single cluster attachment behavior of 3.9 nm-CSC observed on a triangular DNA origami framework. CSC attached to the DNA template are highlighted with white arrows. **b,** Schematics and HS-AFM images of collective assembly behavior of 3.9 nm-CSC observed on a two-layer honeycomb structured Y-Z DNA lattice. Free CSC are highlighted with blue arrows; assembled CSC are highlighted with white arrows. Scale bars, 50 nm.

Y-Z DNA frameworks and observed periodic ordering of the CSC specifically along the Z-axis of the Y-Z DNA lattice, with a center-to-center Y-axis distance of 8.2 nm (**Fig. 4b**). Ordered attachment of the CSC to the DNA nanostructures was accomplished within 5 min even when highly dilute CSC dispersions were used. This resulted in loss of the native framework Z-axis periodicity (35.7 nm), which was readily resolved in the HS-AFM images of the undecorated samples (**Supplementary Video 2**).

Cryo-EM studies further confirmed that the fidelity of template replication was increased by commensurate matching of cluster size and framework architecture. For example, attachment of 3.9 nm-CSC to a 24-helix DNA framework resulted in specific ordering of the clusters onto adjacent DNA helices in the honeycomb lattice (**Fig. 5a**). Interestingly, attachment of the CSC gave rise to uniform expansion of the DNA lattice with the inner pore size increasing from ~2.6 nm to a cluster matching size of ~3.1 nm. In contrast, larger CSC exhibited random attachment along the z-axis, indicating that the DNA framework was not sufficiently flexible to adapt to CSC larger than 3.9 nm (**Fig. 5b to 5d**).

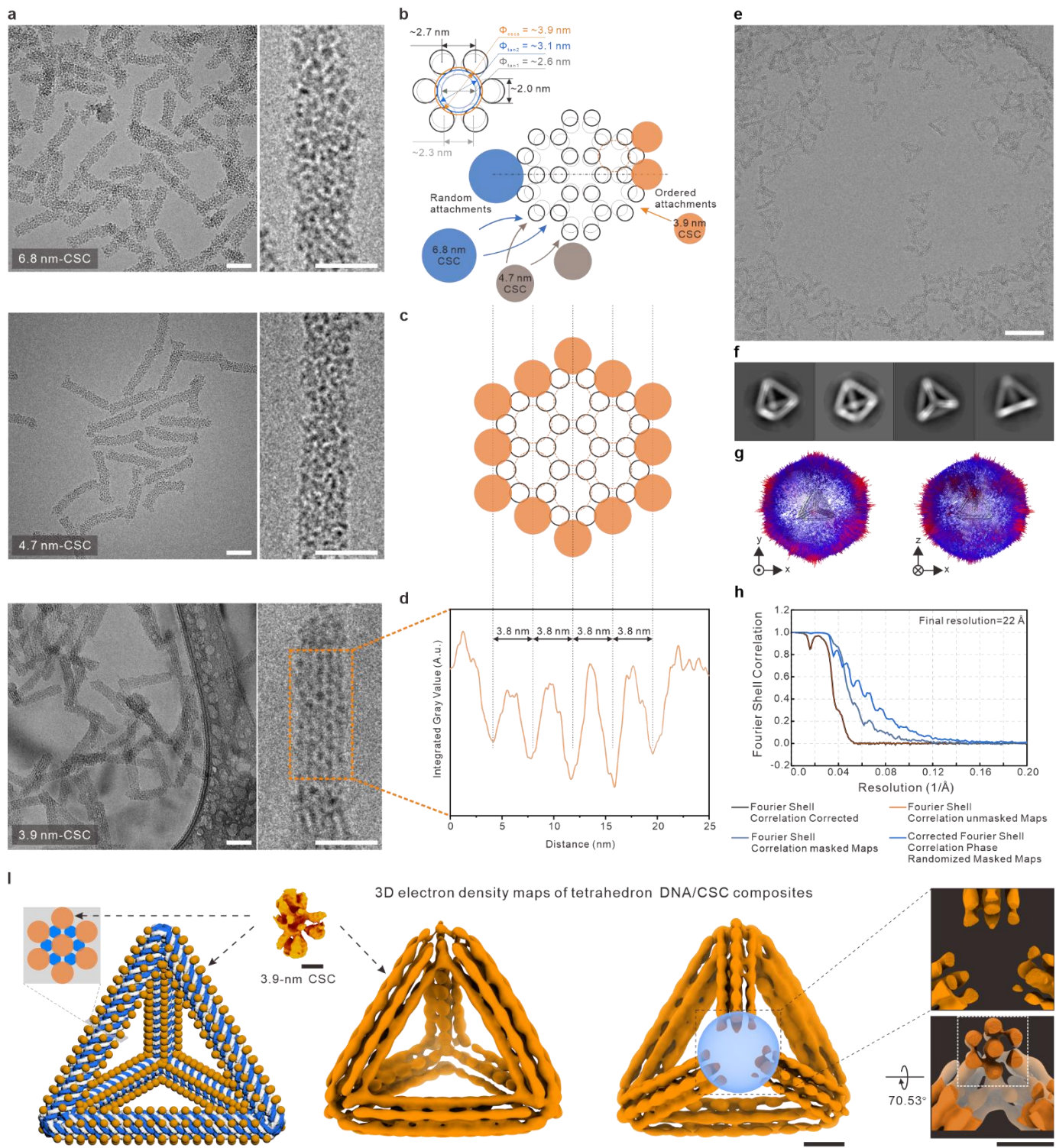

**Fig. 5 | Size dependency of CSC during silicification by cluster attachment. a,** Exemplary cryo-EM images for typical CSC (6.8 nm, 4.7 nm and 3.9 nm) assembled onto a honeycomb latticed 24-helix DNA framework. Of which, only 3.9 nm-CSC showed ordered attachment. Scale bars, 50 nm (left images), 25 nm (right images). **b,** Schematic of the attachment behavior of CSC with different sizes (drawn to scale). Top left inset; lattice expansion of the DNA framework after CSC attachment was illustrated. Where the theoretical diameter of dsDNA is ~2.0 nm and the diameter of internal tangent circle $\Phi_{tan1}$ (gray line) for ideal honeycomb structured DNA origami (dotted gray lines) is ~2.6 nm. However, the measured diameter of internal tangent circle $\Phi_{tan2}$ (blue line) is ~3.1 nm. Accordingly, the center-to-center distance of adjacent dsDNA strands increases from ~2.3 nm to ~2.7 nm to accommodate 3.9 nm-CSC. Orange circles ($\Phi_{CSC}$ = ~3.9 nm) show the relationship between 3.9 nm-CSC and the DNA framework. **c,** Schematic of a section of the CSC/24-helix DNA framework composite. **d,** Line profile of integrated gray value analysis across the CSC/24-helix DNA framework shown in (**a**) (lower right panel, orange box) showed a ~3.8 nm interplanar spacing of the CSC. **e,** Exemplary micrograph of 3.9 nm-CSC-bound tetrahedron DNA framework. Scale bar, 100 nm. **f,** Representative 2D class averages. **g,** Histogram representing the orientational distribution of particles. **h,** Fourier shell correlation plots. **i,** Left, schematics of a 3.9 nm-CSC-bound tetrahedron DNA framework. Middle, 3D electron density maps of the DNA/3.9 nm-CSC nanocomposite determined by means of single-particle reconstruction from cryo-EM data. Right, interceptions generated by the blue sphere highlight the sections that are perpendicular to the long axis of the tetrahedron edges. Scale bars, 10 nm; for 3.9 nm-CSC, 2 nm.

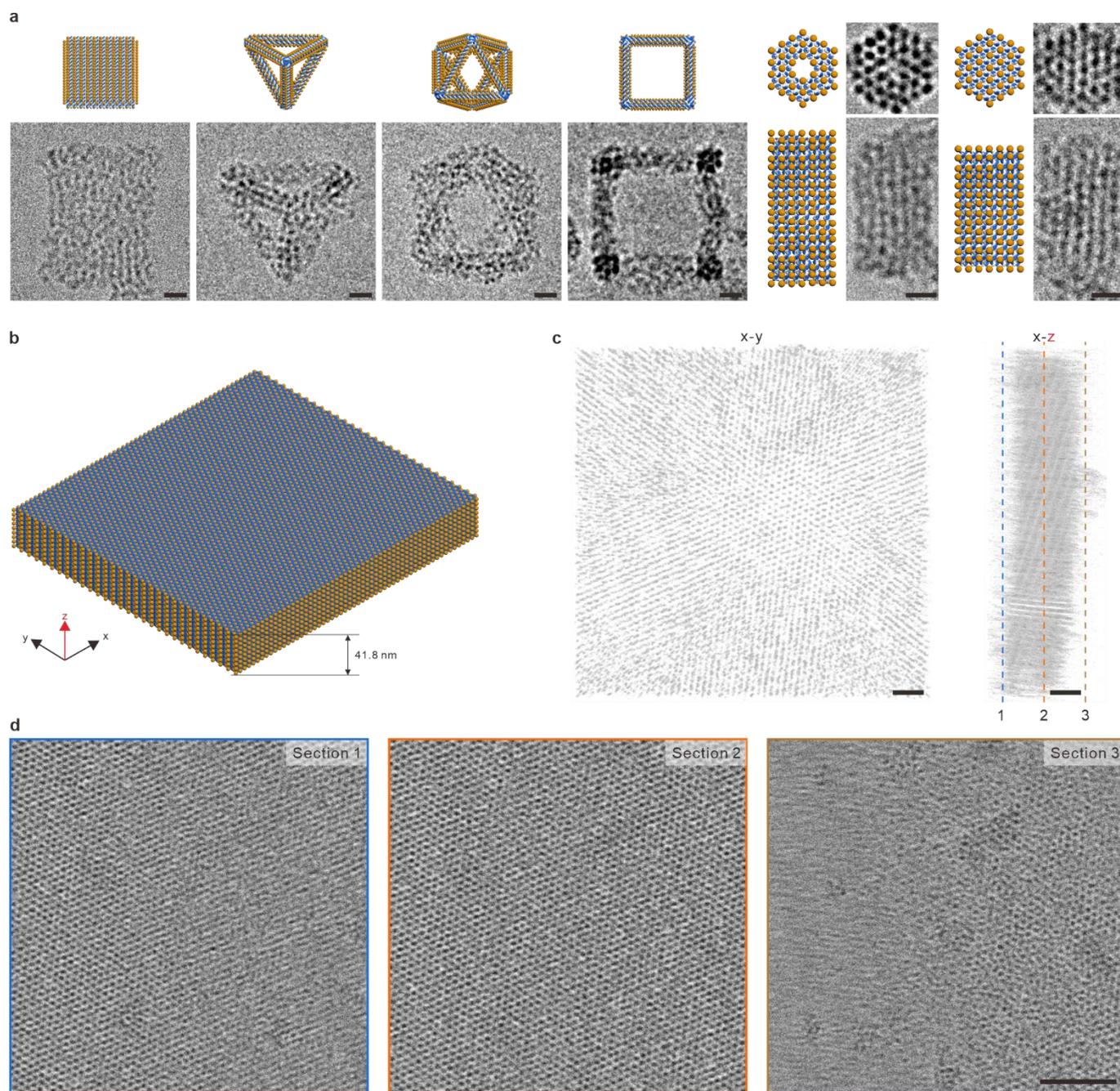

**Fig. 6 | DNA framework-programmed CSC attachments. a,** Representative DNA/CSC composites, including 3.9 nm-CSC bound double-layer, tetrahedron, octahedron, cube, 48-helix and 54-helix DNA frameworks. Scale bars, 10 nm. **b-d,** Schematic illustration, Cryo-ET images and sections of *p6mm* DNA/CSC superstructure. Scale bars, **c** and **d**, 20 nm, **e**, 50 nm.

In other studies, single-particle reconstruction of the cryo-EM data of a CSC-decorated tetrahedron DNA framework at a resolution 2.2 nm revealed that the attached CSC were organized preferentially onto the grooves of adjacent DNA helices with an inter-cluster distance of ~4.1 nm, and also located within the nanopores surrounded by six DNA helices **(Fig. 5e to 5i, Supplementary Fig. 8 and Supplementary Video 3)**. These data suggested that high fidelity CSC attachment onto programmable DNA nanostructures is achieved by precise dimensional matching and enhanced by adaptive changes in the template microstructure. Given the above observations, we propose that the self-adapting nature of the silification mechanism arises from the flexibility and structural dynamics of both the programmable DNA lattice and ultrastable CSC. This cooperativity enables high fidelity attachment and enables limited levels of mismatch between the cluster and DNA nanopore sizes to be tolerated during the assembly process.

Lastly, we explored the generality of our approach by using a diversity of DNA origami nanostructures such as multihelix-rods, planar frameworks, and 3D architectures. In each case, ordered attachment of the CSC was observed giving rise to a series of organized silicified DNA superstructures (**Fig. 6a Supplementary Fig. 9 to 16**). Similar results were also observed for micrometer-scale DNA frameworks. For example, a modified single-stranded tile assembly technique was used to construct a *p6mm* DNA 2D lattice (**Fig. 6b, Supplementary Fig. 17 to 20**). High-resolution cryo-EM images recorded perpendicular to the Z-axis of the undecorated DNA lattice or DNA/CSC superstructure, along with the reconstructed Cryo-ET data revealed a high precision silicification process (**Fig. 6c and 6d, Supplementary Video 4 and 5**). The fast Fourier transform (FFT) analyses (data not shown) corresponding to the $d_{1000}$ (= $d_{0100}$ = $d_{0010}$ = 4.0 nm) spacing for the *p6mm* DNA lattice and the $d_{C1000}$ (= $d_{C0100}$ = $d_{C0010}$ = 4.1 nm) spacing for the *p6mm* DNA/CSC superstructure were almost identical, indicating that the original spatial information encoded in the DNA lattice was inherited by the silicified replica.

**Discussion**

In conclusion, our results demonstrate a unified approach to DNA framework-programmed silicification by precise spatial attachment of positively charged silica nanoclusters with tailored molecular composition, charge, size, and stability (**Extended Data Fig. 5**). We employ a spontaneous self-capping mechanism of alkoxide co-condensation to arrest the growth of pre-nucleation silica clusters to generate discrete nanoscale building blocks (CSC) capable of high-fidelity DNA framework silicification at single-cluster resolution. Significantly, growth of the CSC is inhibited due to spontaneous surface segregation of the positively charged side chains of TMAPS that increases interfacial hydrophilicity and attenuates polycondensation. Based on this mechanism, molecular tuning of the CSC is achieved by systematic changes in the TMAPS/TEOS molar ratio, leading to a range of clusters with different morphological, structural and physiochemical properties.

Having established a systematic approach to silica cluster stabilization, we use a range of complex DNA nanostructures as programmable attachment templates to probe the assembly dynamics of the CSC. Silica cluster attachment is facilitated by electrostatic interactions with the DNA templates and leads to silicified superstructures that can exhibit superlattice ordering under conditions of size matching and structural adaptation. The latter mechanism involves cooperative adjustments of both the DNA framework and cluster morphology, for example via stretching or compression of DNA crossovers and reconfiguration of the siloxane network, thereby minimizing the total energy of the silicified DNA superlattices.

Our results have direct implications for the programmed construction of new-generation hybrid materials through the systematic integration of sol-gel chemistry and DNA nanotechnology. In particular, the ability to generate silica/DNA superstructures with controllable surface-to-surface distances less than 1.0 nm surpasses the precision of previously reported DNA nanotechnology-enabled strategies[43-46]. More indirectly, our work could offer insights into the mechanisms of biological silicification. For example, we observed that the properties of the CSC facilitate attachment with high specificity both at the surface and inside of the DNA frameworks, suggesting that similarly constructed silica-based building blocks could be responsible for the fine-scale elaboration of organic matrices involved in biosilicification[47-50]. In this regard, we note that our simple *p6mm* DNA/CSC superstructure highly resembles the natural hybrid silica/protein superstructure of the demosponge *T. aurantium*, both from the perpendicular and longitude cut of the structure[5]. Despite the fact that these two hybrid superstructures comprise different constituents - a cationic silicatein framework and anionic silicic acid precursors in the demosponge, and anionic DNA framework and cationic silica clusters in the DNA/CSC superstructure - they demonstrate identical hexagonal pattering of the silica building blocks and similar superlattice parameters ($d$ = 5.95 nm and 4.1 nm, respectively)[5,28]. Thus, it seems feasible that molecular tuning of organic framework-programmed silicification by silica cluster attachment is a potentially generic pathway in the formation of ordered amorphous silica minerals[24-27]. In conclusion, our studies provide new research directions for biomineralization, biomimetic materials chemistry and DNA nanotechnology.

**References**

1. Lowenstam, H. A. W., S. On biomineralization. *Paleobiology* **16**, 521-526 (1989).
2. Gong, Y. U. T. *et al.* Phase transitions in biogenic amorphous calcium carbonate. *Proceedings of the National Academy of Sciences* **109**, 6088-6093 (2012).
3. Lupulescu, A. I. & Rimer, J. D. In situ imaging of silicalite-1 surface growth reveals the mechanism of crystallization. *Science* **344**, 729-732 (2014).
4. James, J. D. Y. *et al.* Crystallization by particle attachment in synthetic, biogenic, and geologic environments. *Science* **349**, 6760-6769 (2015).
5. Görlich, S. *et al.* Natural hybrid silica/protein superstructure at atomic resolution. *Proceedings of the National Academy of Sciences of the United States of America* **117**, 31088–31093 (2020).
6. Wegst, U. G. K., Bai, H., Saiz, E., Tomsia, A. P. & Ritchie, R. O. Bioinspired structural materials. *Nature Materials* **14**, 23-36 (2015).
7. Meyers, M. A., McKittrick, J. & Chen, P. Y. Structural biological materials: critical mechanics-materials connections. *Science* **339**, 773-779 (2013).
8. Lakes, R. Materials with structural hierarchy. *Nature* **361**, 511-515 (1993).
9. Meldrum, F. C. & Colfen, H. Controlling mineral morphologies and structures in biological and synthetic systems. *Chemical Reviews* **108**, 4332-4432 (2008).
10. Reznikov, N., Bilton, M., Lari, L., Stevens, M. M. & Kröger, R. Fractal-like hierarchical organization of bone begins at the nanoscale. *Science* **360**, 507-517 (2018).
11. Gordon, L. M. *et al.* Amorphous intergranular phases control the properties of rodent tooth enamel. *Science* **347**, 746-750 (2015).
12. Li, X., Xu, Z.-H. & Wang, R. In situ observation of nanograin rotation and deformation in nacre. *Nano Letters* **6**, 2301-2304 (2006).
13. Yong, L. *et al.* Radially oriented mesoporous TiO2 microspheres with single-crystal-like anatase walls for high-efficiency optoelectronic devices. *Science Advances* **1**, e1500166 (2015).
14. Sommerdijk, N. A. J. M. & de With, G. Biomimetic CaCO3 Mineralization using Designer Molecules and Interfaces. *Chemical Reviews* **108**, 4499-4550 (2008).
15. Liu, Z. M. *et al.* Crosslinking ionic oligomers as conformable precursors to calcium carbonate. *Nature* **574**, 394-398 (2019).
16. Zhao, M. *et al.* Pressure-driven fusion of amorphous particles into integrated monoliths. *Science* **372**, 1466-1470 (2021).
17. Mao, L. B. *et al.* Synthetic nacre by predesigned matrix-directed mineralization. *Science* **354**, 107-110 (2016).
18. Palmer, L. C., Newcomb, C. J., Kaltz, S. R., Spoerke, E. D. & Stupp, S. I. Biomimetic systems for hydroxyapatite mineralization inspired by bone and enamel. *Chemical Reviews* **108**, 4754-4783 (2008).
19. Changyu, S. *et al.* Repair of tooth enamel by a biomimetic mineralization frontier ensuring epitaxial growth. *Science Advances* **5**, eaaw9569 (2019).
20. Cho, K.-S., Talapin, D. V., Gaschler, W. & Murray, C. B. Designing PbSe nanowires and nanorings through oriented attachment of nanoparticles. *Journal of the American Chemical Society* **127**, 7140-7147 (2005).
21. Jawaid, A. M., Asunskis, D. J. & Snee, P. T. Shape-controlled colloidal synthesis of rock-salt lead selenide nanocrystals. *ACS Nano* **5**, 6465-6471 (2011).
22. Mirabello, G., Lenders, J. J. M. & Sommerdijk, N. A. J. M. Bioinspired synthesis of magnetite nanoparticles. *Chemical Society Reviews* **45**, 5085-5106 (2016).
23. Guomin, Z. *et al.* Self-similar mesocrystals form via interface-driven nucleation and assembly. *Nature* **590**, 416-422 (2021).
24. Hildebrand, M. Diatoms, biomineralization processes, and genomics. *Chemical Reviews* **108**, 4855-4874 (2008).
25. Van Soest, R. W. M. *et al.* Global diversity of sponges (Porifera). *PLOS ONE* **7**, e35105 (2012).
26. Takahashi, K., Hurd, D. C. & Honjo, S. Phaeodarian skeletons: their role in silica transport to the deep sea. *Science* **222**, 616-618 (1983).
27. Guerriero, G., Stokes, I., Valle, N., Hausman, J.-F. & Exley, C. Visualising silicon in plants: histochemistry, silica sculptures and elemental imaging. *Cells* **9**, 1066-1085 (2020).
28. Werner, P., Blumtritt, H. & Natalio, F. Organic crystal lattices in the axial filament of silica spicules of Demospongiae. *Journal of Structural Biology* **198**, 186-195 (2017).
29. Tiancong, Z., Ahmed, E., Xiaomin, L. & Dongyuan, Z. Single-micelle-directed synthesis of mesoporous materials. *Nature Reviews Materials* **4**, 775-791 (2019).
30. Ma, K. *et al.* Self-assembly of highly symmetrical, ultrasmall inorganic cages directed by surfactant micelles. *Nature* **558**, 577-580 (2018).
31. Sumper, M. Biomimetic patterning of silica by long-chain polyamines. *Angew. Chem. Int. Ed.* **43**, 2251-2254 (2004).
32. Liu, X. G. *et al.* Complex silica composite nanomaterials templated with DNA origami. *Nature* **559**, 593-598 (2018).
33. Liu, B., Cao, Y. Y., Huang, Z. H., Duan, Y. Y. & Che, S. N. Silica biomineralization via the self-assembly of helical biomolecules. *Advanced Materials* **27**, 479-497 (2015).
34. Hench, L. L. & West, J. K. The sol-gel process. *Chemical Reviews* **90**, 33-72 (1990).
35. Carcouet, C. C. M. C. *et al.* Nucleation and growth of monodisperse silica nanoparticles. *Nano Letters* **14**, 1433-1438 (2014).
36. Gebauer, D., Kellermeier, M., Gale, J. D., Bergstrom, L. & Colfen, H. Pre-nucleation clusters as solute precursors in crystallisation. *Chemical Society Reviews* **43**, 2348-2371 (2014).
37. Robert, P. R. & John, A. T. Accurate assessment of mass, models and resolution by small-angle scattering. *Nature* **496**, 477-481 (2013).

38  Rambo, R. P. & Tainer, J. A. Characterizing flexible and intrinsically unstructured biological macromolecules by SAS using the Porod-Debye law. *Biopolymers* **95**, 559-571 (2011).
39  Protsak, I. S. *et al.* A 29Si, 1H, and 13C solid-state NMR study on the surface species of various depolymerized organosiloxanes at silica surface. *Nanoscale Research Letters* **14**, 160 (2019).
40  Herman, C. *et al.* Solution state structure determination of silicate oligomers by 29SI NMR spectroscopy and molecular modeling. *Journal of the American Chemical Society* **128**, 2324-2335 (2006).
41  Aizenberg, J. *et al.* Skeleton of Euplectella sp.: Structural hierarchy from the nanoscale to the macroscale. *Science* **309**, 275-278 (2005).
42  Sumper, M. A phase separation model for the nanopatterning of diatom biosilica. *Science* **295**, 2430-2433 (2002).
43  Park, S. Y. *et al.* DNA-programmable nanoparticle crystallization. *Nature* **451**, 553-556 (2008).
44  Macfarlane, R. J. *et al.* Nanoparticle superlattice engineering with DNA. *Science* **334**, 204-208 (2011).
45  Liu, W. *et al.* Diamond family of nanoparticle superlattices. *Science* **351**, 582-586 (2016).
46  Wang, S. *et al.* The emergence of valency in colloidal crystals through electron equivalents. *Nature Materials* **21**, 580-587 (2022).
47  Kroger, N., Deutzmann, R. & Sumper, M. Polycationic peptides from diatom biosilica that direct silica nanosphere formation. *Science* **286**, 1129-1132 (1999).
48  Yang, S. H., Park, J. H., Cho, W. K., Lee, H. S. & Choi, I. S. Counteranion‐Directed, Biomimetic Control of Silica Nanostructures on Surfaces Inspired by Biosilicification Found in Diatoms. *Small* **5**, 1947-1951 (2009).
49  Ehrlich, H. *et al.* Mineralization of the metre-long biosilica structures of glass sponges is templated on hydroxylated collagen. *Nature chemistry* **2**, 1084-1088 (2010).
50  Maldonado, M. *et al.* Cooperation between passive and active silicon transporters clarifies the ecophysiology and evolution of biosilicification in sponges. *Science advances* **6**, eaba9322 (2020).

**Supplementary Information** is available in the online version of the paper.

**Acknowledgements** We thank Huan Wang, Kevin Chan, Zhi He, Yichong Lao and Dong Zhang for helpful discussions. This work was partially supported by the National Key R&D Program of China (2021YFA1201200, 2021YFF1200404), the National Natural Science Foundation of China (22322205, 92056117, U1967217), the National Center of Technology Innovation for Biopharmaceuticals (NCTIB2022HS02010), the National Independent Innovation Demonstration Zone Shanghai Zhangjiang Major Projects (ZJZX2020014), Shanghai Artificial Intelligence Lab (P22KN00272), and the Starry Night Science Fund of Zhejiang University Shanghai Institute for Advanced Study (SN-ZJU-SIAS-003).

**Author Contributions** X.J., X.L. and C.F. conceived the research, designed the experiments. R.Z., J.H. designed the molecular simulations. R.Z., X.L. S.M. and C.F. supervised the research. X.J. performed the experiments, supported by H.W.. J.H., D.B., and M.L. performed the full-atomic MD and hybrid MC/MD simulations. J.H. carried out potential of mean force calculations. Z.L. developed the Python scripts for SST tile sequence generation and MS analysis. Y.L. and M.P. analyzed cryo-EM data. J.C. and Y.L. assisted with the analysis of MAS-NMR and MS data. Y.X. assisted with the analysis of SAXS data. L.L. performed the statistics of CSC. Q.X. assisted with the HS-AFM experiments. Y.W. assisted with design of DNA lattices. X.J., X.L., S.M. and C.F. interpreted the data. R.Z., X.L., S.M. and C.F. wrote the manuscript. All authors edited and commented on the manuscript.

**Author Information**

## Materials and Methods

**Materials.** N-trimethoxysilylpropyl-N,N,N-trimethylammonium chloride (TMAPS) (50% in methanol, cat. no. T2796) and tetraethyl orthosilicate (TEOS) (cat. no. T0100) were purchased from TCI, Japan. Staple DNA strands were purchased from Sangon Biotech (Shanghai, China). Scaffold DNA strands (5,250 nt, 7,249 nt, 8,064 nt and 10,004 nt) were purchased from Bioruler (cat. no. B3009, B3007, B3005 and B3003). All DNA strands were stored at −20 °C after being dissolved in ultrapure water. Other chemicals were purchased from Sinopharm and Sigma-Aldrich. Carbon-coated TEM grids were purchased from Beijing Zhongjingkeyi Technology Co., Ltd (Beijing, China). Lacey grids were purchased from Ted Pella. C-Flat-1.2/1.3-4 C grids were purchased from Protochips. AFM tips were purchased from Bruker and Olympus. All the reagents were used as received without further purification.

**Preparation of CSC.** TMAPS were added dropwise into TE-$MgCl_2$ buffer (5 mM Tris, 1 mM EDTA, 12 mM $MgCl_2$, pH=8.0). The mixture was then shaken at 800 rpm for 10 min in an Eppendorf Thermomixer, after which TEOS was added, followed by shaking for another 5 days. The stoichiometric ratios of CSC were summarized in **Supplementary Table 2**. All CSC were stored at 25 °C before further use.

**Preparation of DNA frameworks.** All DNA structures were designed with the help of caDNAno software. To fold the DNA origami, staple strands were mixed with the scaffold strands in a molar ratio of 5:1 in TE-$MgCl_2$ buffer. The mixture was then annealed in a PCR thermocycler (Eppendorf) using the following cooling protocol, 65 °C 20 min; 60 to 40 °C at 1 °C per 50 min; 40 to 25 °C at 1 °C per 30 min. To remove excess staples, the annealed mixture was mixed with PEG buffer (5 mM Tris, 1 mM EDTA, 500 mM NaCl, 15% w/v PEG, $M_w$: 8,000 g/mol) in a 1:1 volumetric ratio and centrifuged at 10,000 rcf for 15 min. The DNA pellet was redissolved in TE-$MgCl_2$ buffer and shaken for 12 hours at 800 rpm, 37 °C. The purified DNA origami was diluted to 100 ng/μL with TE-$MgCl_2$ buffer and quantified by UV-visible spectroscopy. All purified DNA origami were stored at 4 °C before further use.

To assemble the 2D DNA lattices, unpurified ssDNA strands were mixed in an equimolar stoichiometric ratio from a 100 μM stock in TE buffer (5 mM Tris, pH=8.0, 1 mM EDTA) supplemented with 40 mM $MgCl_2$. The cooling protocol was as following, 70 °C 10 min, 50-25 °C, at 1 °C per 6 h.

For SST design, the sequences of DNA tiles should be orthogonal and satisfy the following rules. Each sequence should avoid continuous bases of more than 3 nt. E.g., 'AAA' is permitted, while 'AAAA', 'AAAAA' are prohibited. The GC content should range from 45% to 55% to ensure sequence stability. For different sequences in the same group of DNA tiles, every two sequences should avoid identical sub-sequence of bases more than 8 nt. A Python script that integrates the above rules was provided to automatically generate DNA tile sequences of given lengths (See code availability).

**SCA process.** A typical DNA framework solution (100 ng/μL) was quickly added into the CSC dispersion (6.8 nm-CSC, ~7.1 mg/mL; 4.7 nm-CSC, ~9.8 mg/mL; 3.9 nm-CSC, ~15.4 mg/mL; 1.3 nm-CSC, ~26.5 mg/mL) in a 1:1 volumetric ratio. The mixture was shaken at 800 rpm, 25 °C for 10 min. After that, the mixture was centrifuged at 10,000 rcf, 25 °C for 15 min. The supernatant was carefully removed with a pipet under UV lamps, and the sediments (DNA/CSC composites) were resuspended in water. At last, the DNA/CSC composites in pure water were shaken at 800 rpm, 25 °C for 10 min. The DNA/CSC composites should be characterized within 24 h after they are formed.

**Further silicification.** Additional 2.5 μL 5.6% ammonia (v/v, in water) and 2.0 μL 5.0% TEOS (v/v, in ethanol) were added into 100 μL DNA/CSC dispersion. The mixture was shaken at 800 rpm, 25 °C for 1 day. After that, the mixture was centrifuged at 10,000 rcf for 15 min, followed by washing with pure water and ethanol 3 times, respectively.

**Agarose gel analysis.** DNA origami and DNA/CSC assembly were electrophoresed on a 1.0 % agarose gel containing 0.2×GelRed in TAE-$Mg^{2+}$ buffer (40 mM Tris, 2 mM EDTA, 12.5 mM $MgAc_2$, pH=8.0) for 1 h at 100 V bias in an ice-water-cooled gel box. The electrophoresed agarose gels were scanned using a G:BOX Chemi XL1.4 gel image-analysis system at a resolution of 300 DPI.

**DLS and Zeta potential measurements.** Dynamic light scattering experiments of CSC were conducted using Zetasizer Ultra (Malvern Instruments). 100 μL CSC (~2 mM) was pipetted into a disposable cuvette. The hydrodynamic size and ζ-potential value of one sample group were averaged over all parallel measurements (hydrodynamic size, N = 90; ζ-potential, N = 15) and were summarized in **Supplementary Table 2 and 3**.

**HS-AFM characterization.** HS-AFM experiments were conducted using tapping mode HS-AFM (RIBM, Japan) at 25 °C. A silicon nitride cantilever (9 μm long, 2 μm wide and 130 nm thick; BL-AC10DS, Olympus, Tokyo, Japan) with nominal spring constants of 0.1 N/m and a resonance frequency of 1,500 kHz was used.
To capture the dynamic assembly process of CSC, 5 μL of purified triangle DNA origami (5 ng/μL) or Y-Z DNA lattice (100 ng/μL) was incubated on freshly cleaved mica for 5 min. After successful visualization of DNA frameworks, 3.9 nm-CSC with a final diluted ratio of 1:20 (~0.8 mg/mL) in TE-$MgCl_2$ buffer was pumped into the liquid cell (0.5 mL/h for DNA origami, 1.0 mL/h for Y-Z DNA lattice). HS-AFM images were captured at a rate of 100 Hz (DNA origami, 2.56 s/frame) and 50 Hz (DNA lattice, 5.12 s/frame) with scan areas of 250 $nm^2$ and 300 $nm^2$.

**Negative-staining TEM.** For TEM imaging of DNA frameworks, 10 μL of purified DNA origami (5 ng/μL) or unpurified DNA lattice (100 ng/μL) was adsorbed onto glow discharged, carbon-film-coated copper grid for 10 min. The sample drop was wicked from the copper grid with filter paper and the grid was washed three times with pure water. Then the sample was stained for 30 s using a 0.75% aqueous uranyl formate solution. After this, the excess solution was removed with filter paper and the grid was washed three times with pure water. The copper grid was dried at 25 °C. For TEM imaging of DNA/CSC composites, 10 μL of freshly prepared DNA/CSC composites dispersion was adsorbed onto glow discharged, carbon-film-coated copper grids for 10 min. The grid was then washed three times by pure water and dried at 25 °C. All images were acquired using a Talos L120C G2 operated at 120 kV.

**Cryo-EM characterization.** 3 μL of the sample (CSC, DNA origami, DNA lattice or DNA/CSC composites) was piped onto glow discharged lacey grid or C-Flat-1.2/1.3-4 C grid. Then the sample was plunge-frozen using a Vitrobot Mark IV with the temperature of 20 °C, the humidity of 90%, wait time of 60 s, blot time of 10 s, blot force of −1 and drain time of 0 s. Images of 3.9 nm-CSC were collected using an FEI Titan Krios G3i TEM (Thermo Fisher Scientific) operated at 300 kV that equipped with a K2 direct electron detector (Gatan) with a pixel size of 0.822 Å. The CSC datasets have a defocus range from 0.8 to 2.5 μm. Each micrograph was dose-fractioned to 25 frames with 2.4 s exposure time for each frame. The total accumulated dose of each micrograph is 60.0 e-/Å$^2$. The images of DNA origami and DNA/CSC composites were collected using a Glacios TEM (Thermo Fisher Scientific) operated at 200 kV. The images were collected on a Falcon III detector, with defocus ranges from −1.0 to −2.0 μm or −0.8 to −2.6 μm, respectively. Each micrograph was dose-fractioned to 16 frames with 2.5 s exposure time for each frame. The total accumulated dose of each micrograph is 40.0 e-/Å$^2$.

A total of 2,236 cryo-EM images of 3.9 nm-CSC were collected on a K2 detector. Motion correction was performed on the dose-fractioned image stacks using RELION's own MotionCor2 with dose weighting[51]. The CTF parameters of each image were determined with CTFFIND-4.1[52]. Particle picking, 2D classification, 3D initial model, 3D classification, 3D auto-refine and PostProcess were performed with RELION-4.0[53]. An overview of the data processing procedure is shown in **Supplementary Table 5**. After two rounds of 2D classification and two rounds of 3D classification with exhaustive angular searches, a total of 32,564 particles that belong to the CSC were processed with 3D auto-refine and solvent-masked post-processing. A cryo-EM map of the CSC was finally calculated from 32,564 particles at an overall resolution of 8.1 Å. The resolution estimation was based on the gold-standard Fourier shell correlation (FSC) 0.143 criterion[54]. The cryo-EM datasets of the DNA origami and DNA/CSC composites were processed similarly to that of the CSC.

Single-particle style image processing of typical DNA lattices and their corresponding DNA/CSC composites (including contrast transfer function estimation, particle picking, particle extraction and 2D alignment and averaging) were also accomplished using the Relion software package.

Cryo-tomograms were acquired using a Glacios TEM (Thermo Fisher Scientific) operated at 200 kV. Images were collected with EPU tomography with a defocus range from −2 to −3 μm at a calibrated magnification of 120,000, corresponding to a magnified pixel size of 1.2 Å. Movies were comprised of 4 frames and 5 s exposure time. The session was set up as single directional tilting in increments of 2 - 3° up to 60° and the dose rate was set to ~1.03 e/pixel/s. Cryo-tomograms were processed with the IMOD 4.9 routine.

**NMR analysis.** For unhydrolyzed TEOS and TMAPS, 1 mL of TMAPS or TEOS solution was measured on a 700 MHz Bruker Advance NEO operating at 176 MHz for $^{13}$C and 139 MHz for $^{29}$Si. For CSC, hydrolyzed TMAPS and TEOS, 1 mL of freshly prepared samples were desalted by using a NAP-10 column and redispersed into 1 mL pure water, followed by subsequent freeze-drying. The MAS NMR measurements were carried out using a 3.2 mm BL3.2 HXY MAS probe head with a sample rotation rate of 10 kHz, in a magnetic field of 14.09 Tesla on a 600 MHz Bruker Avance NEO operating at 150 MHz for $^{13}$C and 119 MHz for $^{29}$Si. Q8M8 was used for $^{29}$Si NMR shift calibration. The intensities of the spectral components $T^0$, $T^1$, $T^2$, $T^3$ and $Q^0$, $Q^1$, $Q^2$, $Q^3$, $Q^4$ were obtained by deconvolution of the $^{29}$Si MAS NMR spectra in terms of Gaussian line shapes using DMfit2011. The chemical shifts and corresponding intensities percentages of spectral components of CSC were summarized in **Supplementary Table 4.**

**MS analysis.** Freshly prepared CSC dispersions were directly used for analysis. For MALDI MS measurement, a Shimadzu MALDI-7090 was used with the following parameters, Tuning Linear, Power139, P. Ext at 3,000.00 (bin189), Ion Gate Blanking: 500.00, Laser Diameter 200.

**MS spectra Analysis.** For MALDI MS spectra analysis, the molecule fragment ions always have one positive charge. Therefore, the m/z peak value ($N$) equals the corresponding molecule fragment ion's total molecular weight ($MW_{TOTAL}$). Given that a single CSC is formed through condensation of the certain number of TMAPS ($a$) and TEOS ($b$) molecules, meanwhile losing $m$ H$_2$O, the function of $N$ can be described as follow,

$$N = MW_{TOTAL} = a * MW_{TMAPS} + b * MW_{TEOS} - m * MW_{H2O} \quad (1)$$

in which $(2a + 2b)/2 < m < (3a + 4b)/2$, according to the lowest crosslinking degree of 2 from NMR data. The exact mass values of these molecules are $MW_{TMAPS}$ = 180.105, $MW_{TEOS}$ = 95.988 and $MW_{H2O}$ = 18.010, respectively. Based on the above restrictions, we wrote a script in Python (See Code availability) to screen out all possible combinations of $a$, $b$ and $m$. Then, the final molecular formula of a typical CSC was figured out based on the comparative analysis of this value and the experimental $N$ value.

**Hybrid MC/MD simulations.** We built the structures of TMAPS and TEOS using Avogadro software[55]. The force field parameters involving silicone atoms used were as described previously[56]. The molecular topology of TMAPS and TEOS were created using the antechamber tool[57]. Specifically, the Antechamber tool[5] was used to assign the atomic type of TMAPS and TEOS using the GAFF force field[58] and the partial charges on the atoms were assigned using the AM1-BCC method[59]. Initially, three simulation systems were built and the first system consisted of 1,001 TEOS molecules. The second and third systems were composed of 1,001 TEOS and TMAPS molecules with ratios of 5:2 and 5:8, respectively. These ratios of the three simulated systems were in line with the experimental conditions. These molecules were randomly distributed in cubic simulation boxes of 10×10×10 nm$^3$ and simulation boxes were solvated with the TIP3P explicit water model[60]. All molecular dynamic simulations were performed using the GROMACS 2020[61] simulation package. The temperature (T = 300 K) and pressure (P = 1 atm) were maintained using a stochastic velocity rescaling thermostat[62] and Parrinello–Rahman barostat[63], respectively. Periodic boundary conditions were applied to all systems in all directions, and using of the LINCS algorithm enabled a standard integration time step of 2 fs[64]. All systems were simulated for 0.1 ns for each step. Short-range electrostatic and *van der Waals* interactions were calculated at a cut-off distance of 1.2 nm, while long-range electrostatic interactions were treated via the particle mesh Ewald (PME) method[65]. All the simulation snapshots were rendered with VMD[66]. In addition to the analysis tools of GROMACS, MDAnalysis[67] was also extensively used for the analysis of the simulated trajectories.

The hybrid MC/MD workflow was created and implemented using a Python pipeline. After the initial energy minimization and MD simulation, the distances among silicon atoms of TEOS and TMAPS were calculated. Molecule pairs with silicon atoms within 6 angstroms were counted and a *Monte Carlo* step using metropolis algorithm was used to determine whether the dehydration reaction was accepted for the counted molecule pairs. Specifically, a random number was generated and if the random number was smaller than the reaction probability, the dehydration reaction of the closest hydroxyl groups was accepted. The reaction probability was determined by the specific molecules (i.e., TEOS and TMAPS) and described in detail below.

From the NMR experiments, a silica unit with $n$ Si-O-Si bonds can be inferred, denoted here as $Q^n$. As a result, $Q^0$ stands for a monomer with no bonds, while $Q^4$ symbolizes a silica unit with bonds at all four of its vertices with four other silica units. While TEOS is viewed as $Q^0$, TMAPS is treated as $T^0$ due to its propyl-N, N, N-trimethylammonium moiety. We have set different reaction probabilities ($P(T^m Q^n)$) based on the $T^m$ of TMAPS and $Q^n$ of TEOS molecules. The reaction probability of $T^m$ and $Q^n$ is similar to the previous kinetic *Monte Carlo* algorithm by Malani et al.[68],

$$P(T^m Q^n) = \frac{1}{1+(m+2)} \cdot \frac{1}{1+(n+1)} \cdot S \quad (2)$$

where $S$ is the scaling factor and in this study S=0.1 was used. After the *Monte Carlo* step, the system topology was modified accordingly using the automatic Python code. The next round of MD simulations was then conducted until the specified 1,000 of runs were reached.

**Code availability.** The Python scripts for SST tile sequence generation and MS analysis can be downloaded from https://github.com/zimu-liii/SST-sequence-generator.git and https://github.com/zimu-liii/MS-analysis.git.

51 Zheng, S. Q. *et al*. MotionCor2: anisotropic correction of beam-induced motion for improved cryo-electron microscopy. *Nature Methods* **14**, 331-332 (2017).
52 Rohou, A. & Grigorieff, N. CTFFIND4: Fast and accurate defocus estimation from electron micrographs. *Journal of Structural Biology* **192**, 216-221 (2015).
53 Kimanius, D., Dong, L., Sharov, G., Nakane, T. & Scheres, S. H. W. New tools for automated cryo-EM single-particle analysis in RELION-4.0. *Biochemical Journal* **478**, 4169-4185 (2021).
54 Scheres, S. H. W. & Chen, S. Prevention of overfitting in cryo-EM structure determination. *Nature Methods* **9**, 853-854 (2012).
55 Hanwell, M. D. *et al*. Avogadro: an advanced semantic chemical editor, visualization, and analysis platform. *Journal of Cheminformatics* **4**, 17 (2012).
56 Lin, P.-H. & Khare, R. Molecular simulation of cross-linked epoxy and epoxy-POSS nanocomposite. *Macromolecules* **42**, 4319-4327 (2009).
57 Wang, J. M., Wang, W., Kollman, P. A. & Case, D. A. Automatic atom type and bond type perception in molecular mechanical calculations. *J Mol Graph Model* **25**, 247-260 (2006).
58 Wang, J., Wolf, R. M., Caldwell, J. W., Kollman, P. A. & Case, D. A. Development and testing of a general amber force field. *Journal of Computational Chemistry* **25**, 1157-1174 (2004).
59 Jakalian, A., Jack, D. B. & Bayly, C. I. Fast, efficient generation of high-quality atomic charges. AM1-BCC model: II. Parameterization and validation. *Journal of Computational Chemistry* **23**, 1623-1641 (2002).
60 Jorgensen, W. L., Chandrasekhar, J., Madura, J. D., Impey, R. W. & Klein, M. L. Comparison of simple potential functions for simulating liquid water. *Journal of Chemical Physics* **79**, 926-935 (1983).
61 Abraham, M. J. *et al*. GROMACS: High performance molecular simulations through multi-level parallelism from laptops to supercomputers. *Softwarex* **1**, 19-25 (2015).
62 Bussi, G., Donadio, D. & Parrinello, M. Canonical sampling through velocity rescaling. *Journal of Chemical Physics* **126** (2007).
63 Parrinello, M. & Rahman, A. Polymorphic transitions in single crystals: A new molecular-dynamics method. *Journal of Applied Physics* **52**, 7182-7190 (1981).
64 Hess, B., Bekker, H., Berendsen, H. J. & Fraaije, J. G. LINCS: a linear constraint solver for molecular simulations. *Journal of Computational Chemistry* **18**, 1463-1472 (1997).
65 Darden, T., York, D. & Pedersen, L. Particle mesh Ewald: An N⋅log(N) method for Ewald sums in large systems. *Journal of Chemical Physics* **98**, 10089-10092 (1993).
66 Humphrey, W., Dalke, A. & Schulten, K. VMD: visual molecular dynamics. *Journal of Molecular Graphics* **14**, 33-38 (1996).
67 Michaud-Agrawal, N., Denning, E. J., Woolf, T. B. & Beckstein, O. MDAnalysis: a toolkit for the analysis of molecular dynamics simulations. *Journal of Computational Chemistry* **32**, 2319-2327 (2011).
68 Shere, I. & Malani, A. Polymerization kinetics of a multi-functional silica precursor studied using a novel Monte Carlo simulation technique. *Physical Chemistry Chemical Physics* **20**, 3554-3570 (2018).

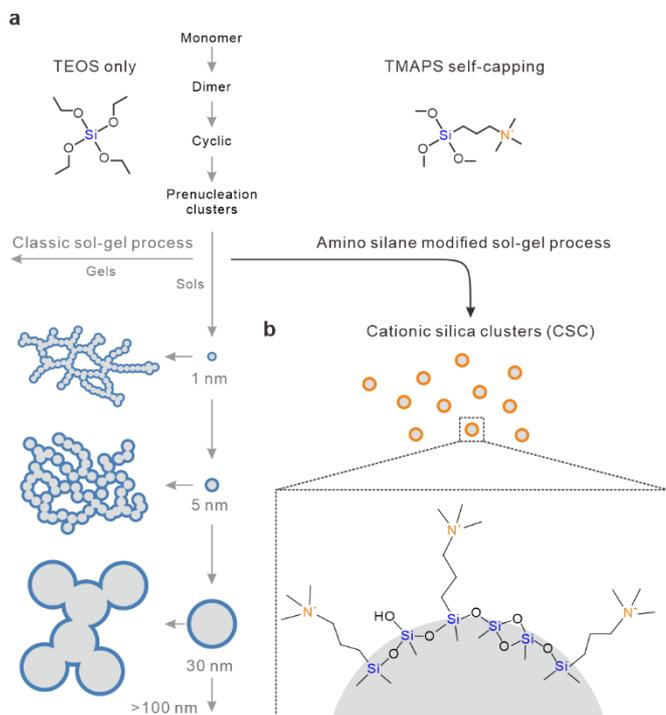

**Extended Data Fig.1 | Comparison between amino silane modified sol-gel process and classic sol-gel process. a,** Scheme showing classical silica sol-gel process (left) and arrested growth of silica pre-nucleation clusters by self-capping to produce ultrastable CSC (right). **b,** Schematic illustration of the self-capped CSC showing spontaneous surface segregation of the positively charged side chains of TMAPS that increases interfacial hydrophilicity and attenuates polycondensation.

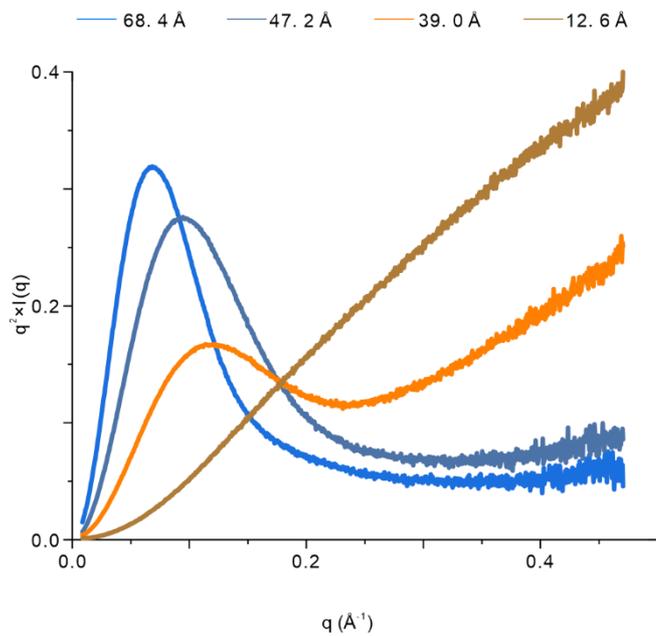

**Extended Data Fig.2 | Kratky plots derived from SAXS profiles.** 3.9 nm-CSC showed characteristic partial parabolic convergence curve.

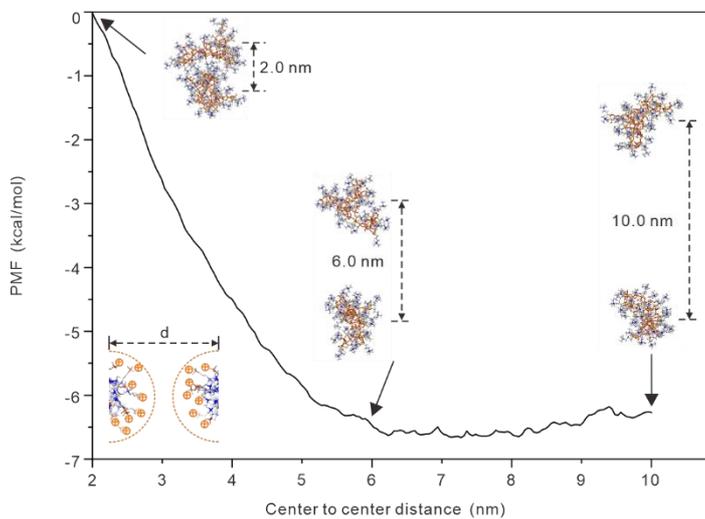

**Extended Data Fig.3 | Potential Mean Force (PMF) analysis of a two-cluster system.** The PMF plot revealed that a ~6.0 nm center-to-center distance was critical for 3.9 nm-CSC stabilization in solution.

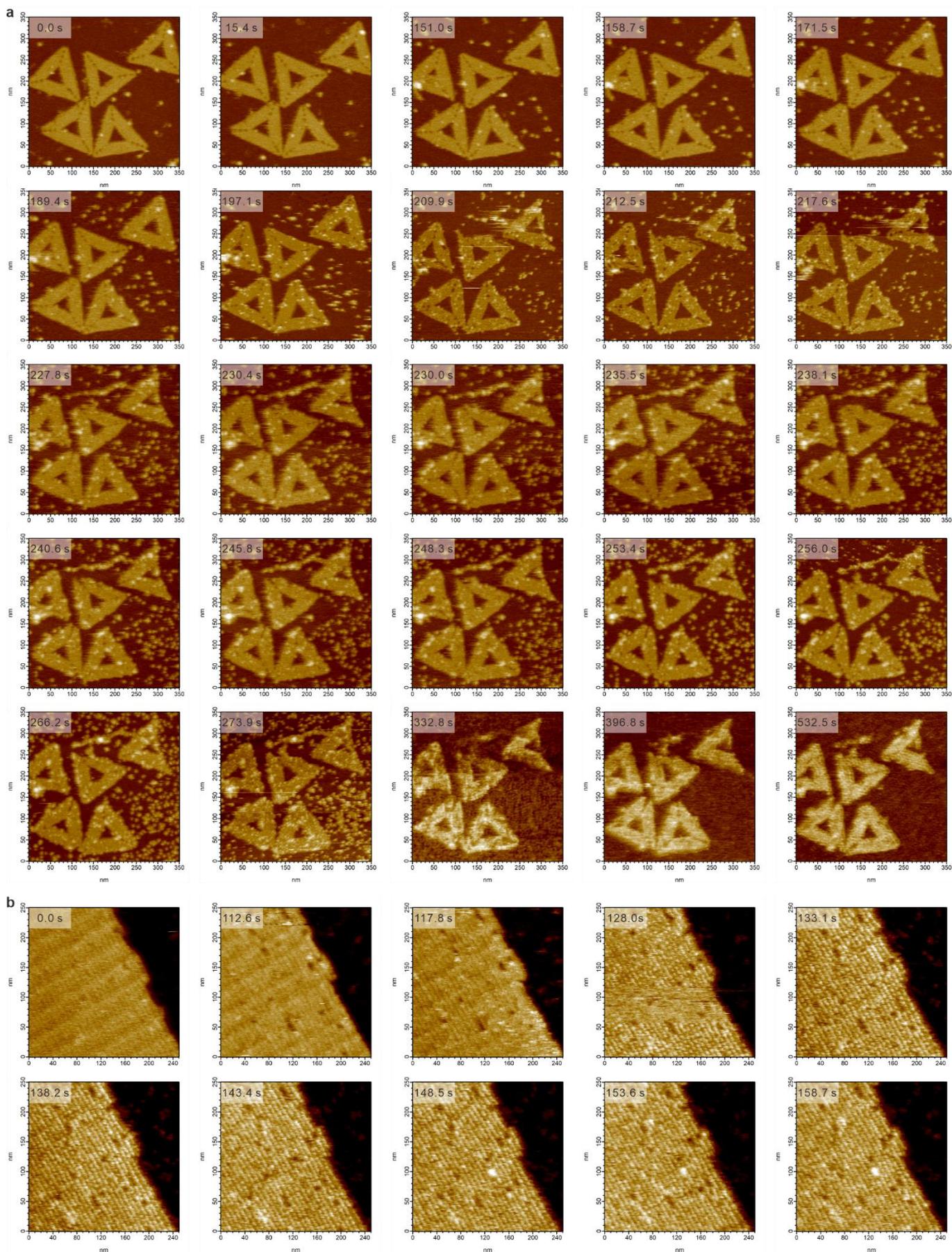

**Extended Data Fig.4 | Sequential in-situ HS-AFM images from Supplementary Videos 1 (a) and 2 (b).**

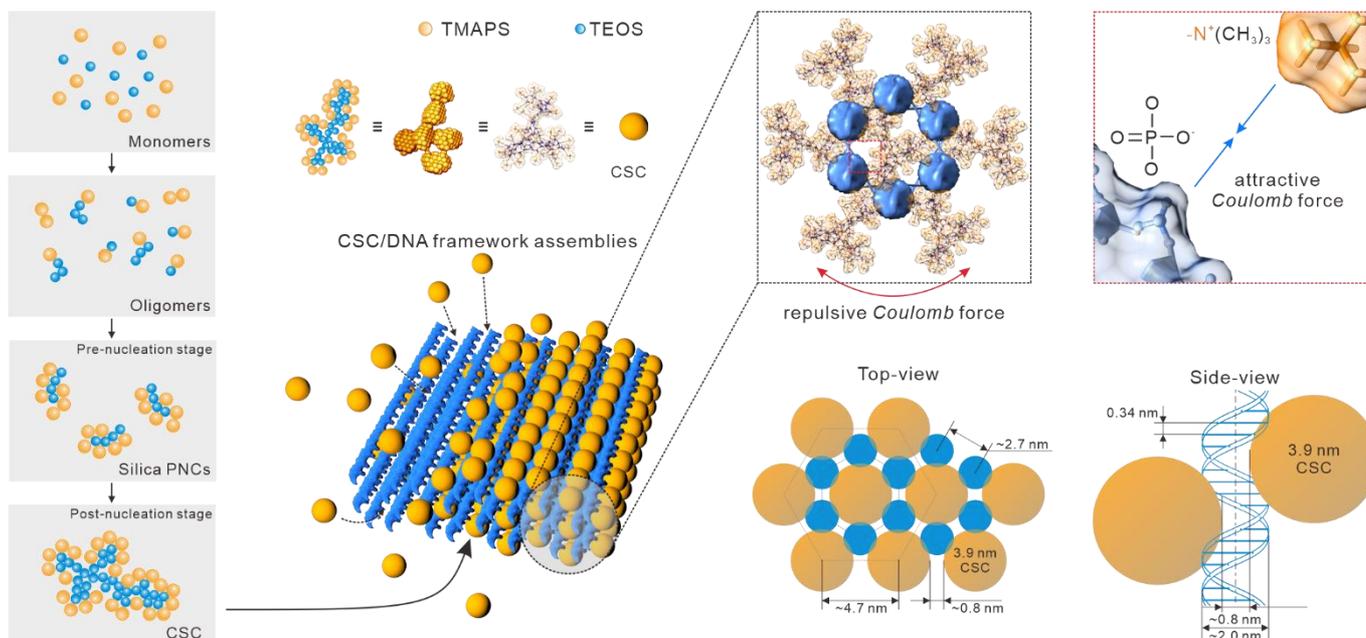

**Extended Data Fig.5 | Schematic illustration summarizing the CSC-directed silicification of programmable DNA frameworks.** We proposed a picture of organic template-directed precise amorphous silica mineralization, which starts with the formation of silica PNCs with molecular weight of ~1.2 kDa and size of ~1.3 nm. Their self-limited aggregation leads to the formation of size-controlled CSCs with typical molecular weight of ~8.2 kDa and size of ~3.9 nm. Further, these CSCs are orderly arranged by Coulomb forces on substrate frameworks.

# Supplementary Information for

# Molecular tuning of DNA framework-programmed silicification by cationic silica cluster attachment

Xinxin Jing[1,2,8], Haozhi Wang[1,8], Jianxiang Huang[3,4,8], Yingying Liu[1,8], Zimu Li[1], Jielin Chen[1], Yiqun Xu[1], Lingyun Li[1], Yunxiao Lin[1], Qinglin Xia[5], Muchen Pan[5], Yue Wang[5], Mingqiang Li[1], Ruhong Zhou[3,4,*], Xiaoguo Liu[1,*], Stephen Mann[6,7,*] and Chunhai Fan[1,*]

1. School of Chemistry and Chemical Engineering, New Cornerstone Science Laboratory, Frontiers Science Center for Transformative Molecules, Zhangjiang Institute for Advanced Study and National Center for Translational Medicine, Shanghai Jiao Tong University, Shanghai 200240, China.
2. Institute of Molecular Medicine, Shanghai Key Laboratory for Nucleic Acid Chemistry and Nanomedicine, Renji Hospital, School of Medicine, Shanghai Jiao Tong University, Shanghai, China.
3. Institute of Quantitative Biology, College of Life Sciences, and Department of Physics, Zhejiang University, Hangzhou, China.
4. Shanghai Institute for Advanced Study, Zhejiang University, Shanghai, China.
5. Division of Physical Biology, CAS Key Laboratory of Interfacial Physics and Technology, Shanghai Institute of Applied Physics, Chinese Academy of Sciences, Shanghai 201800, China.
6. School of Materials Science and Engineering, Shanghai Jiao Tong University, Shanghai 200240, China.
7. Max Planck-Bristol Centre for Minimal Biology, School of Chemistry, University of Bristol, Bristol BS8 1TS, United Kingdom.
8. These authors contribute equally, Xinxin Jing, Haozhi Wang, Jianxiang Huang and Yingying Liu.
*e-mails: rz24@columbia.edu, liuxiaoguo@sjtu.edu.cn, s.mann@bristol.ac.uk, and fanchunhai@sjtu.edu.cn

# Contents

# S1. Supplementary data for CSC

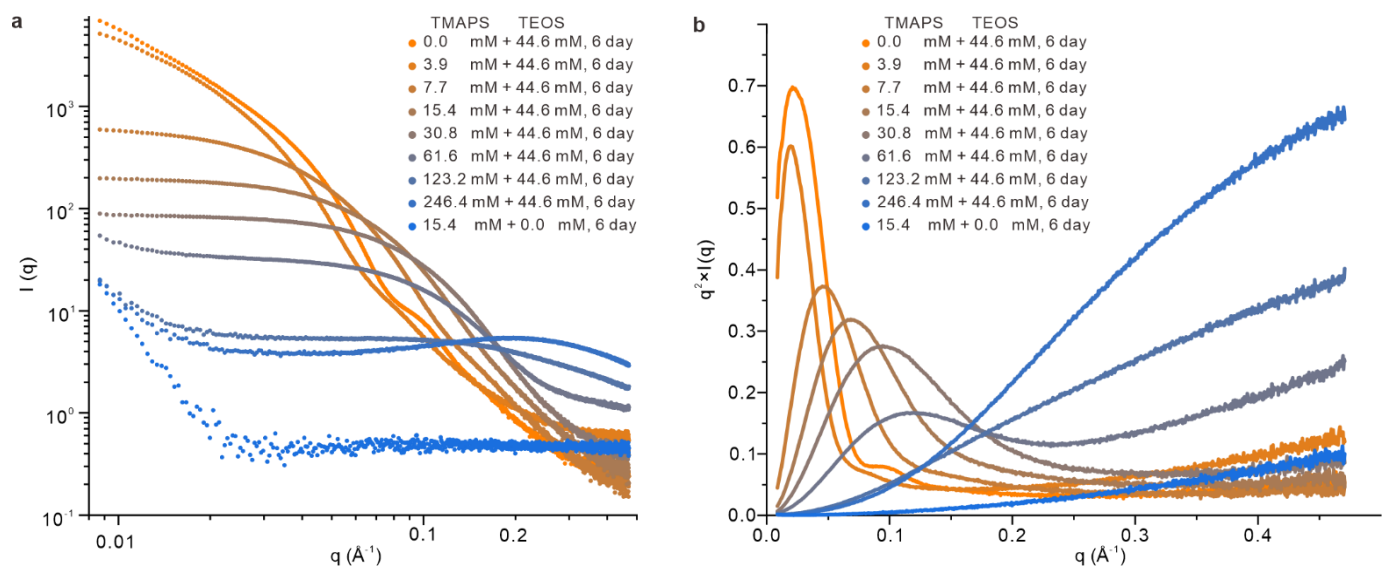

**Supplementary Fig.1 |  SAXS analysis of CSC prepared with different TMAPS/TEOS ratios.**
**a.** The slope of the SAXS profiles in the low-$q$ region decreased as the TAMPS concentrations increased, indicating a negative correlation between CSC size and TMAPS concentration. **b.** *Kratky* plots showed continuous changes of curve shapes from parabolic convergence, partial parabolic convergence to linear divergence, revealing a decrease in the compactness of CSC with increasing TMAPS concentration. The *Kratky* plot of 15.4 mM TMAPS + 0.0 mM TEOS exhibited a linear divergence feature, suggesting that CSC formation was inhibited in the absence of TEOS.

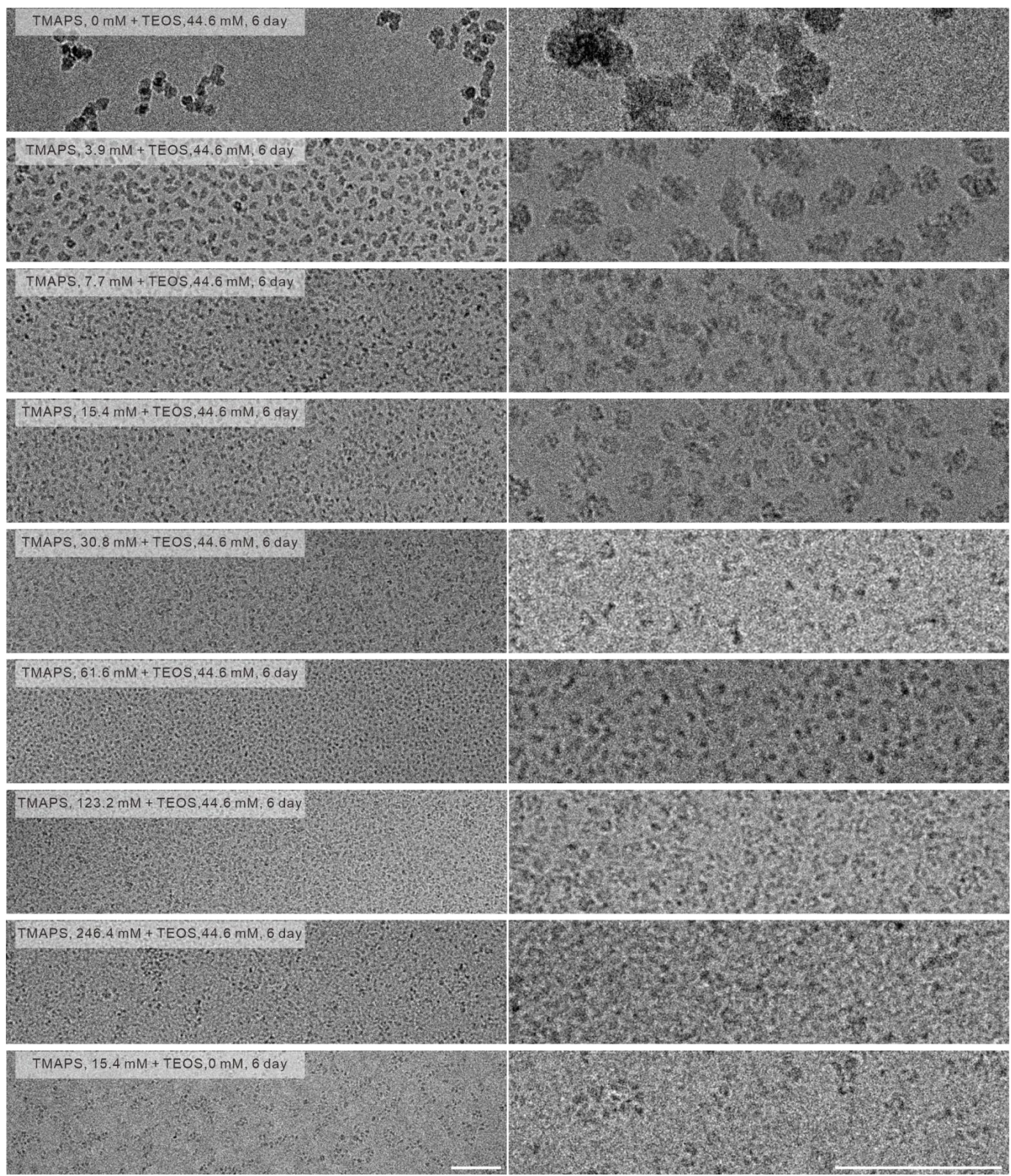

**Supplementary Fig.2 |    Cryo-EM images of CSC prepared with different TMAPS/TEOS ratios.**
Corresponding cryo-EM images of samples characterized by SAXS in Fig. S1. The images showed that the sizes of the CSC decrease as the concentration of TMAPS increases (TEOS = 44.6 mM). Inhomogeneous silica clusters occurred when the TMAPS concentration was higher than 123.2 mM, or in the absence of TEOS. Scale bars, 50 nm. See also Supplementary Figure 3.

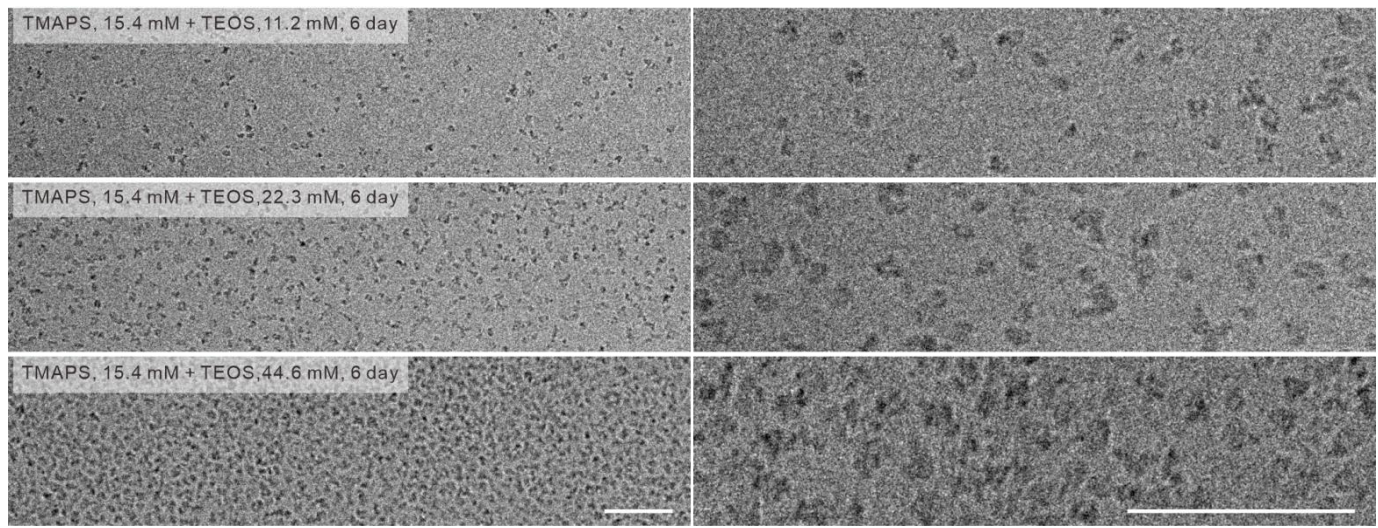

**Supplementary Fig.3 | Cryo-EM images of CSC prepared with different TMAPS/TEOS ratios.**
The results revealed that the sizes of CSC mainly depended on the concentrations of TMAPS rather than TEOS. As TEOS concentration increased from 11.2 mM to 22.3 mM and 44.6 mM, there was no significant change in the sizes of the CSC. An increase in the particle number was observed as the TEOS concentrated increased. Scale bars, 50 nm.

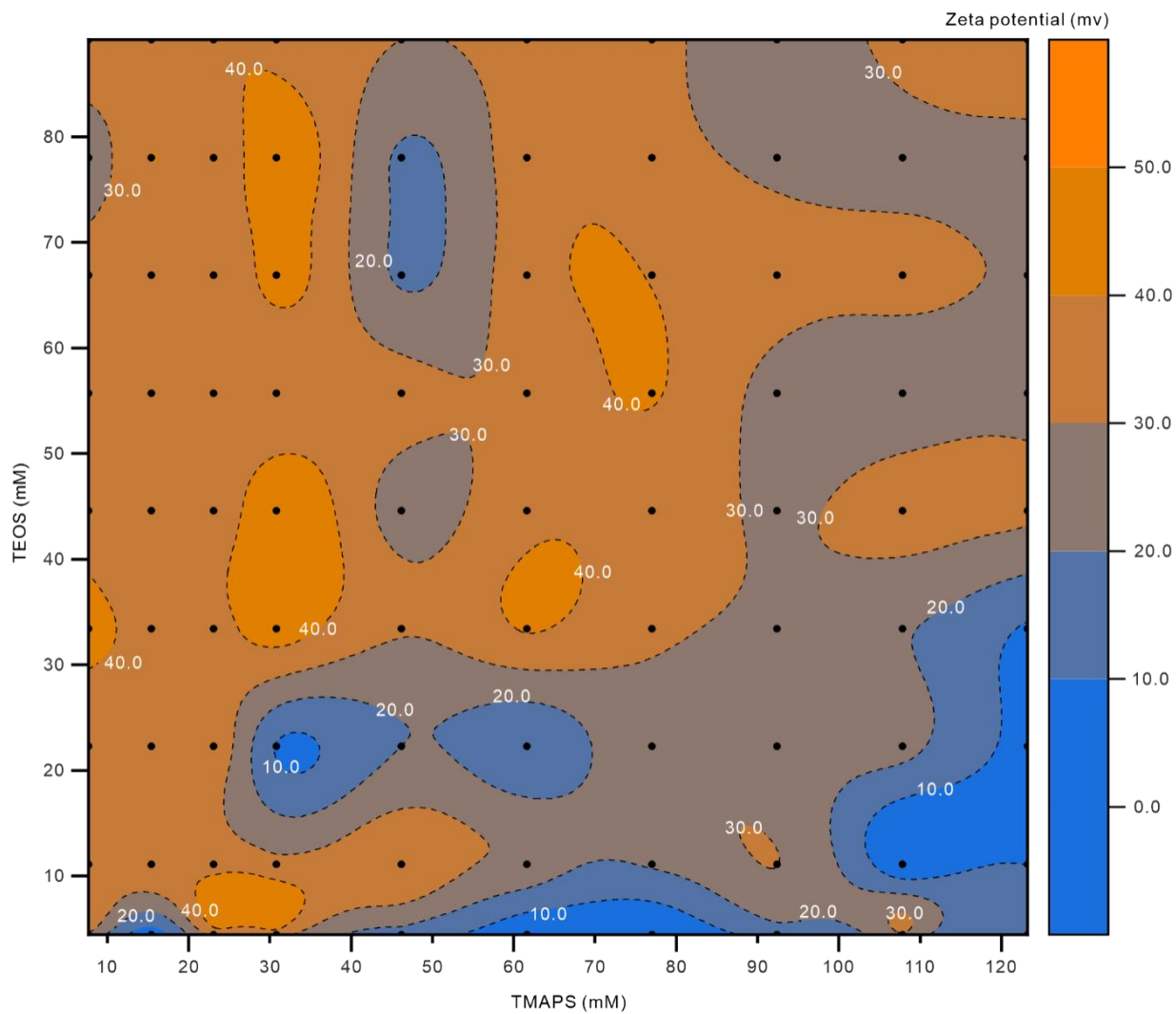

**Supplementary Fig.4 | ζ-potential phase diagram of CSC.**
Most CSC exhibited positive ζ-potentials in the range of 30 mV to 40 mV, owing to the pendent trimethylammonium side chain (-$N^+(CH_3)_3$) of TMAPS. The ζ-potentials below 10 mV occurred frequently when the concentration of TMAPS was extremely high (123.2 mM or 246.2 mM). This result mainly derived from the inhibitory effect of high concentrations of TMAPS on the formation of silica pre-nucleation clusters. The black dots are recorded experimental data. Dotted contour lines indicate the boundaries of different ζ-potentials, as marked by white letters.

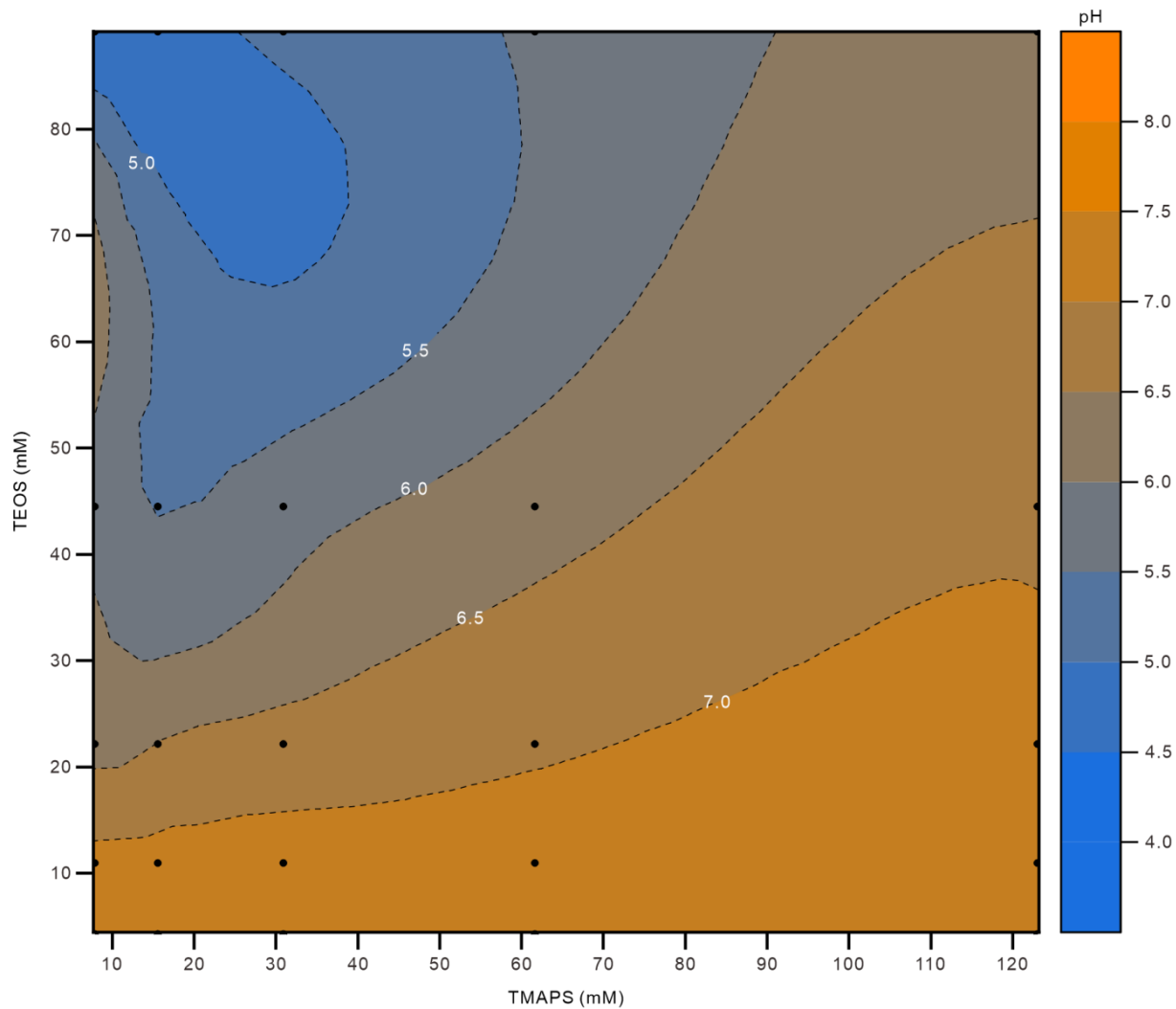

**Supplementary Fig.5 | pH binary phase diagram of CSC.**
A gradual decrease of pH was observed as the concentration of TMAPS decreased or the concentration of TEOS increased due to consumption of hydroxide ions during the sol-gel process. The initial mixtures of TMAPS+TEOS in TE-MgCl$_2$ buffer all have a pH of 8.0. The black dots are recorded experimental data. Dotted contour lines indicated the boundaries of different pH, as marked by white letters.

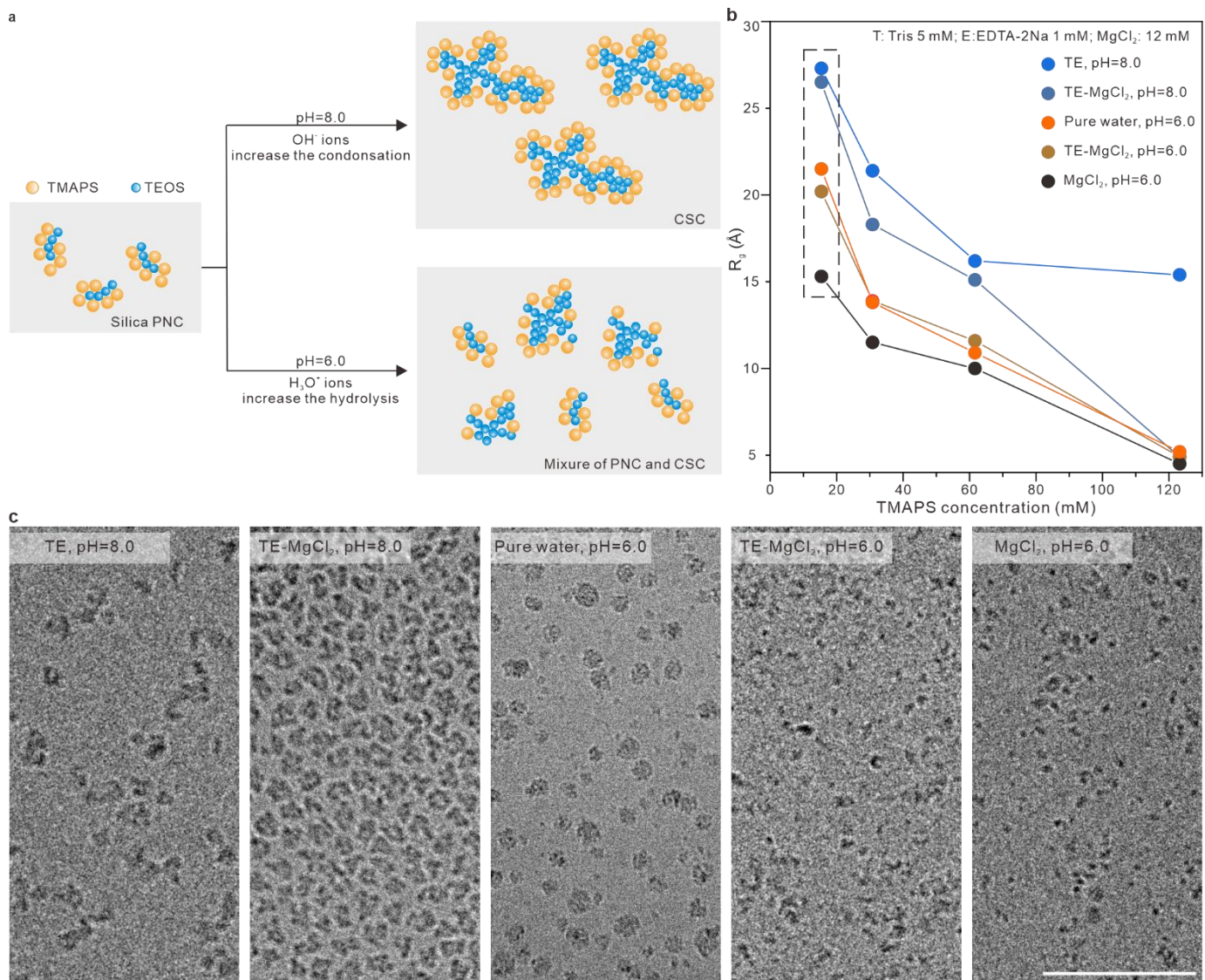

**Supplementary Fig.6 | CSC formation under different pH conditions.**

**a, b.** Schematics and corresponding SAXS $R_g$ data of CSC formation under different pH conditions. Previous studies indicated that $H_3O^+$ ions in the solution increase the hydrolysis rate, whereas $OH^-$ ions increase the condensation rate[33]. Homogeneous CSC are produced only when the pH of solution is ~8.0. In contrast, at pH ~6.0, a mixture of silica pre-nucleation cluster (PNC) and CSC are produced due to the hydrolysis rate being higher than the condensation rate. **c.** Five typical samples (dashed box in **b**) were imaged by cryo-EM. In general, CSC were only produced in TE-MgCl$_2$ buffer (pH=8.0). TE buffer (pH=8.0) without MgCl$_2$ induced further aggregation of the CSC, while buffers at pH 6.0 resulted in inhomogeneous clusters with reduced dimensions compared with the CSC. Scale bar, 50 nm.

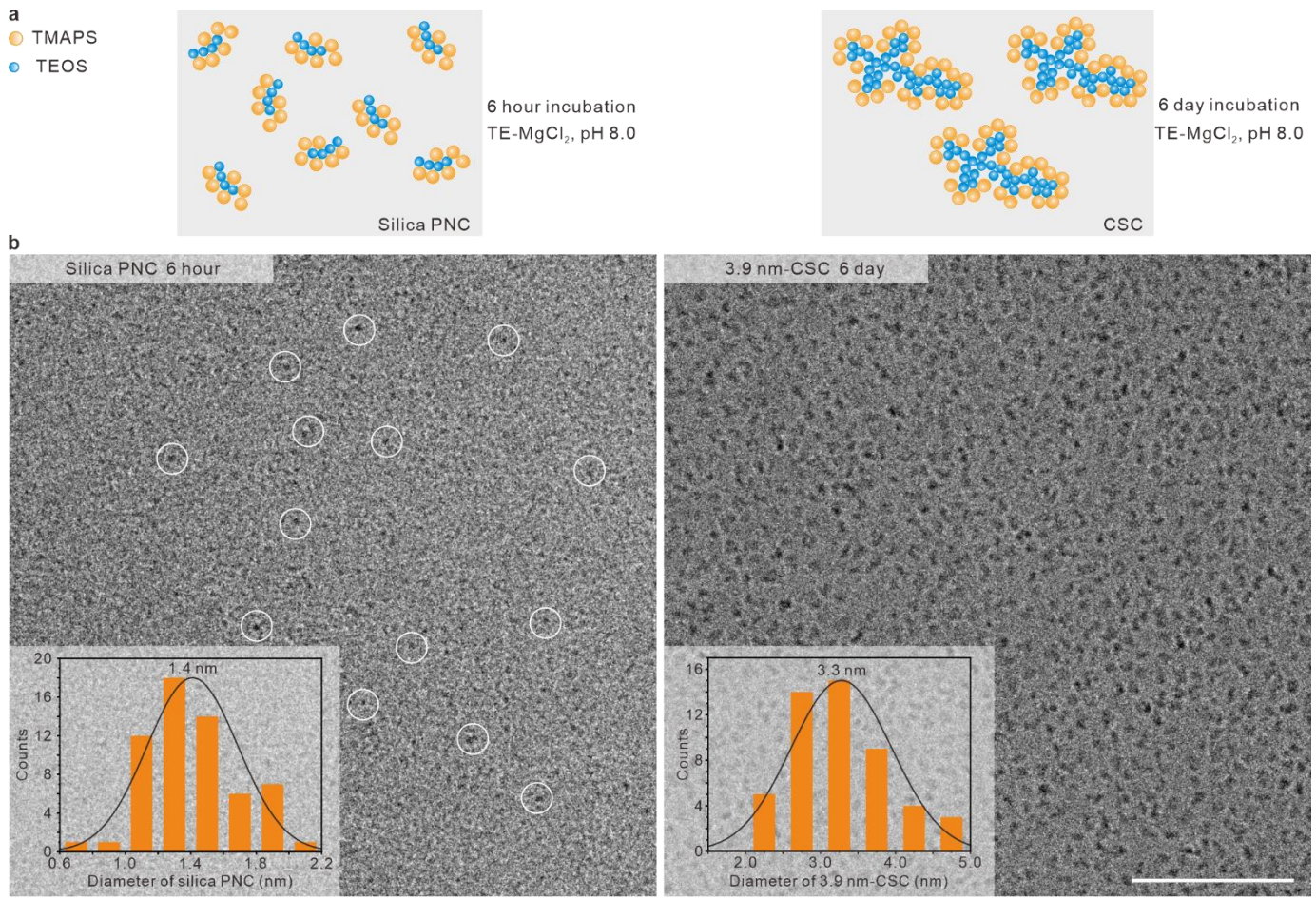

**Supplementary Fig.7 | Cryo-EM images of silica prenucleation clusters (PNCs) and CSC.**
**a.** Schematics of the formation of silica PNCs after 6-hour incubation and CSC after 6-day incubation. **b.** Cryo-EM images of silica PNCs and CSC. Small silica PNCs can be clearly observed in the early growth stage of CSC (left panel). Representative silica PNCs were marked by white circles. The bottom left insets showed the statistics of the silica PNCs and CSC sizes; ~1.4 nm for silica PNCs and ~3.3 nm for CSC. Scale bar, 50 nm.

**Supplementary Table 1 | Averaged sizes of CSC using different characterization methods.**

| Characterization method | Diameter (Å) | | | |
| --- | --- | --- | --- | --- |
| | 15.4 mM TMAPS +44.6 mM TEOS | 30.8 mM TMAPS +44.6 mM TEOS | 61.6 mM TMAPS +44.6 mM TEOS | 123.2 mM TMAPS +44.6 mM TEOS |
| DLS | 72.3 | 45.0 | 36.0 | 14.2 |
| Cryo-EM | 60.2 | 50.0 | 32.8 | 27.3 |
| SAXS ($2\times(5/3)^{0.5}\times R_g$) | 68.4 | 47.2 | 39.0 | 12.6 |
| Name of CSC | 6.8 nm-CSC | 4.7 nm-CSC | 3.9 nm-CSC | 1.3 nm-CSC |

**Supplementary Table 2 | DLS diameters (nm) of CSC prepared with different stoichiometric ratios.**

| TEOS(mM) \ TMAPS(mM) | 0.0 | 3.9 | 7.7 | 15.4 | 23.1 | 30.8 | 46.2 | 61.6 | 77.0 | 92.4 | 107.8 | 123.2 |
|---|---|---|---|---|---|---|---|---|---|---|---|---|
| 0.0 | 0.0 | 0.8 | 0.7 | 0.8 | \ | 0.7 | \ | 0.8 | \ | \ | \ | 0.8 |
| 2.2 | 0.7 | 0.7 | 0.7 | 0.7 | \ | 0.7 | \ | 0.7 | \ | \ | \ | 0.8 |
| 4.5 | 60.7 | 16.7 | 7.8 | 5.7 | 4.3 | 4.7 | 0.7 | 0.7 | 0.8 | 0.8 | 0.7 | 0.8 |
| 11.2 | 18.0 | 11.9 | 6.7 | 5.6 | 4.8 | 4.8 | 4.7 | 3.9 | 0.7 | 0.9 | 0.7 | 0.7 |
| 22.3 | 22.2 | 14.3 | 7.9 | 5.6 | 4.8 | 4.8 | 3.8 | 4.2 | 3.4 | 0.8 | 0.7 | 0.7 |
| 33.5 | \ | \ | 6.6 | 5.3 | 5.0 | 4.7 | 3.0 | 4.5 | 4.4 | 5.3 | 0.8 | 0.8 |
| 44.6 | 34.1 | 23.4 | 7.6 | 5.5 | 4.4 | 5.7 | 3.3 | 4.6 | 4.8 | 4.3 | 0.6 | 0.8 |
| 55.8 | \ | \ | 8.9 | 6.0 | 4.6 | 4.9 | 3.7 | 3.1 | 4.5 | 4.3 | 0.8 | 1.0 |
| 66.9 | \ | \ | 6.8 | 6.1 | 5.1 | 5.9 | 4.0 | 4.1 | 4.2 | 4.5 | 0.9 | 0.8 |
| 78.1 | \ | \ | 9.3 | 6.4 | 5.2 | 5.8 | 4.0 | 4.5 | 4.4 | 3.8 | 3.7 | 4.5 |
| 89.2 | 31.3 | 16.3 | 9.3 | 6.4 | 6.6 | 6.1 | 4.0 | 4.5 | 3.2 | 3.8 | 3.9 | 4.4 |

**Supplementary Table 3 | ζ-potential (mV) of CSC prepared with different stoichiometric ratios.**

| TMAPS(mM)<br>TEOS(mM) | 7.7 | 15.4 | 23.1 | 30.8 | 46.2 | 61.6 | 77.0 | 92.4 | 107.8 | 123.2 |
|---|---|---|---|---|---|---|---|---|---|---|
| **4.5** | 28.1 | 4.9 | 33.8 | 35.2 | 15.0 | 1.8 | 0.3 | 14.3 | 29.1 | 11.5 |
| **11.2** | 32.4 | 37.6 | 37.0 | 33.2 | 38.5 | 27.2 | 21.1 | 30.1 | 1.9 | 9.9 |
| **22.3** | 35.1 | 34.6 | 37.8 | 9.5 | 20.0 | 11.3 | 26.9 | 25.1 | 23.5 | 7.6 |
| **33.5** | 41.7 | 37.7 | 36.6 | 44.7 | 30.9 | 40.8 | 31.6 | 28.8 | 21.0 | 9.0 |
| **44.6** | 37.2 | 32.7 | 33.0 | 47.7 | 25.5 | 37.2 | 34.7 | 29.1 | 34.5 | 31.9 |
| **55.8** | 32.6 | 37.9 | 34.8 | 34.5 | 33.6 | 32.0 | 40.7 | 26.6 | 24.4 | 26.8 |
| **66.9** | 35.1 | 37.8 | 30.7 | 43.5 | 18.4 | 36.3 | 39.7 | 33.9 | 32.9 | 29.2 |
| **78.1** | 23.0 | 40.1 | 33.2 | 47.1 | 18.4 | 36.1 | 35.0 | 27.2 | 26.8 | 27.4 |
| **89.2** | 40.0 | 36.0 | 39.7 | 38.7 | 30.0 | 34.8 | 34.0 | 20.9 | 34.4 | 38.1 |

**Supplementary Table 4 | Integrated intensity ratios of $^{29}$Si NMR peaks in different-size CSC.**

| $^{29}$Si NMR peak | Chemical shift (ppm) | Integrated intensity ratios (%) | | | | | | |
|---|---|---|---|---|---|---|---|---|
| | | TEOS only | 11.9-nm CSC | 6.8-nm CSC | 4.7-nm CSC | 3.9-nm CSC | 1.3-nm CSC | TMAPS only |
| $T^0$ | −41 | \ | \ | \ | \ | \ | \ | \ |
| $T^1$ | −50 | \ | \ | \ | \ | 0.4 | \ | 0.2 |
| $T^2$ | −59 | \ | 15.0 | 6.8 | 36.0 | 22.7 | 57.9 | 59.5 |
| $T^3$ | −68 | \ | 27.1 | 27.6 | 34.1 | 31.5 | 30.7 | 40.3 |
| $Q^0$ | −73 | \ | \ | \ | \ | \ | \ | \ |
| $Q^1$ | −82 | \ | \ | \ | \ | \ | \ | \ |
| $Q^2$ | −91 | \ | 0.5 | 2.3 | \ | 2.2 | \ | \ |
| $Q^3$ | −100 | 20.5 | 43.4 | 45.9 | 22.7 | 34.0 | 11.4 | \ |
| $Q^4$ | −109 | 79.5 | 14.0 | 17.4 | 7.2 | 9.2 | \ | \ |

# S2. Supplementary data for CSC-directed silicification of programmable DNA frameworks

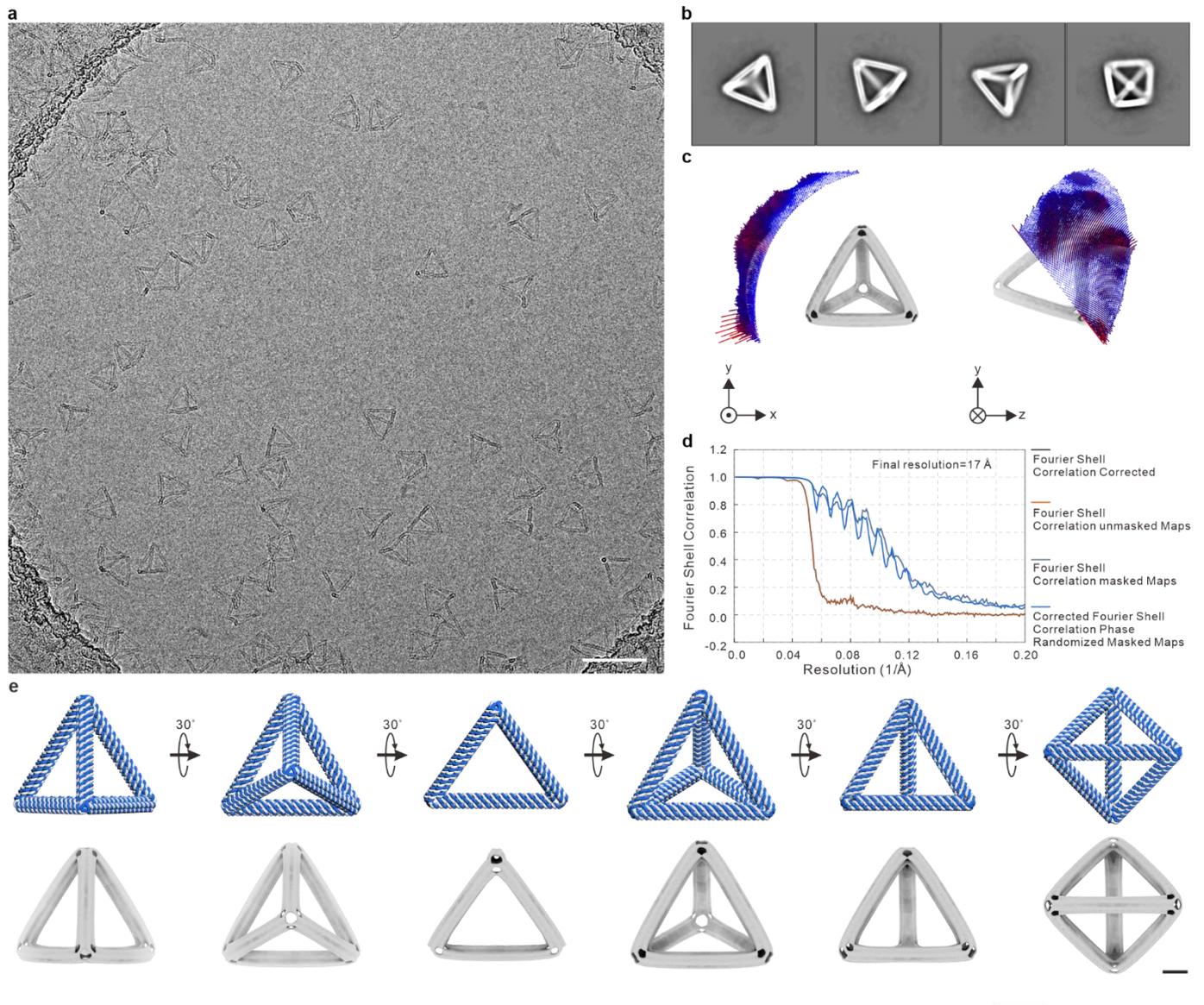

**Supplementary Fig.8 |   Cryo-EM single-particle analysis of native tetrahedron DNA origami.**
**a.** Exemplary micrograph. Scale bar, 100 nm. **b.** Representative 2D class averages. **c.** Histogram representing the orientational distribution of particles. To reduce computational complexity, T symmetry was used in the 3D classifications and refinements. **d.** Fourier shell correlation plot. **e.** Six different views of the electron density map. Scale bar, 10 nm.

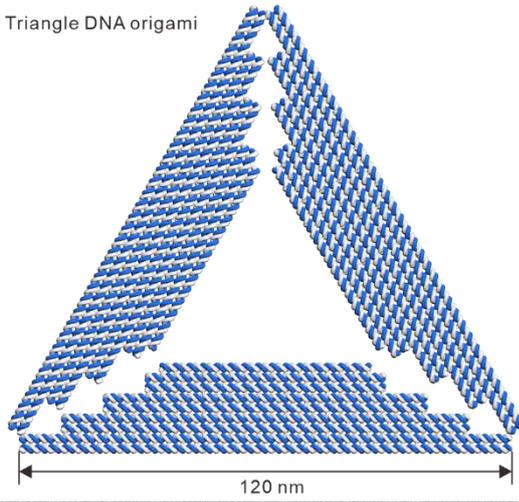
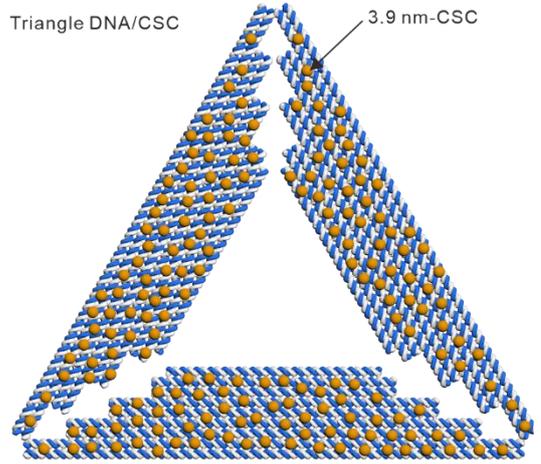
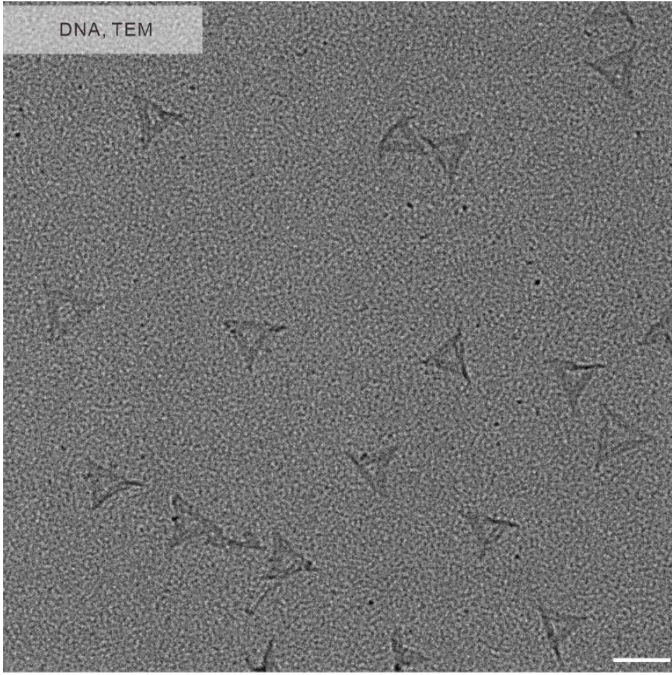
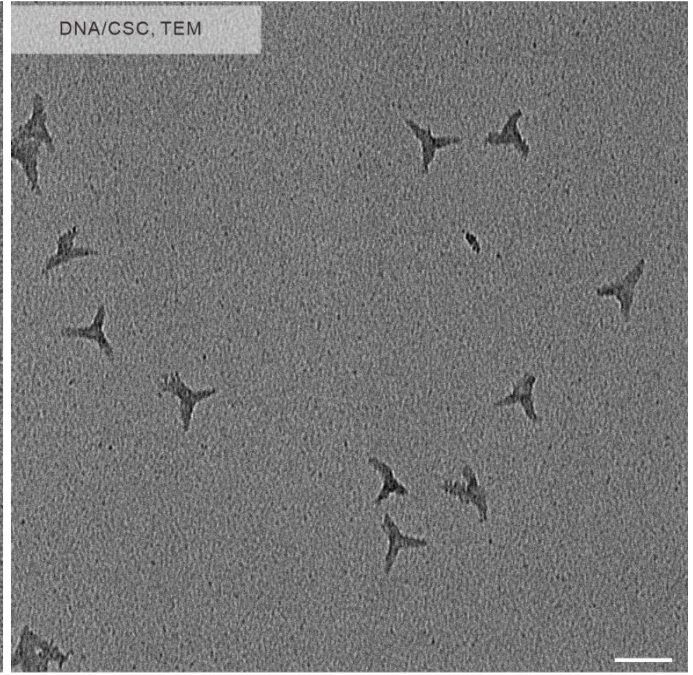
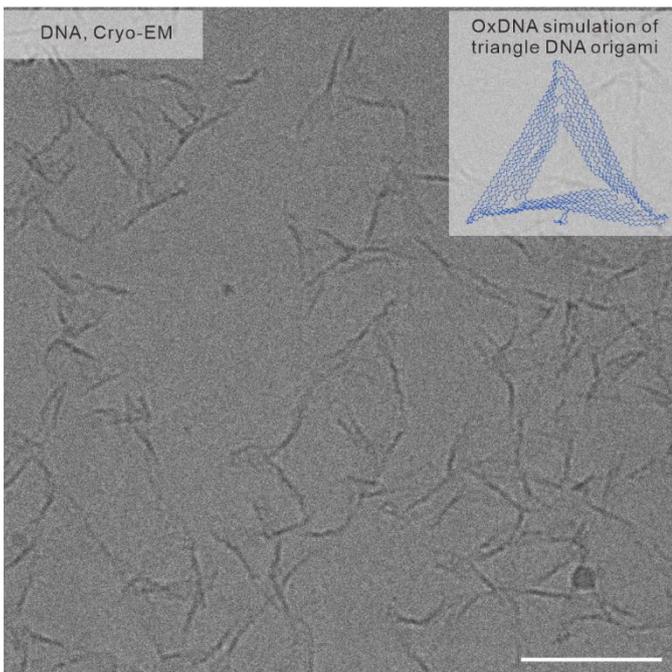
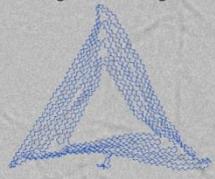
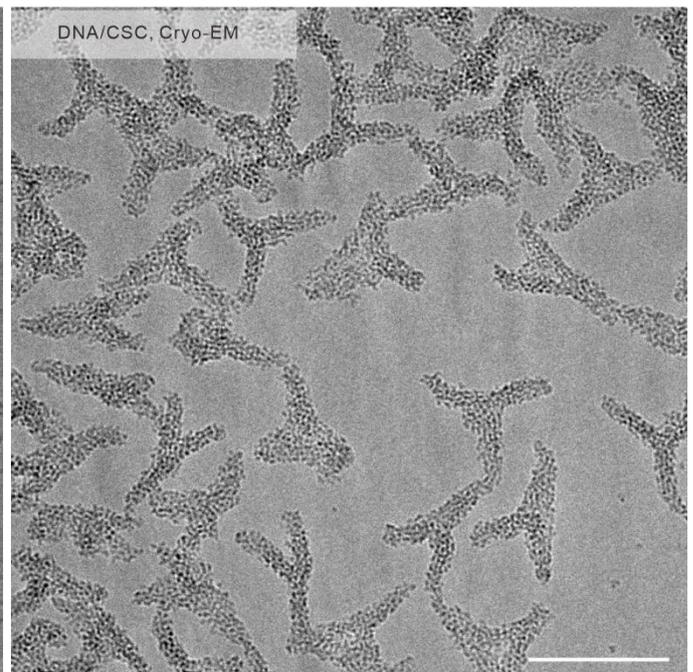

**Supplementary Fig.9 | TEM and cryo-EM images of triangle DNA origami sample group.**
The inset in the cryo-EM image is the OxDNA simulation of triangle DNA origami, illustrating the twisted conformation of the initially planar triangle DNA origami, consistent with the cryo-EM images. Scale bars, 100 nm.

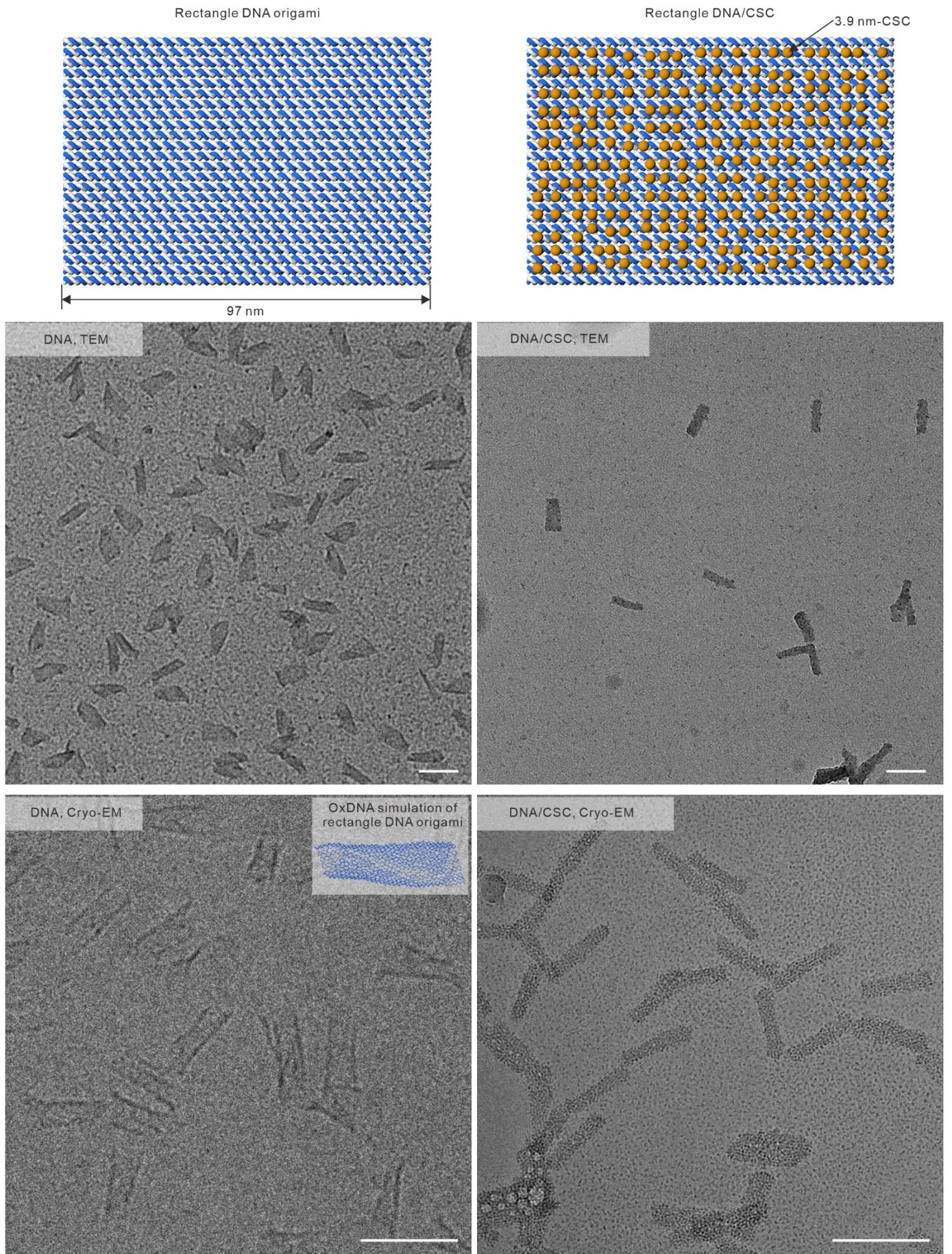

**Supplementary Fig.10 | TEM and cryo-EM images of rectangle DNA origami sample group.**
Rectangle DNA origami and DNA/CSC superstructure have a twisted conformation. Inset showed a barrel-shaped conformation of rectangle DNA origami by OxDNA simulation that was consistent with the cryo-EM images. Scale bars, 100 nm.

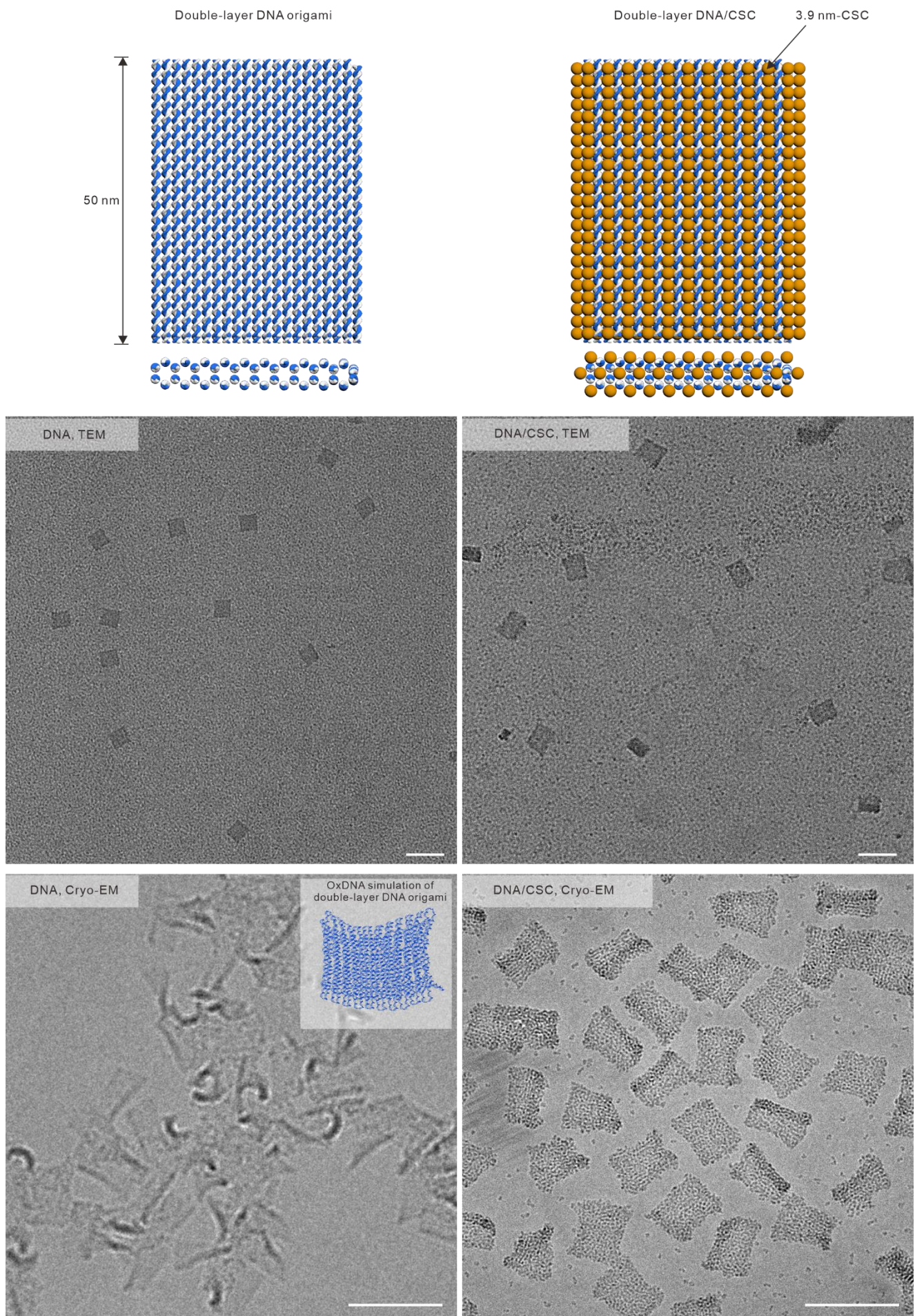

**Supplementary Fig.11 | TEM and cryo-EM images of double-layer DNA origami sample group.**
Inset showed a twisted conformation of double-layer DNA origami by OxDNA simulation, consistent with the cryo-EM images. Scale bars, 100 nm.

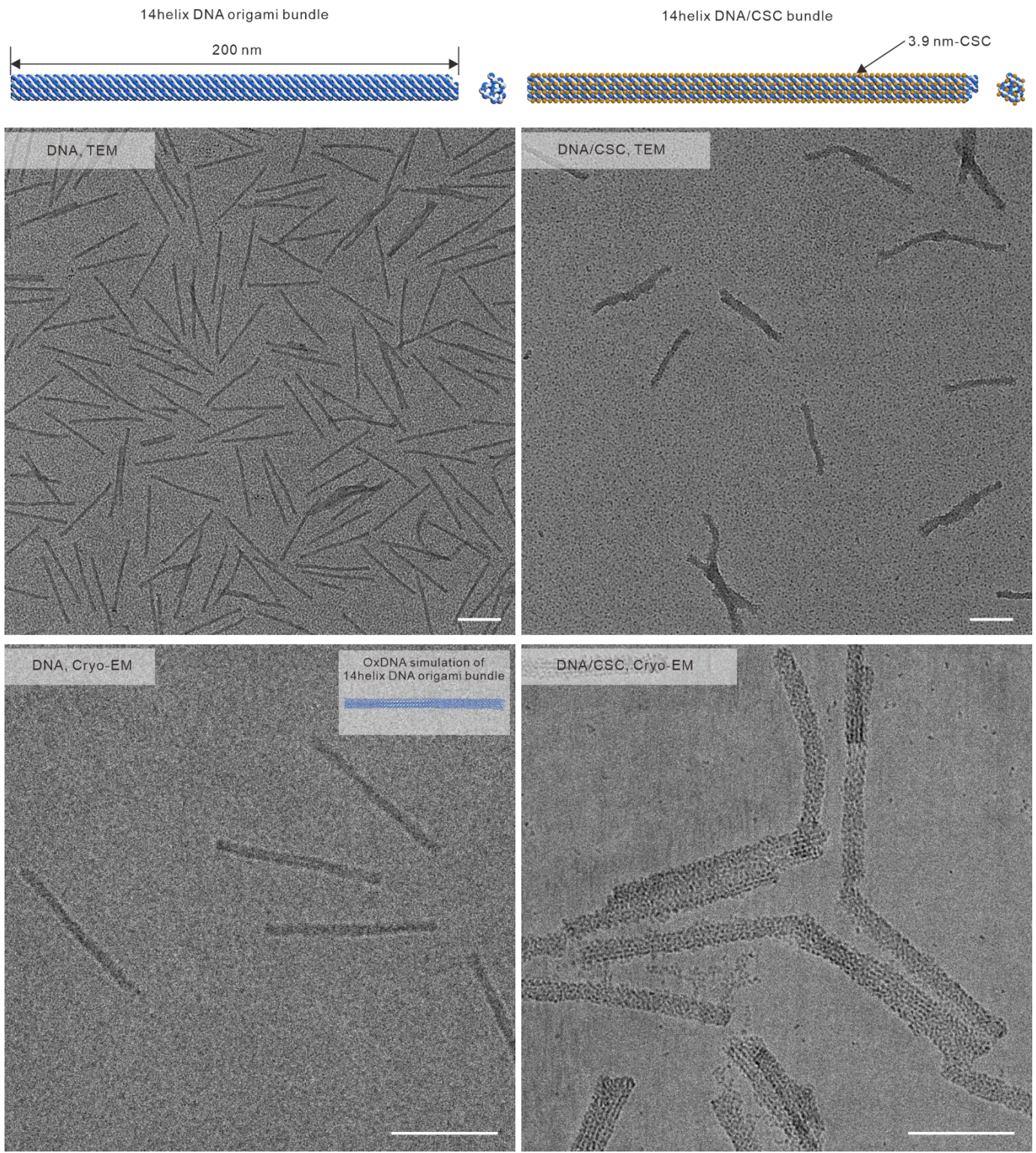

**Supplementary Fig.12 | TEM and cryo-EM images of 14-helix DNA origami sample group.**
Inset showed rigid conformation of 14helix DNA origami bundle by OxDNA simulation, consistent with the cryo-EM images. Scale bars, 100 nm.

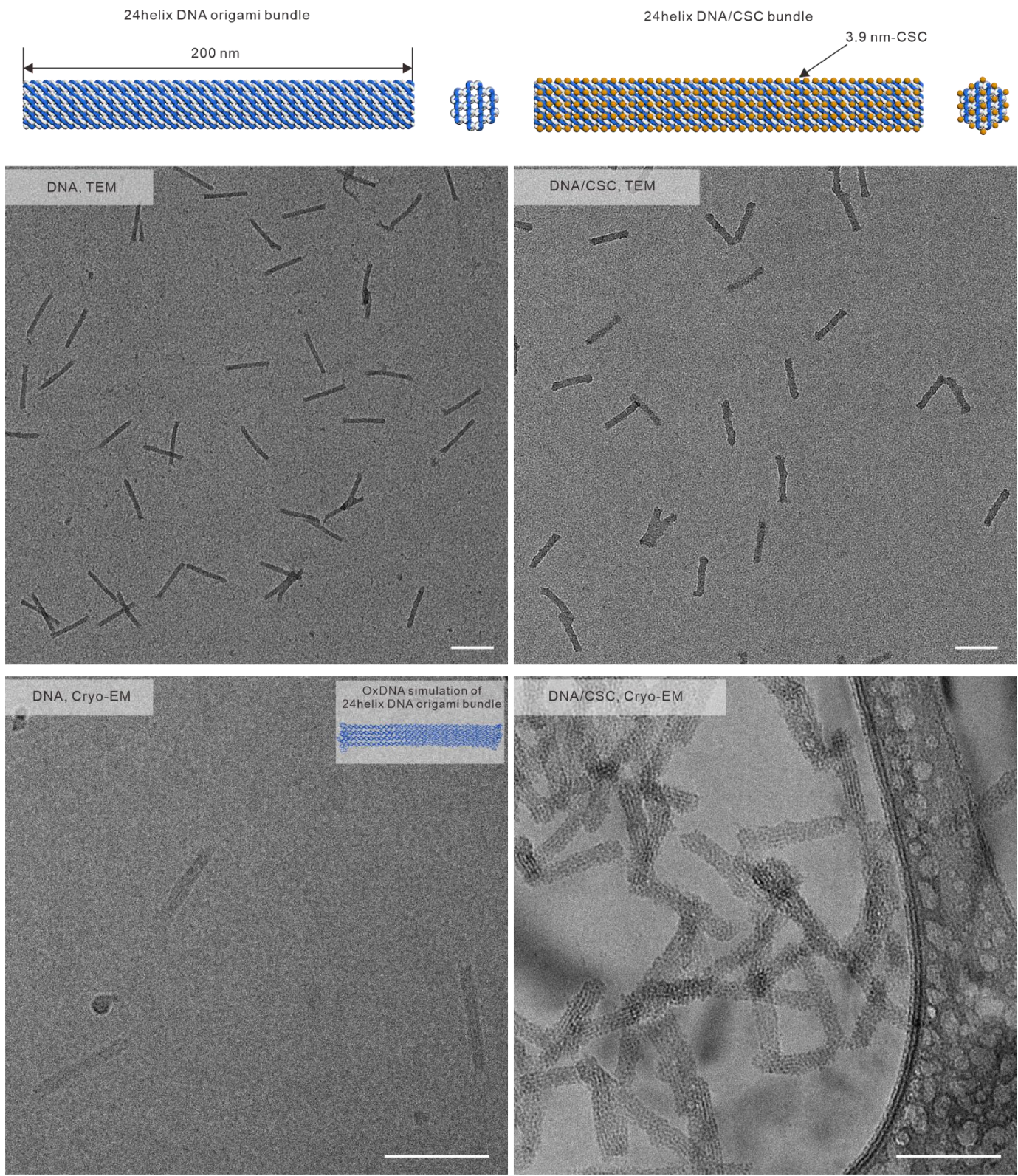

**Supplementary Fig.13 | TEM and cryo-EM images of 24-helix DNA origami sample group.**
Inset showed rigid conformation of 24helix DNA origami bundle by OxDNA simulation, consistent with the cryo-EM images. Scale bars, 100 nm.

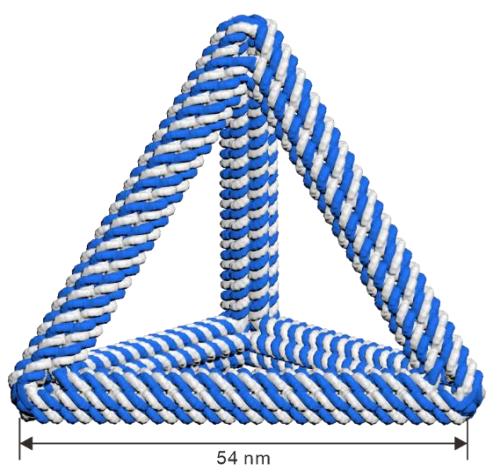
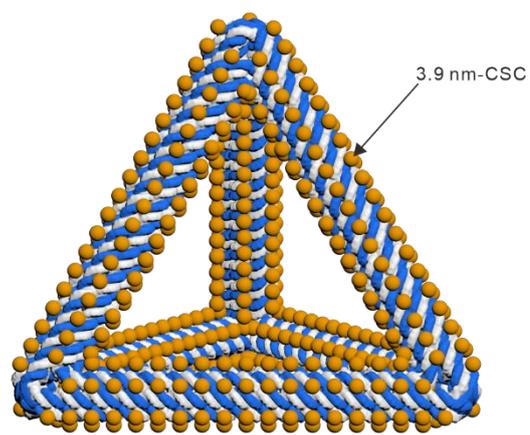
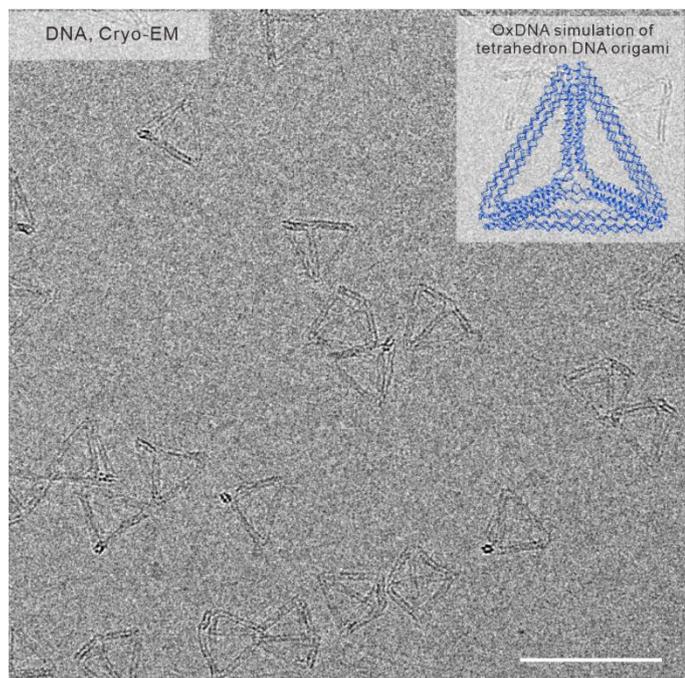
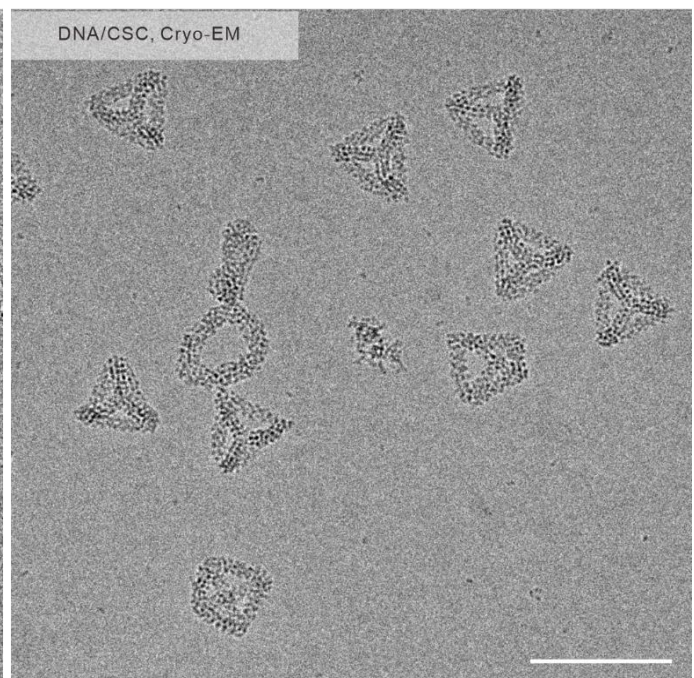

**Supplementary Fig.14 | Cryo-EM images of tetrahedral DNA origami sample group.**
Inset showed rigid conformation of tetrahedron DNA origami by OxDNA simulation, consistent with the cryo-EM images. Scale bars, 100 nm.

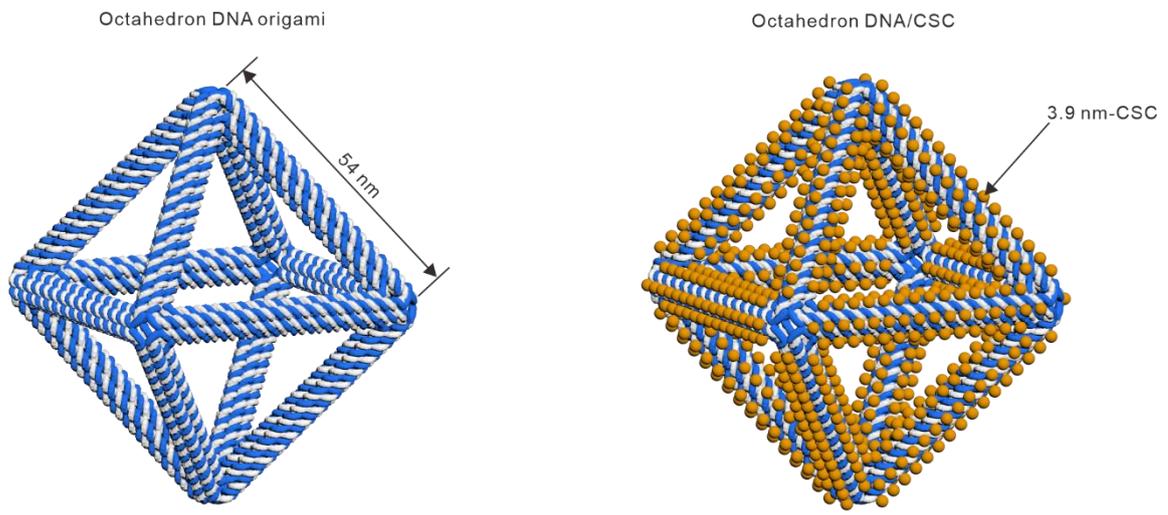

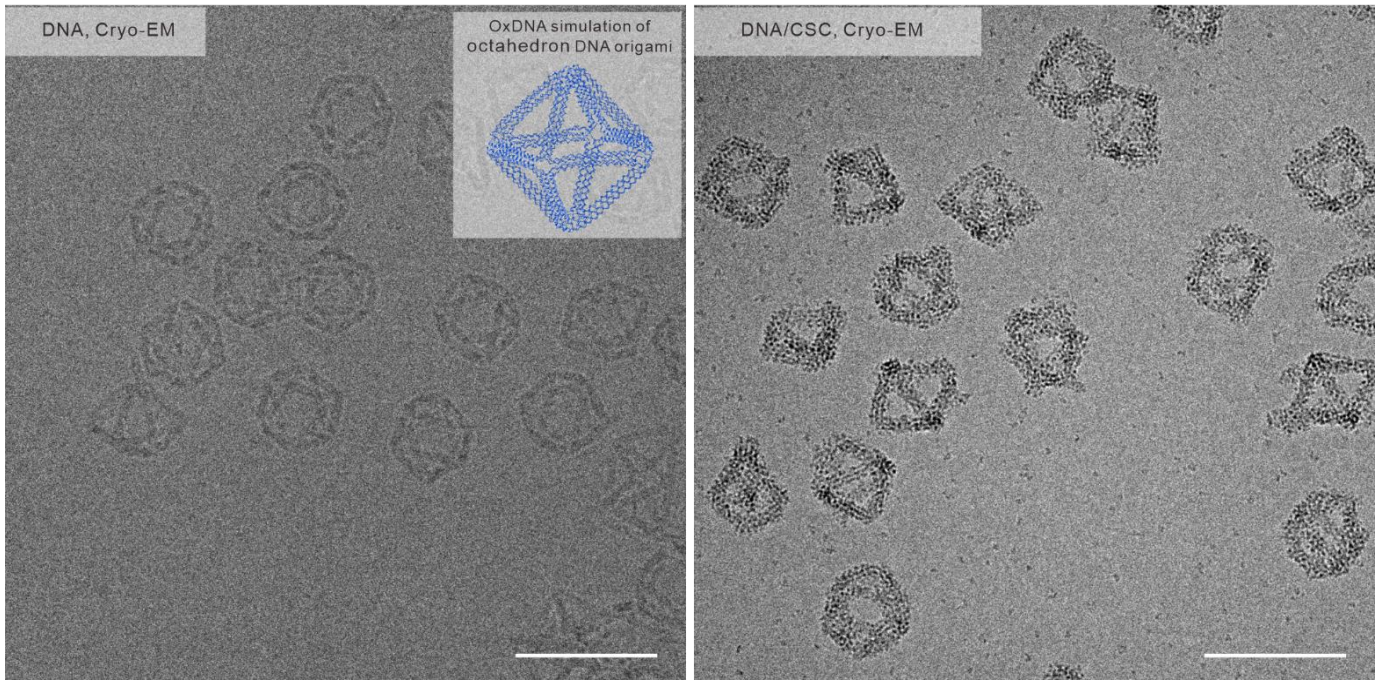

**Supplementary Fig.15 | Cryo-EM images of octahedral DNA origami sample group.**
Inset showed rigid conformation of octahedron DNA origami by OxDNA simulation, consistent with the cryo-EM images. Scale bars, 100 nm.

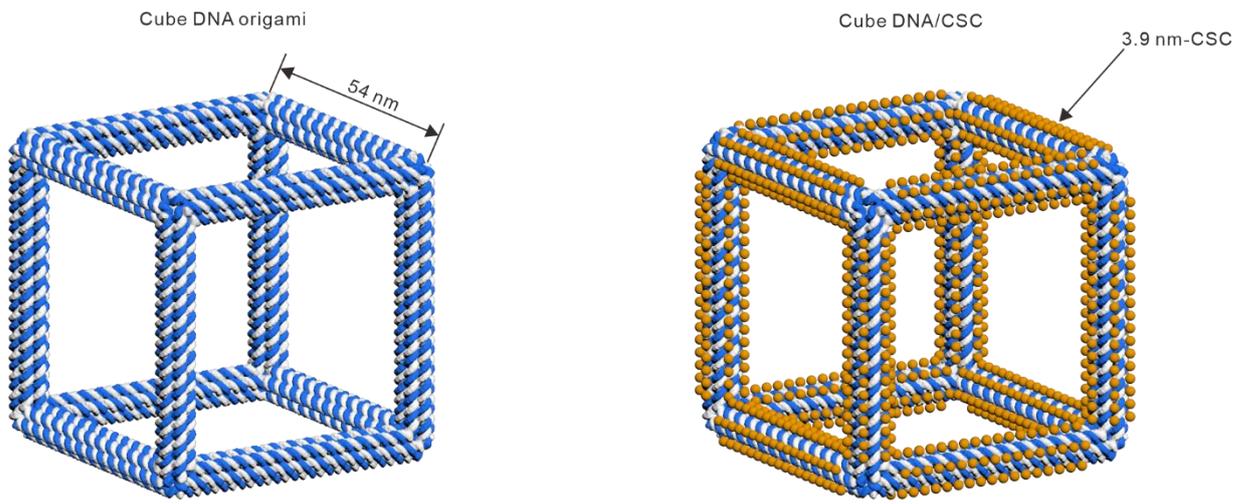
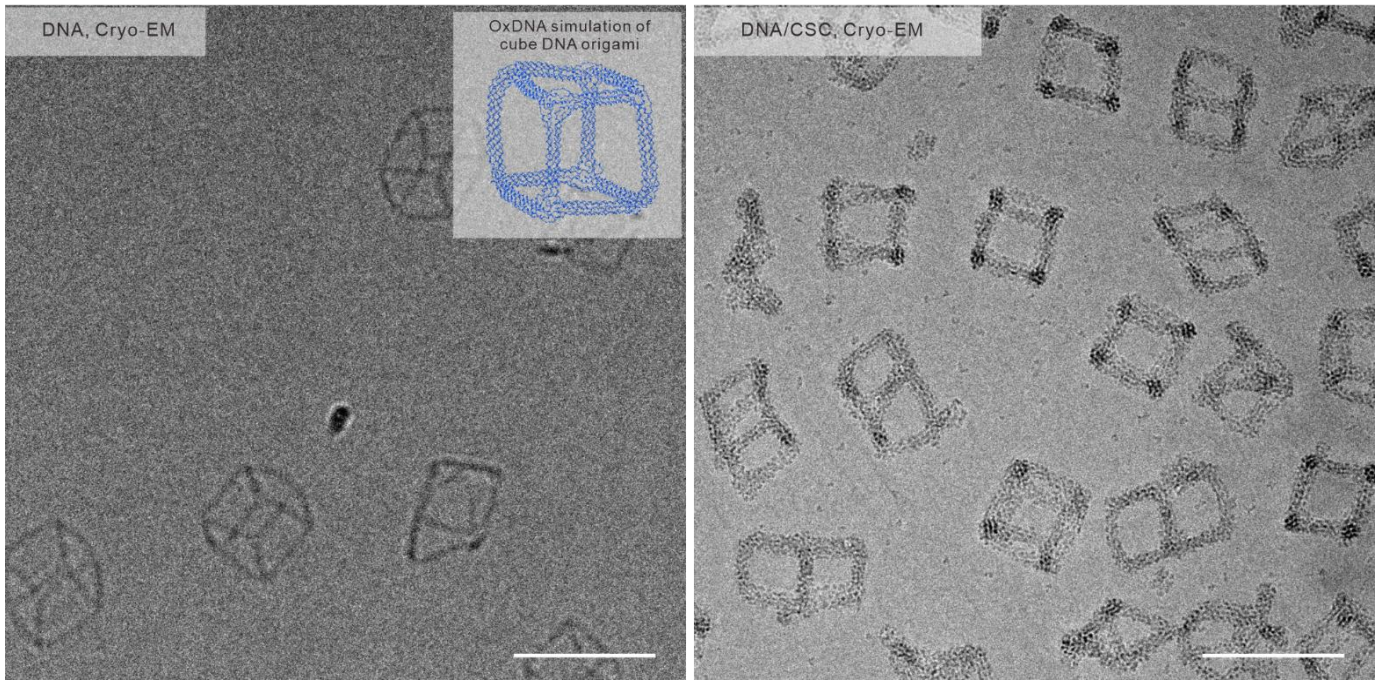

**Supplementary Fig.16 | Cryo-EM images of cubic DNA origami sample group.**
Inset showed rigid conformation of cube DNA origami by OxDNA simulation, consistent with the cryo-EM images. Scale bars, 100 nm.

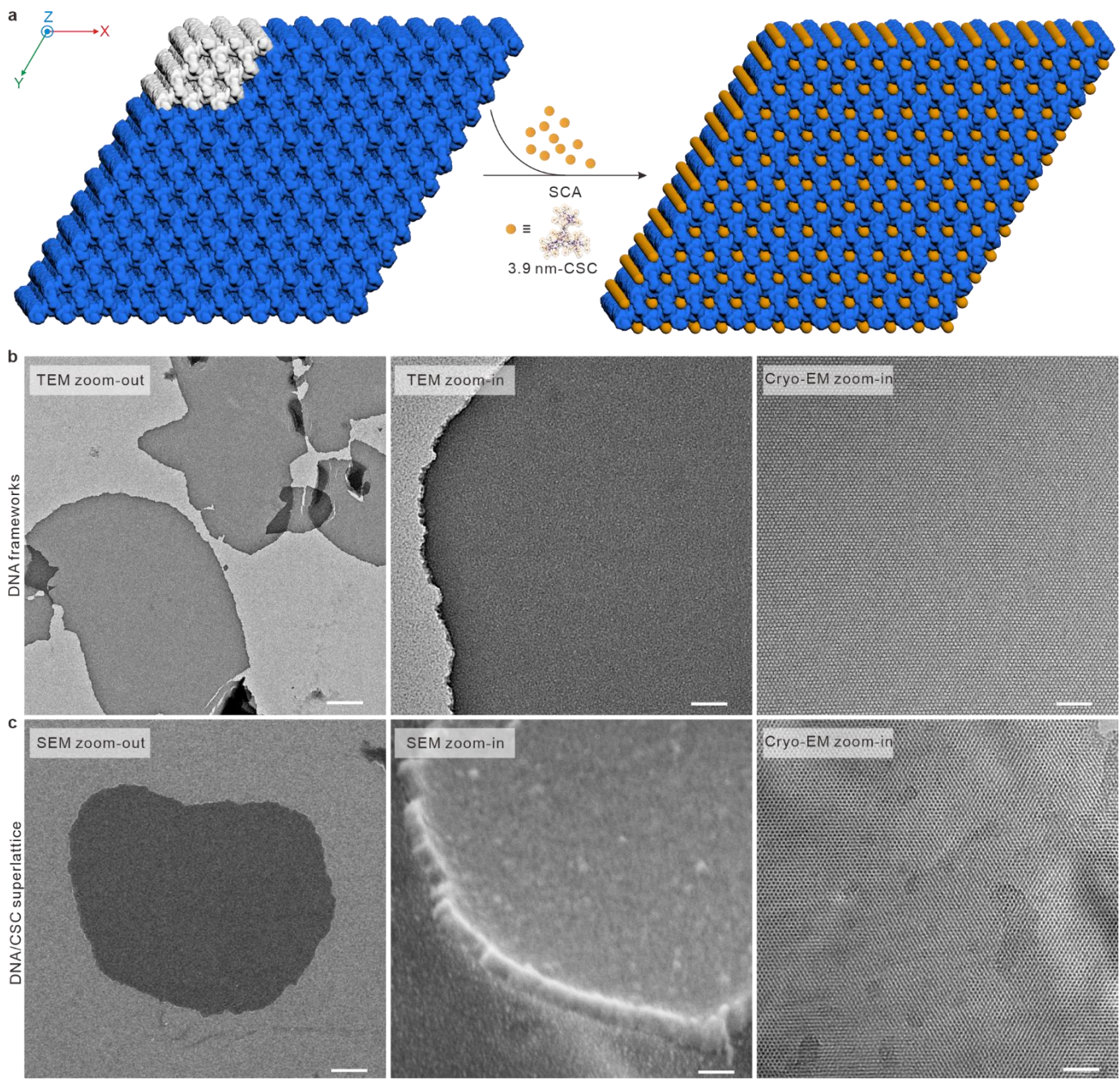

**Supplementary Fig.17 | TEM/SEM/cryo-EM images of *p6mm* 2D DNA and DNA/CSC superstructure.**
**a.** Formation of *p6mm* 2D DNA/CSC superstructure by silica cluster attachment (SCA). **b.** TEM/cryo-EM images of *p6mm* 2D DNA lattice. **c.** SEM/cryo-EM images of *p6mm* 2D DNA/CSC superstructure. Scale bars, from left to right, 500 nm, 100 nm and 40 nm.

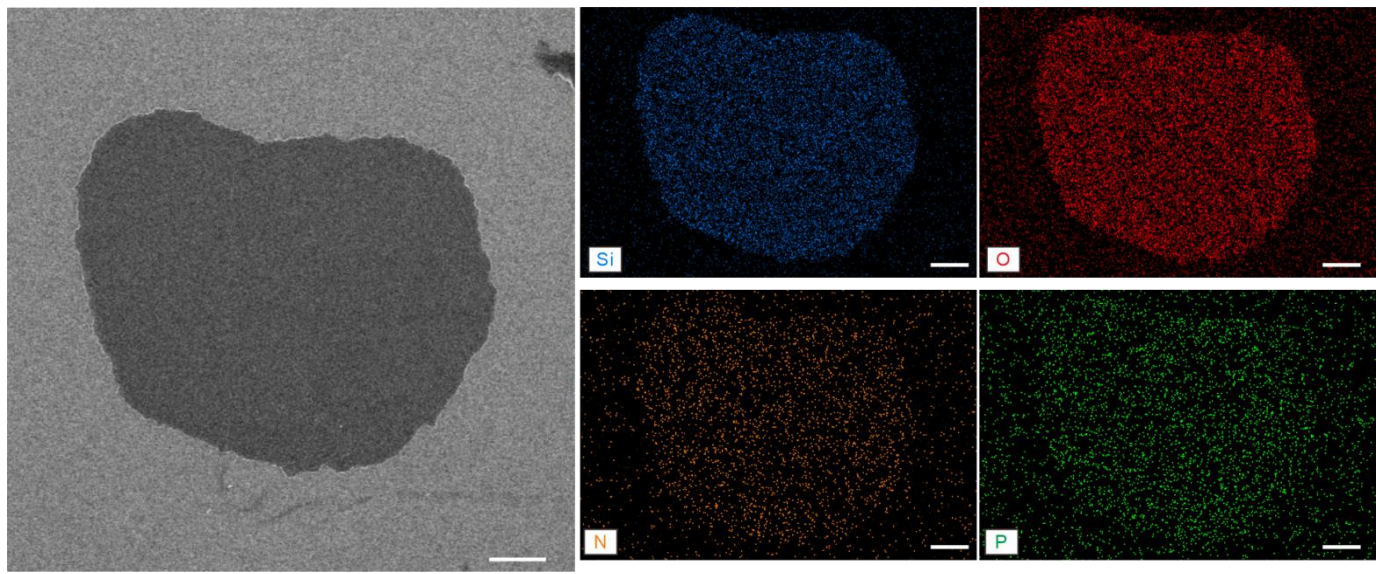

**Supplementary Fig.18 | EDS mapping of *p6mm* 2D DNA/CSC superstructure.**

EDS mapping showed uniform distributions of Si, O, N and P elements in *p6mm* 2D DNA/CSC superstructure. Scale bars, 500 nm.

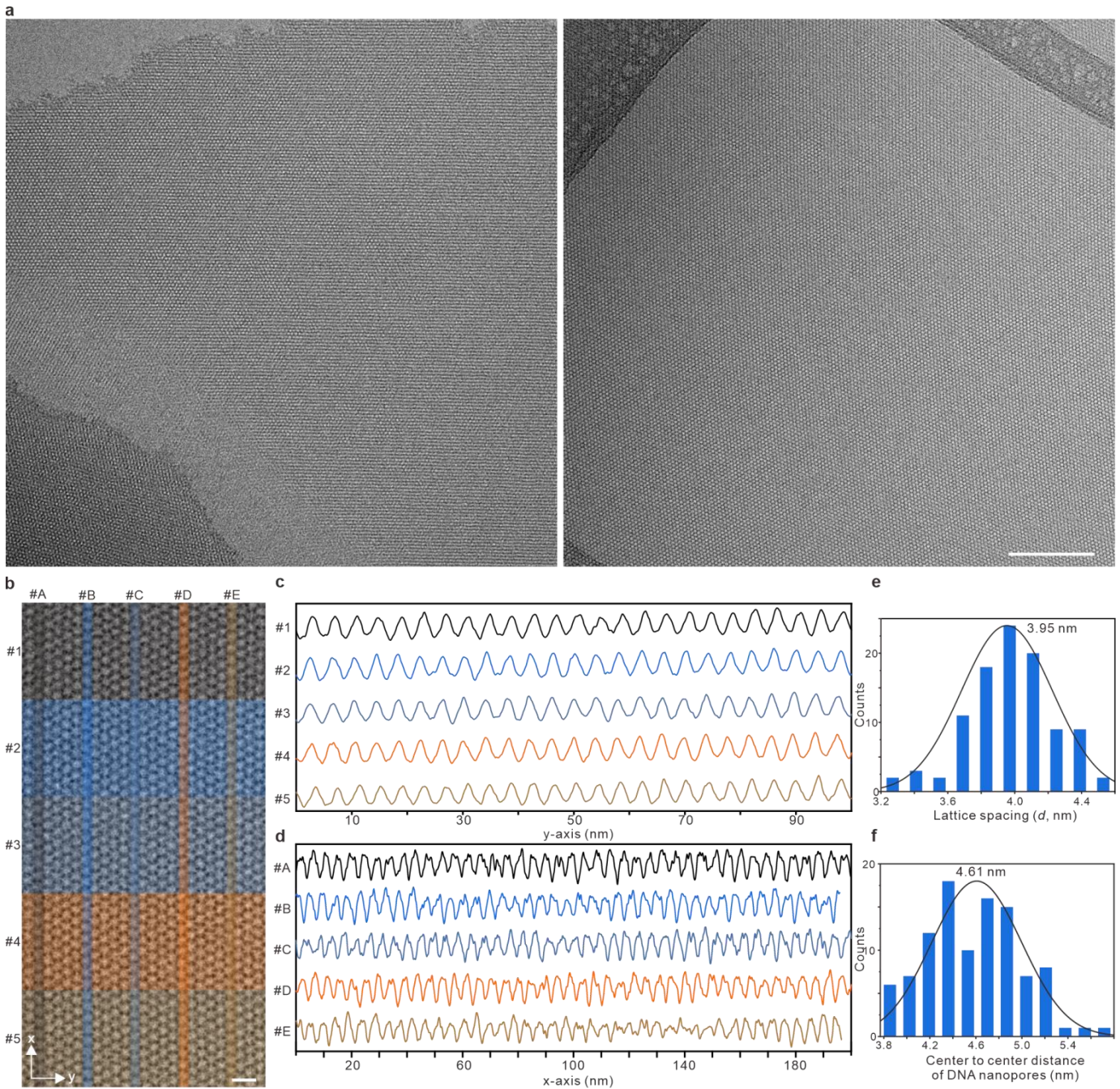

**Supplementary Fig.19 | Detailed characterization of *p6mm* 2D DNA lattice.**
**a.** Zoom-out cryo-EM images of *p6mm* 2D DNA lattice. Scale bar, 100 nm. **b.** Zoom-in cryo-EM image section of *p6mm* 2D DNA lattice with labeled scan-lines. Scale bar, 10 nm. **c and d.** Corresponding cryo-EM line-scan profiles along the y and x axis. **e and f.** Histograms of lattice spacing (*d*) and center-center distance of DNA nanopores.

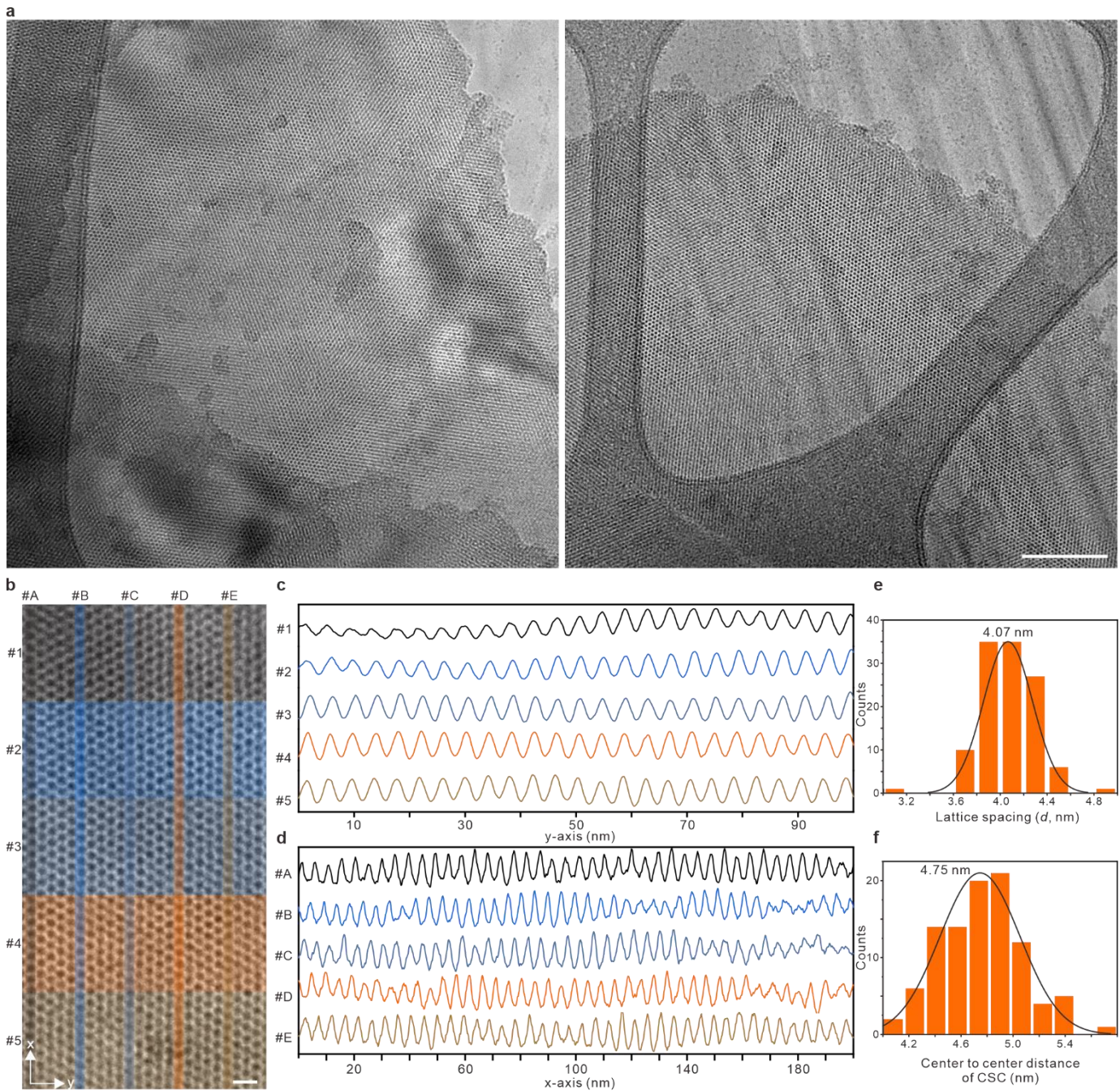

**Supplementary Fig.20 | Detailed characterization of *p6mm* 2D DNA/CSC superstructure.**
**a.** Zoom-out cryo-EM images of simple 2D DNA/CSC superstructure. Scale bar, 100 nm. **b.** Zoom-in cryo-EM image section of *p6mm* 2D DNA/CSC superstructure with labeled scan-lines. Scale bar, 10 nm. **c and d.** Corresponding cryo-EM line-scan profiles along the y and x axis. **e and f.** Histograms of lattice spacing (*d*) and center-center distance of CSC.

**Supplementary Table 5 | Cryo-EM Data Collection and Refinement Statistics.**

|  | CSC | DNA origami | DNA/CSC composites |
|---|---|---|---|
| **Microscope** | Titan Krios | Glacios | Glacios |
| **Detector** | K2 | Falcon III | Falcon III |
| **Voltage (kV)** | 300 | 200 | 200 |
| **Electron exposure (e⁻/Å²)** | 60 | 40 | 40 |
| **Defocus range (μm)** | −0.8 to −2.5 | −1.0 to −2.0 | −0.8 to −2.6 |
| **Pixel size (Å)** | 0.822 | 2.500 | 2.000 |
| **Symmetry imposed** | C1 | T | C1 |
| **Initial particle images (no.)** | 316,325 | 199,847 | 140,554 |
| **Final particle images (no.)** | 32,564 | 62,638 | 38,401 |
| **Map resolution (Å) FSC threshold: 0.143** | 8.1 | 17.0 | 22.0 |

# S3. Design of DNA frameworks

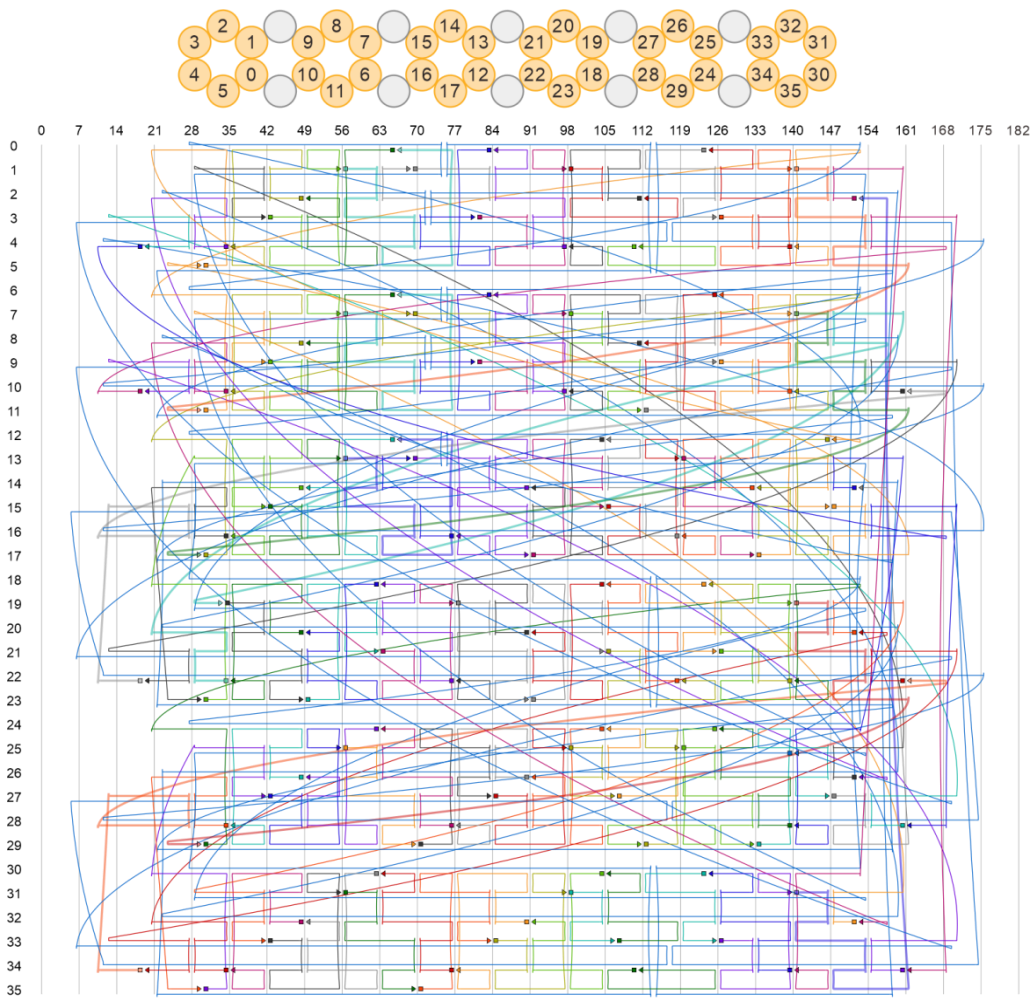

**Supplementary Fig.21 | Strand diagram of tetrahedral DNA origami.**
Top, cross-section view in caDNAno format. Bottom, strand diagram of tetrahedral DNA origami. The numbers on the left indicate the DNA helices. The numbers on the top indicate the position of the base pairs.

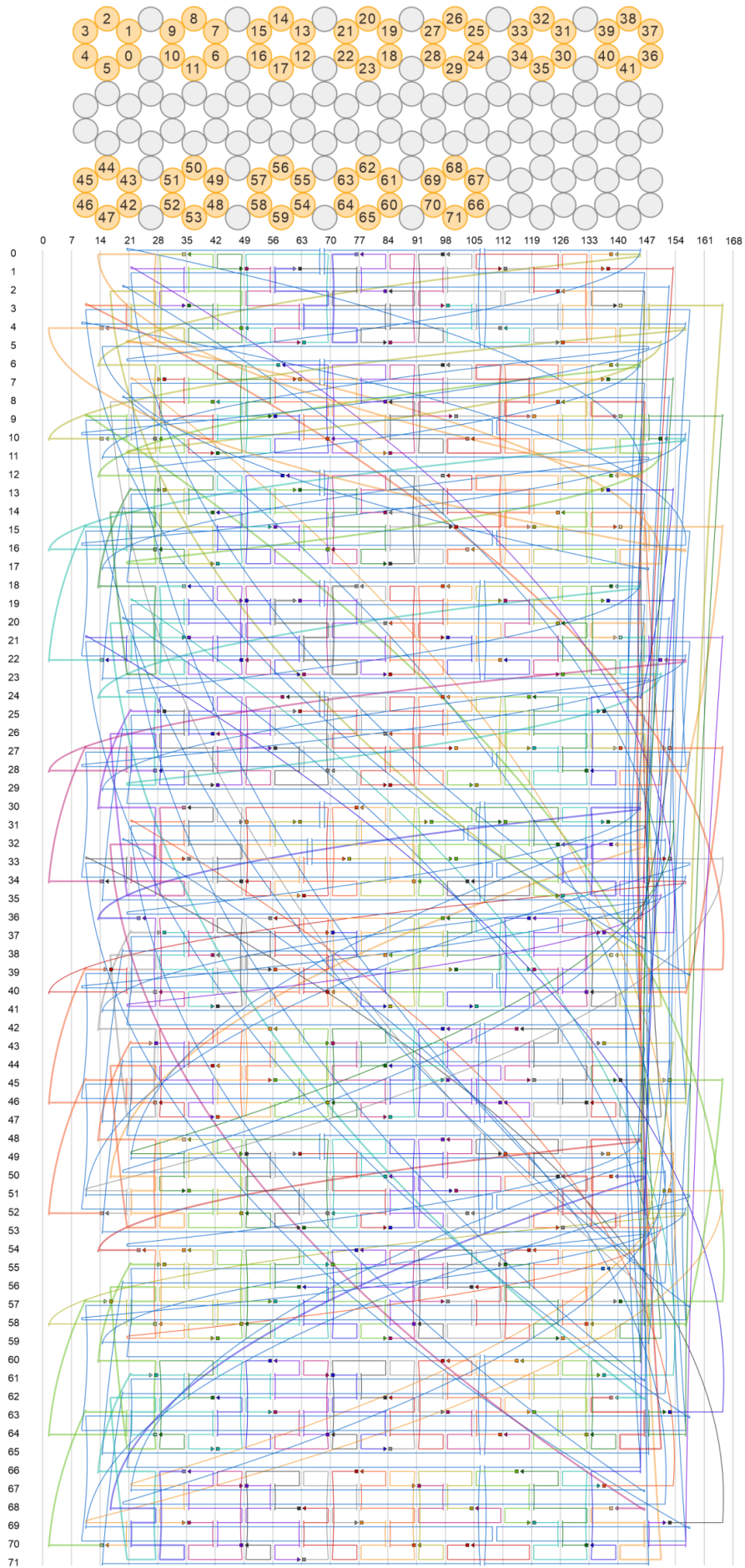

**Supplementary Fig.22 | Strand diagram of octahedral DNA origami.**

Top, cross-section view in caDNAno format. Bottom, strand diagram of octahedral DNA origami. The numbers on the left indicate the DNA helices. The numbers on the top indicate the position of the base pairs.

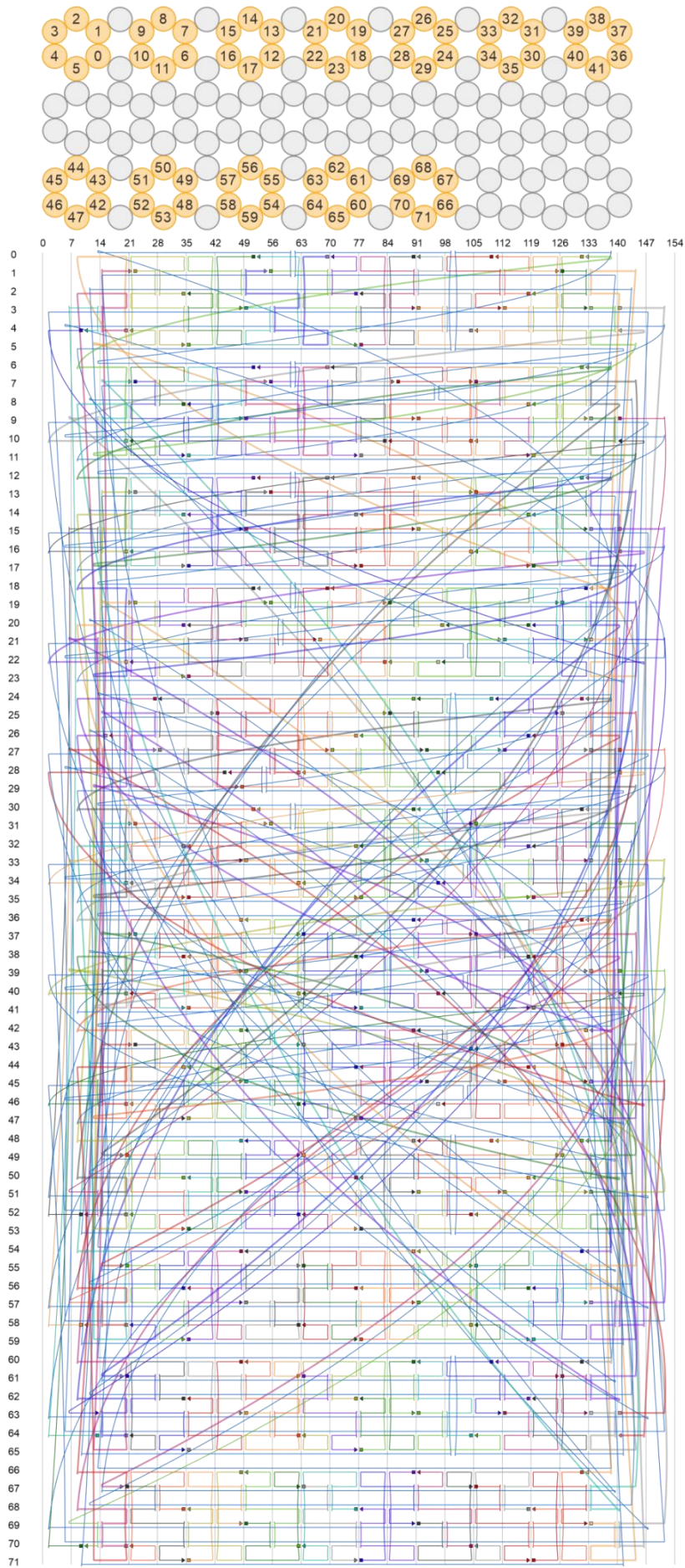

**Supplementary Fig.23 | Strand diagram of cubic DNA origami.**
Top, cross-section view in caDNAno format. Bottom, strand diagram of cubic DNA origami. The numbers on the left indicate the DNA helices. The numbers on the top indicate the position of the base pairs.

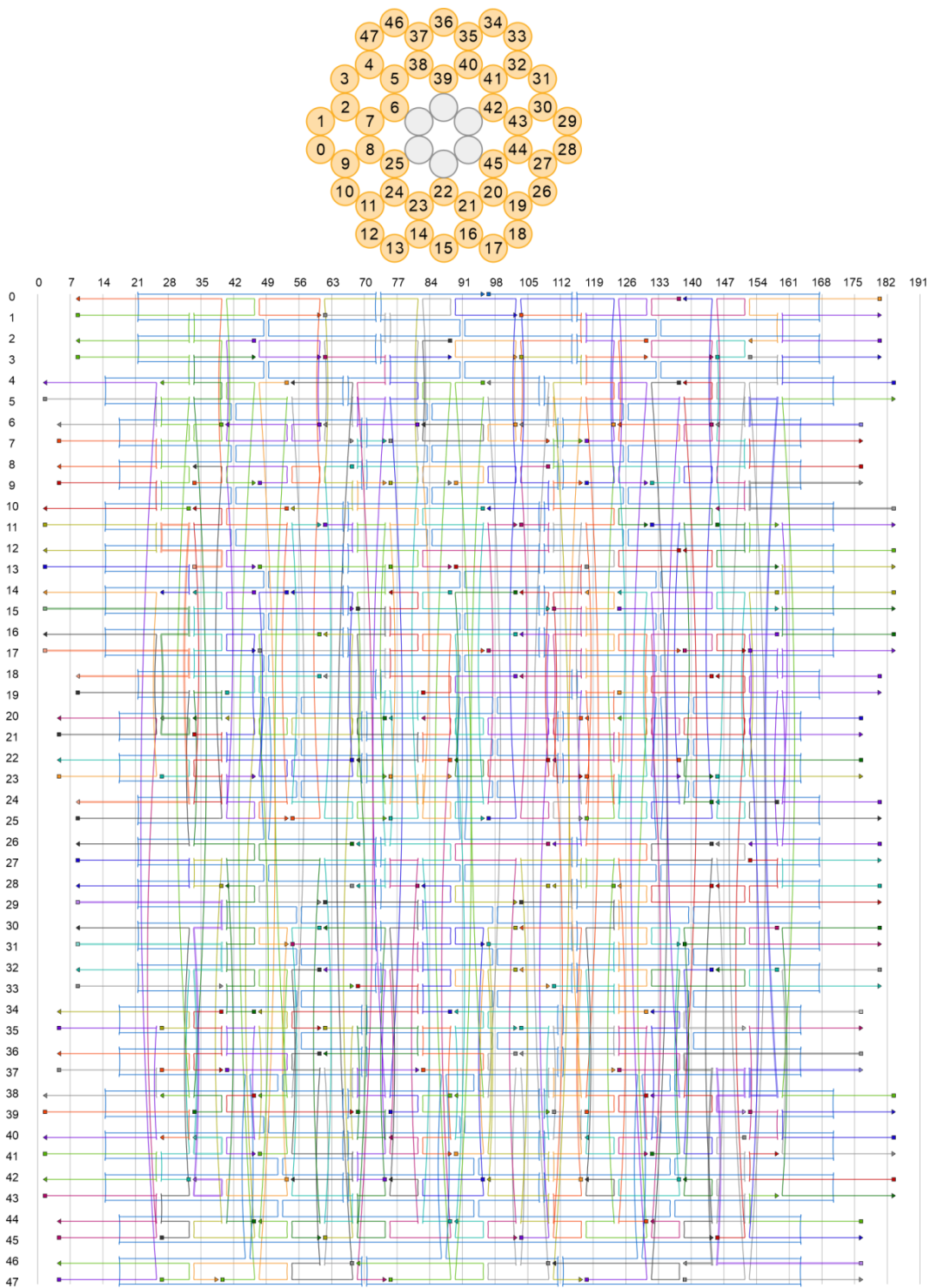

**Supplementary Fig.24 | Strand diagram of 48-helix DNA origami.**
Top, cross-section view in caDNAno format. Bottom, strand diagram of 48-helix DNA origami. The numbers on the left indicate the DNA helices. The numbers on the top indicate the position of the base pairs.

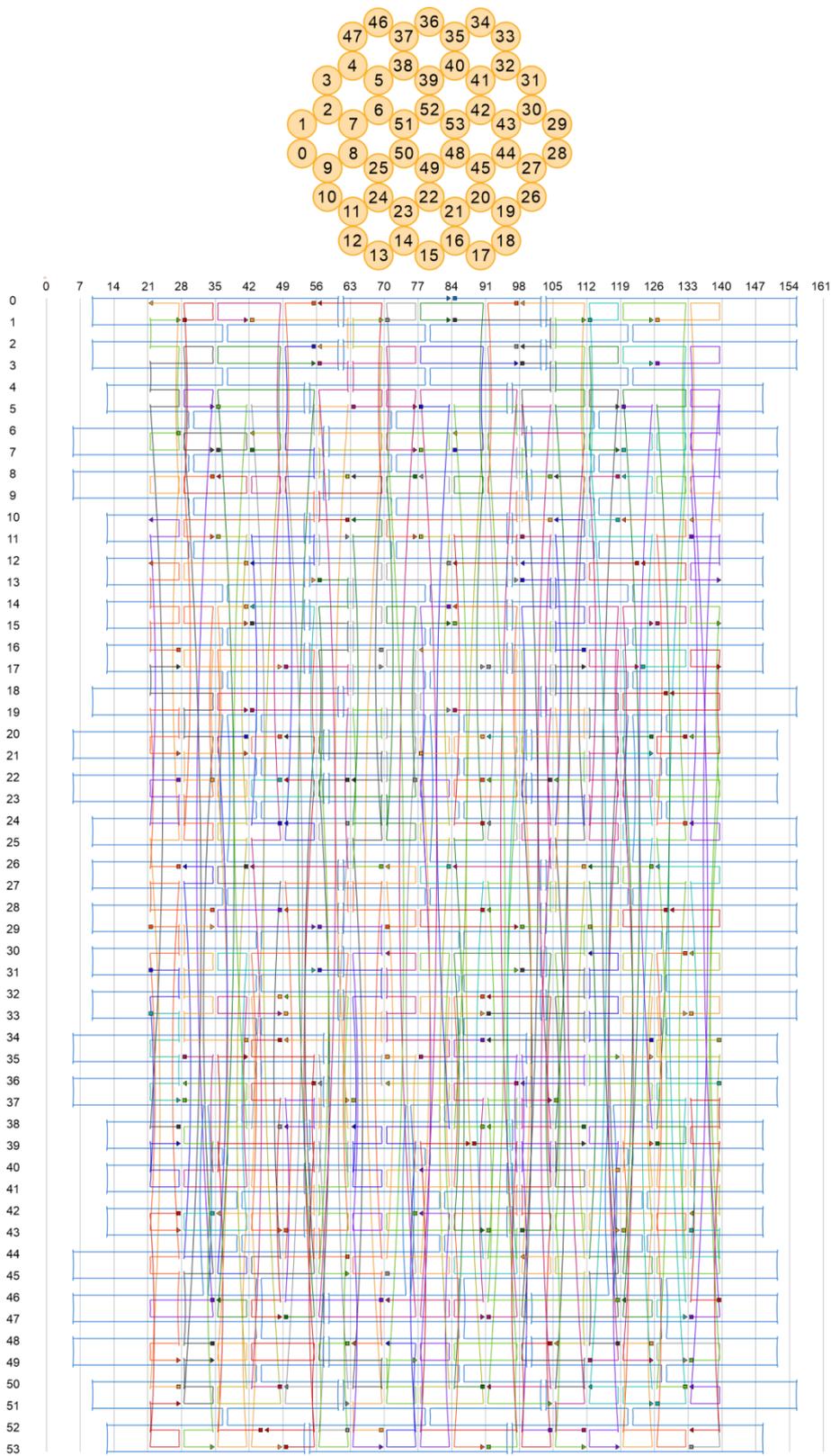

**Supplementary Fig.25 | Strand diagram of 54-helix DNA origami.**
Top, cross-section view in caDNAno format. Bottom, strand diagram of 54-helix DNA origami. The numbers on the left indicate the DNA helices. The numbers on the top indicate the position of the base pairs. Since 54-helix DNA origami belongs to simple *p6mm* lattice with 7-nt domain. The SAXS data for the *p6mm* DNA lattice with 7-nt domain (Figure 4, Supplementary Fig.16 and Supplementary Table 7) were obtained from this DNA origami.

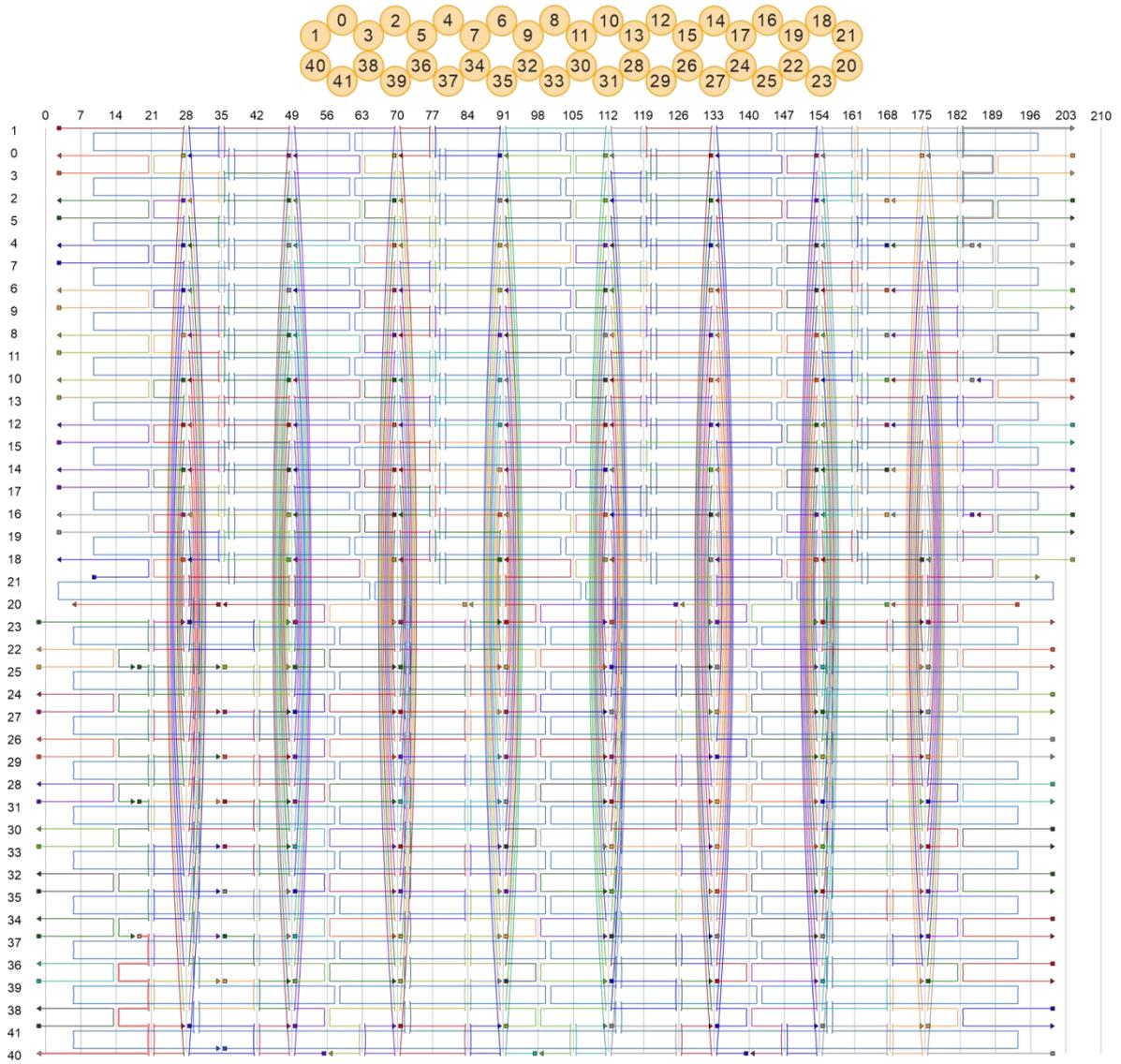

**Supplementary Fig.26 | Strand diagram of 2-layer DNA origami.**

Top, cross-section view in caDNAno format. Bottom, strand diagram of 2-layer DNA origami. The numbers on the left indicate the DNA helices. The numbers on the top indicate the position of the base pairs.

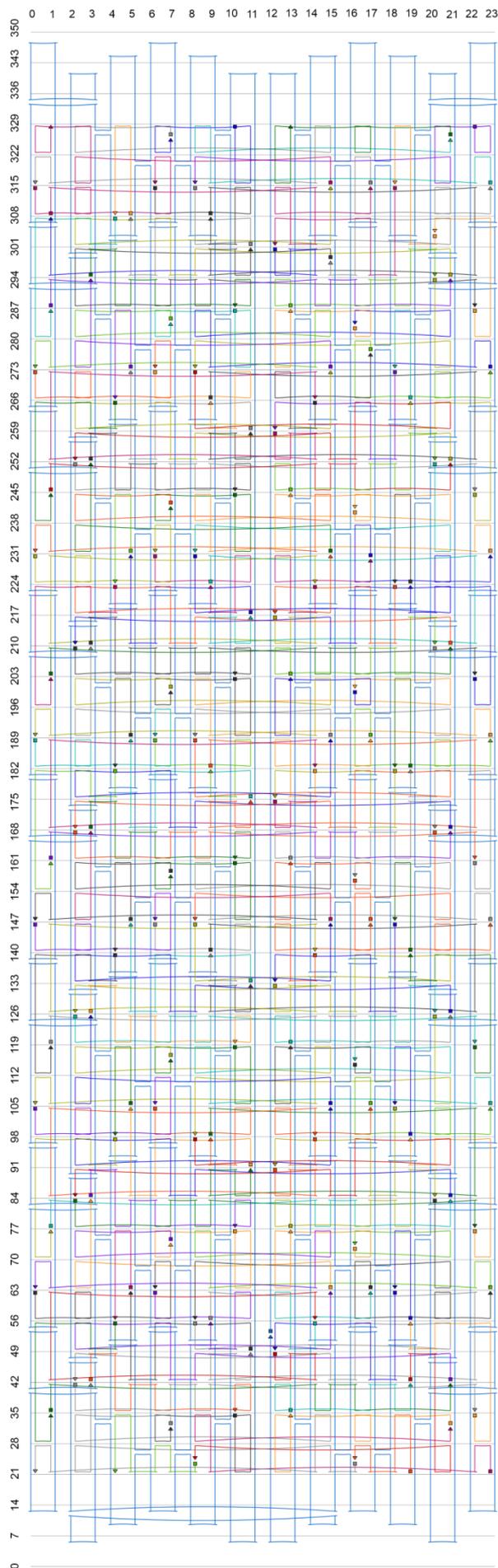
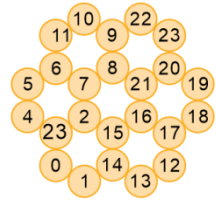

**Supplementary Fig.27 | Strand diagram of 24-helix DNA origami.**
Right, cross-section view in caDNAno format. Left, strand diagram of 24-helix DNA origami. The numbers on the top indicate the DNA helices. The numbers on the left indicate the position of the base pairs.

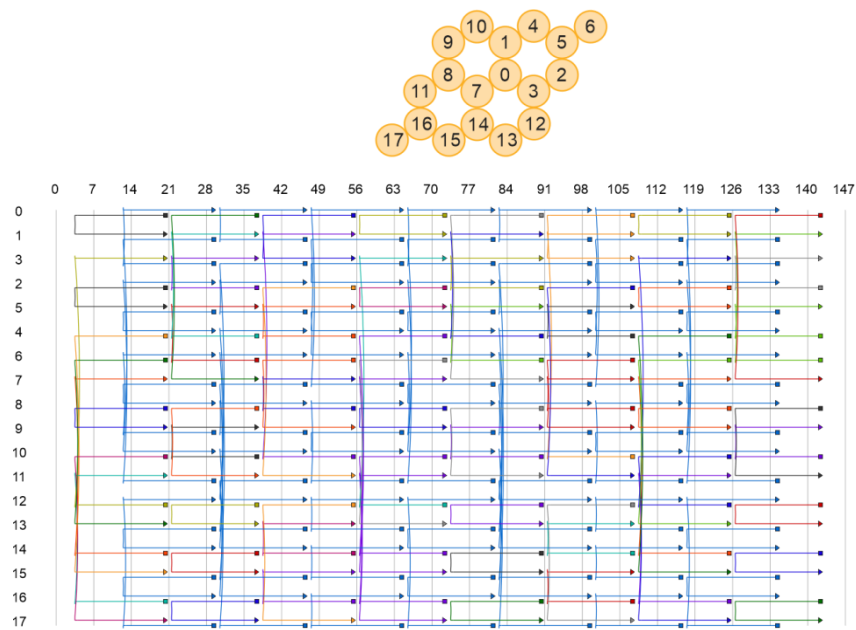

**Supplementary Fig.28 | Strand diagram of *p6mm* 2D DNA lattice.**

Top, cross-section view in caDNAno format. Bottom, strand diagram of *p6mm* 2D DNA lattice with 9-nt domain. The numbers on the left indicate the DNA helices. The numbers on the top indicate the position of the base pairs.

# S4. DNA sequences

**Supplementary Table 6 | DNA sequences of tetrahedron DNA origami (Scaffold 5250)**

| Name | Sequences |
| --- | --- |
| TDN-126-1 | ACGGGCGCGCTTTGGTGAAATTGTTATCACGTGGAGCTC |
| TDN-126-2 | GTTTATTCTTAGATCCCTTTCATCCCGGGCGATAAAGATA |
| TDN-126-3 | TTTAACTATACCCGTTCAATTCCACACAACAGTCCACTGAAAAAC |
| TDN-126-4 | CCGCCATGGATCCAAGTGAGGGTTAATTGCGCGATGGGGGTA |
| TDN-126-5 | TGAACATTCAGGCACTGGAAGCTCCCTCGACGCTCCAGCGCACCGCCTTTCGCCACC |
| TDN-126-6 | ATAGAAAAAGGGCATTAAAGAACGTGGCGCTCACAGCTCGCACACAGTCGATTCAGTGA |
| TDN-126-7 | TCCCGGTGACATCGCGCCGCCACCGCGTCCAACGTCAAGTGC |
| TDN-126-8 | GGAAAGAGTAAACTTTTAAATTAAAAATGATATAATCATGCC |
| TDN-126-9 | AAGGCGGACCTATCTTTGGTCATGAGATCAATTCAATTATAA |
| TDN-126-10 | CGAGCGGATTTCGTTCATCCATATTGACTCCCGTCGTGTCAAAATGCC |
| TDN-126-11 | TCCGCCCCTAGCATCACAAAAATCGTGCGCTCTCCTGTTCTTGCCGCTT |
| TDN-126-12 | GCGTTTTCCCCGTTGGCAGCAGCCACTTTTAACAG |
| TDN-126-13 | TCGTTATCGCGCCACCGGAGAACTTCATTATCAAATTAATCA |
| TDN-126-14 | GACGCTTCGAGTTTTTTTTGGCAACACCGCAGGCTCCAGAAGTCAG |
| TDN-126-15 | AACAATTCATCACCCCAGCTGCATTAATTGAGCGGCACAAGA |
| TDN-126-16 | GAACATTCCAGCGGGTTGCGCTCACTGCTAGGGGTGCTTCGA |
| TDN-126-17 | CCACCACTTACGGCAAATTTGCGTATTGGGCTTTTCAATTCGAT |
| TDN-126-18 | GACTTATCTGGACAGAAAATAAACAAACCGCTTTCCGTTCC |
| TDN-126-19 | TTTTACTTAGTTTTATATTATTGAAGCAGAGGCGGGAATTTCGCCGAGAATTTCTGGATA |
| TDN-126-20 | GACAAACACCAACCCATGTCGTGGGTTATTGTCTCAGAATCGGCAAGGTG |
| TDN-126-21 | TTAGTAGAGAAAACGAATGATTGGTGCATCCCACGCCAACGC |
| TDN-126-22 | ATGAGCTCGAGTCATTATCGGGCGGAATTGGCCAGCCAGTCG |
| TDN-126-23 | CCAGAATATTTTTTTTTTCGATCGCTTGTTTATTCACTAATTCGC |
| TDN-126-24 | GAACCAGAACTCGTAGTCCGTTGATTCATGATATTCGCTTCC |
| TDN-126-25 | TTTGAATTGTCGTGTACCACCGAGATAAGGGTGTGAATAATA |
| TDN-126-26 | ACATTCTAATTGCCTCCAGTAAGCCTTTCTTTCAG |
| TDN-126-27 | CTCTTCCGCTCTTCGGCGCGGGCCAGTGTTATCGCGAGGTGT |
| TDN-126-28 | GCGGGGATTTATCATGGCGAAAAGTCATTGAAGT |
| TDN-126-29 | GGAAACCGTATTTATGCAATTGTTTTATCTCTGAAGTAAAAC |
| TDN-126-30 | TCGCTCAGTTGAATTCTGCATCCACCTTTGCTCTCT |
| TDN-126-31 | CGCATACTTGAATGAATGAATATTTTTTTACTATATACATA |
| TDN-126-32 | GTCACTGCATGCTTAGAATCCCAAGGGAATATCTTCCGCGC |
| TDN-126-33 | AATGTGAGCTTTCTTTTTTTTATGAAACCATTGTTCAGATCACTCA |
| TDN-126-34 | AAATCCTCCCAGTAAACCCACTCGTGCACACGCTCGCAGATT |
| TDN-126-35 | CATTTCAGCAGAAACTGATGGTAGTAGTAGCGGAATAAAC |
| TDN-126-36 | AATATTAACTCGATTCACCAGCGTTTCTGCTTACCGATCTTT |
| TDN-126-37 | GGAATTTGAAAAAGATACCGCGAGACCCCCAACTATTGCAA |
| TDN-126-38 | TGTTGGCCCGGCAAGGCCGAGCGCAGAAGGGCGAACAACTCTGGCTTTTTAGCTATC |
| TDN-126-39 | TTCTCAATGGTCTGATACGATACGGGAGGGGGTGAGACATTAA |
| TDN-126-40 | ATGGTGCGCCGACCATGATTGCGTAATCTGATGCTATGGTCA |
| TDN-126-41 | TTTAATGCCATGTTTTTTTTATTCAGATTGGACCTGGTCGCCTTT |
| TDN-126-42 | CGTCTATCCACTTTAGGTGACGGTTAATGGTAGAACGGAAGC |
| TDN-126-43 | ACGCGCAACGCCGCGAAAGTCCTACGGTAGCAATAGATCTTC |
| TDN-126-44 | TTGATCTTTTGTTTTTTTTCGGCGTGACAACTTACCTTGCCGGGA |

| TDN-126-45 | TCTACGGAACCGGTGGATATTGAGACATTTTACGACAAAAAC |
| TDN-126-46 | CCAGATTTCGATGTGTACTGCAATTTATTGGTTGTTGTTGAG |
| TDN-126-47 | CAGCCAGTTACCGCATGAGAGTAGATAGATCTCCGGACGAGG |
| TDN-126-48 | CCCAGTGTTTACTTGCACGCAATGGTGTCAAGAGATCATAGA |
| TDN-126-49 | AGCATCTCTGCAATAGGATCTCAAGAAGGTTGGTGAGCAATA |
| TDN-126-50 | ATCCAGTTATCAGCTGGTTTTTTGTTTATAACGCGGTTCCA |
| TDN-126-51 | AAGGATCCCGGAAGACAAACCACCGCTGATCCATCCTTAGAATT |
| TDN-126-52 | TTCTTCGGTGGTCCTGCAACTTTTCTCCATCCAGTCTATTGCTCCTTCG |
| TDN-126-53 | GCAAAAAATAGGGCGACACGGAAATCTGACTCGCTGCGCTCGTGGCTGCGG |
| TDN-126-54 | AGGAAGGAGATAACCGCTCAGTGGAACTTAAACTCACGTTCAGATTTTTTTTTTGCAGG |
| TDN-126-55 | CTCATTCGCATGTGGATTTTTCCGCCGCGCAAGCAACCGGCT |
| TDN-126-56 | GCTTTTCGGTGAGTTTTTTTAACTTAGGTGTTTTAGGTTCAACTCATA |
| TDN-126-57 | GCGACTTTCCGTCCAATCCAGCCCCAACATCCTTTATCTGGC |
| TDN-126-58 | CCTAAAGGTTAAGACATTCTGAGAATAGAAGATGCCTAAAGG |
| TDN-126-59 | GCATTTGGATTCAAAAATCAGCTACTCAACCAAGTCGTAATG |
| TDN-126-60 | TATATTGTTCAAATCGGCGTCAATACGGTATGGCACGAACGT |
| TDN-126-61 | GAGCGCGCGGAACCTTTTCTGTG |
| TDN-126-62 | CTATAGGAGGTGCCTTTTTTAACCAATATTTTGTTGTGGACG |
| TDN-126-63 | CCAGGGTAAGCCGGGCACTGCATAATTCTCTTGCCATGGAAC |
| TDN-126-64 | GAGCCCCAAACGACTGACCCAGCAAAAGCCTCATTGGCGACC |
| TDN-126-65 | CACCCTTTGCCATTTTTTTGAATTGTCAGATTTCAGGGCGAGTTTGG |
| TDN-126-66 | GGCGAGATAAGTTGGAATGCTGCCAGCAACTAAGTCCGCGCC |
| TDN-126-67 | CATCCGTTGTATGCTCTAAATATGAAGAAACAACCGGCCAGT |
| TDN-126-68 | ACTGGTGAGTCAGTAAAGCACTAAATCGTAATATACTCCT |
| TDN-126-69 | AAATCCCTTGTTAAAAATTGTTTTATTTTCAATAT |
| TDN-126-70 | TCATGGTGATAATAAGTGATGAAAACTTGACTCAAGGTAACG |
| TDN-126-71 | GAGTTGCTCTTACTGAGCTTGACGGGGATTTCCCAATACTTT |
| TDN-126-72 | GTTAAATGGCCGAAGTTTTTTGGGGTCGGCGAATTGTTTAGT |
| TDN-126-73 | TAATATTTTATAAATCAAAAGAATGAGATAGGGTTGAGTGGTAAAGCCT |
| TDN-126-74 | ACATAGCCGCAGTGGGAAGAAAGCGAATTGAGCGGG |
| TDN-126-75 | AACCACAATCCTTTCGACTCA |
| TDN-126-76 | ATGATAACAGTTTCGGGTACCTAAATGGTAATCAAATCGGCA |
| TDN-126-77 | AAAAAATAGTTTATTTTTTAAATCACTAGGCGAAAAGAAACTCTC |
| TDN-126-78 | TCATCAGGTACACAGTCACGACGTTGTACGATTTAGTCATGC |
| TDN-126-79 | CAGCTTTTGTTTCCTGTTGGAACGCAGAAACAGCACCAACGCGCG |
| TDN-126-80 | CCAGGCGGAAACCCCTGACGCTTCAGTTACACGTAGAAACGC |
| TDN-126-81 | AACAAGATACGAGCCTCTTCCTGCATTGCCTTTTGTCATTTA |
| TDN-126-82 | ATTCAGGCGCTACATATGGCTTCATTCAATTGCTATAACTAC |
| TDN-126-83 | TAATTCGTCTCCACCGGGTTCAGCAGGTGCTACCTCATAG |
| TDN-126-84 | CTCTTCGTCACGCTTACATGATCCCCCAGTAAGTATCTGCTG |
| TDN-126-85 | TGCAGTCGGTGGCCCAGGCATCG |
| TDN-126-86 | CGAGGCGGCAGAGCATCTCAGTTCGGTGACCTGTCGCTTGAGGATTTTTGGTATT |
| TDN-126-87 | GCAGGAATCTGCGCGTTCGCCAGTTAATGGCGAGTGCGCGTA |
| TDN-126-88 | GGCTACATGCCTGGACTGTTGGGAAGGGCACCCGCTTCCCAA |
| TDN-126-89 | GATTAAATTAAGTTTTTTTCGCCCGAGATTAAACGCCACTCAGCCCG |
| TDN-126-90 | AAGCCAGGTGGAAAGCCAGCTGGCGAAAGGGCGCTCAAAAAA |
| TDN-126-91 | TGTTGCCGCTCCGGCGCGCTTAATGCGCCTGCGCACTGGAAG |
| TDN-126-92 | TGGTGTCACGTTAGAGTTCTTGAAGTAACAATGATTCGCC |
| TDN-126-93 | CTCACGCCGGGAAGGCACCAATGAGCGTAAATGCCTCTTTCA |
| TDN-126-94 | AGCTAGATGTTGTGGGCAAGTGTAGCGGCTATTACCCTTCGC |
| TDN-126-95 | CGATCAAAGTTTGCGACAGTATTTGGTAGTCGGCATGCGGGC |
| TDN-126-96 | GTTTGGGGGCGCGTCTCATCACGGCTGCGTGGCGCTCTCGTC |
| TDN-126-97 | CTCCCTTTGTAGGTGAGGTATGTAGGCGATGATATCTTGACCGG |
| TDN-126-98 | ACCGGATTAGGTCGTTCGCTCCATGCTGTGTGCACGAACCTCCATAGGC |
| TDN-126-99 | GTCCTCCGTTCAGAAGTAAGTTGGCAGAACTTTAAAAGTGCTTGGAAAACG |
| TDN-126-100 | GCGGTTAAATTGTTCGGAAAAGAGTTTTTAGCTC |
| TDN-126-101 | CGCTAGGGGATGTTTTTTTGCTGCAAGGCGATAAGGAAGTTATCAC |
| TDN-126-102 | ACCACCACGATCGGTATTGATCCGCCACCTAGAAGGCAACGT |
| TDN-126-103 | TAGCTGTTCCCTTTCTAGTTCTAGAGCGTGCAAAGCGGAAAT |
| TDN-126-104 | GTAATCGTATTCGGACAGCCAGCTTTGGACAGGACTCTTGGC |
| TDN-126-105 | AGGTGGCTTTCCCCAGGGATCTGGAACTCGGACCAAGTCGATAT |
| TDN-126-106 | GGGGTGCCTTGAGCTAACTCACATCCCGAAAAGTGCCACCTGTAAGCGT |
| TDN-126-107 | ATAAAGTTTGTTCCATGGCCCACTACGTTAACCAT |
| TDN-126-108 | TTCCATCTATGCCTTTTTTTCAGCAC |
| TDN-126-109 | TTCGTCTTCTGATGAGCAAAAGGCCAGCTCCAATCTAAA |
| TDN-126-110 | GCTTTCCTTACCACGTTATCCACAGAATCCAATGCAAGGATC |
| TDN-126-111 | ATTAAGGCCACCTGATCCAGGAACCGTAAAACGCCTTAAAGACAC |
| TDN-126-112 | AACCAGAAGATCCTTGGTCTGACAGTTACAGGGGACGCATCA |
| TDN-126-113 | CATTGAGAAGGGATTCAGCGATCTGTCTTATCAGCAGATCCCTGAGCTTTTGGCGAC |
| TDN-126-114 | GCGCCAGACCCGGTTCCGGTAACTATCGAAAAGGCGGTATTGAACAGCAAAGTTTTCCAG |
| TDN-126-115 | CTGCCACTTGCCGAAGAAGTTTTAAATTTGAGTCCATTCTTC |
| TDN-126-116 | TTCACCTAAGAATGGCGTGGTCGCCTCGCAGTGGTTAACGCA |
| TDN-126-117 | GACTTATCGGCTCTTGGTGAGCAGAGCCCAATCACGTTGCTG |
| TDN-126-118 | GTATATATGAACATGTGAGTAACGCGCACGCGAGAAACGAAAGCG |
| TDN-126-119 | GTGAGGCTAATACGGACAGAGATTGAGCCTAAAAATCACCTG |
| TDN-126-120 | ACCGCTGAGGCCGCAAACACTTCGATAATGTACATTTGGCTT |

| Name | Sequences |
| --- | --- |
| OCTA-126-1 | AACGGCTGTCAGGGACTCCTTATTACGCACCGGAAAAGAGTC |
| OCTA-126-2 | CGAGGTAACCGGCCGCAATAATAACGGGCCACCCATCCGTC |
| OCTA-126-3 | CATTGCAGTATCTTAAAGAAACGCAAAGTTCATCGGATGGCCCATTTTGAACCAT |
| OCTA-126-4 | TCACTTCGCGTCAGTAGCAAACGTAGAACATCTTTGTGGACTCCAACGTACTATC |
| OCTA-126-5 | TCGTAAGAGGGTTGCCTCCCTCAGAGCCAATACCCTTCAGTT |
| OCTA-126-6 | GTCGAACGACTTTTTTCGAGCGTAGCGAGCCGGTGCCCAGAAC |
| OCTA-126-7 | AACTCAACAAAGGGTTATTAGCGTTTGCAATACATCACCACT |
| OCTA-126-8 | GGCCTTTGTTAAGTAAAAGAACTCATAATCAAAATCAGTATGTATTTCGT |
| OCTA-126-9 | CCCAATTAGTTACCGGATAAGAAAATGTGAGTATTATCTTATCCATGCAAGCAAG |
| OCTA-126-10 | TCCATACCGCGCCCATTTTTTTATAGCAAGCAAAATC |
| OCTA-126-11 | TTACCATTTTTTTGAAGGAAACCCTCAGAGCCACTTTTTTTTCACCCTCAGAAGCCT |
| OCTA-126-12 | TAGCCGAGCATTAGGAAGACAAATTTATTCTGCAATGTGCGCCTTTA |
| OCTA-126-13 | ATAGCTAGAGGGTAGAGTCGGCATACAATGAAAA |
| OCTA-126-14 | CAGTGGCTTTTCGAGTAATCTCTTGCTCATATTCCCCTGAAC |
| OCTA-126-15 | TGGACTCGTCTTCAGACCGTACTCAAACTGTTTATGATAACCCA |
| OCTA-126-16 | ATTATTTGGAATGAAACAATAGATAAGTACAACATATAAGCT |
| OCTA-126-17 | ATGGAAACGAGGGATCAATAGAAAATTCGACATTAGCGTAA |
| OCTA-126-18 | CTGCATATGATGAGTGTTAATTTACGAGCATCCGACAAGTGATTT |
| OCTA-126-19 | ACGTGGCACATTTTTTGACAATATTTTGTAACGGTTCATTAA |
| OCTA-126-20 | CGCTATGATAACCGAGAATATAAAGTAGTAGAAATGCCTAC |
| OCTA-126-21 | GATGCAACGACCAGTTTCTCACGACGACGACAATAACCTGAACGCGGCCA |
| OCTA-126-22 | TGGCTGGACCTGAACATTTTGACGCTCAAAGGCAGTTTACCAGCCGCCTG |
| OCTA-126-23 | CTCACTTTACGGGGCGCTGGCATTCGCAAGCGAACCCAATCA |
| OCTA-126-24 | TTTGCCAAATAAAACAGATTCACCAGTCGACAAACAAGAAAA |
| OCTA-126-25 | ATGCAGAACGGCAACAAGTCTGGAAAAACAGAGGAAATGG |
| OCTA-126-26 | AAAAGGGCATATGGGAACAGGAGAGCGCATACCTAAGCGTAAGAAT |
| OCTA-126-27 | TTGAGGGTGTCACAAGCCGCCATTTAACGCCTGACAGGAAATTTTTTAACGCTC |
| OCTA-126-28 | AGTAATTCCCATCCCAGGCGTGGAATGAACGGTGAGTAT |
| OCTA-126-29 | TTTATCCACCGGTAAACCGAATCGTCTATAGAACCCTTCTGCCTGTTGCAAAGAC |
| OCTA-126-30 | ATAATCGGTAATAATTATTTTTGTTTGGAAATTACGACATCATTTTTTTTTACGCAT |
| OCTA-126-31 | ATAATATCTGTCCACGGATTCAGTCGTCGTTCAG |
| OCTA-126-32 | CCGGAACTGTCTGATTGGAGGTCAATGGACTCATGAAGGTAA |
| OCTA-126-33 | TAACAGCACATTGGGGGACATTCTGGCCGAAATGCGTTCAGCTA |
| OCTA-126-34 | CTTCATTTGGAACTCTTACCAGTATAAAATGGTTTAAATCAC |
| OCTA-126-35 | TGCATGTATCCTGACTAGAAAAGCCTGATTTTATTCCCGA |
| OCTA-126-36 | TAACTATATAGGTTGCCAACATGTAATTCATTCAGATAGATGAATTTACGCCTGG |
| OCTA-126-37 | TTCAAAATACCAGGCTCAACAGTAGGGCGAGTAGTCCGGGCGAGCTCGCACCTGG |
| OCTA-126-38 | ATGCTCGGACCGTTGAATTACCTTATGCGTTTAGTAGGTATT |
| OCTA-126-39 | AATTGGTTAATTTTTTTTGGTTGTAACACAACTTTTTCTACGT |
| OCTA-126-40 | ACTGGTGCACTCATAGAAACACCAGAACTTAATTGGAATCGC |
| OCTA-126-41 | AAGAGTATGGTATTGTCGATCCAAATTGGGCTTGAGGCCAACGATCTTGC |
| OCTA-126-42 | CCGAGATCCATTTCTTGGAGGAGCGCAGTCACCCTGCCCCGT |
| OCTA-126-43 | CACTATTTCTTCCTGGCCCTC |
| OCTA-126-44 | CAACGCAATGAGTTACCACTGCGATCCCCGGGAAGTCATTTG |
| OCTA-126-45 | CAACTGGATAATCCCGGCTTT |
| OCTA-126-46 | ACTTTAAACAAATTGCCTCGGTGAAACAGCATTCCATCATAT |
| OCTA-126-47 | GCTCATTCATAATTATTCAGGTTTTATTGTTGACCGTACTCTTTTTTTCTGATGA |
| OCTA-126-48 | GGCTTGCTAACAACGTATTGATGTTGGAACAAGACGCACTGC |
| OCTA-126-49 | GCGTTATTCATTGTCCGTGGCAAAGCAATGCAGTTTTTTCTC |
| OCTA-126-50 | GAATATACCAGTCAGTTTTTTTGACGTTGGGATAGGA |
| OCTA-126-51 | GAGGCATTTTTTTTTTTCGAGCCAGCTGTCTTTCCTTTTTTTTTATCAT |
| OCTA-126-52 | CCATATTCCTGACGTGTTGTGAGTGTGGTTATCAACCGTTAG |
| OCTA-126-53 | AGAAGAAGTTACTCTTTCTAATCAG |
| OCTA-126-54 | AGACCGAATTTCTGCGGCAGTTAATCGACGAGTCGAGAATCG |
| OCTA-126-55 | CATCCTAACAGAAATGATATGAATAAATAAGTTCAAATTTCA |
| OCTA-126-56 | CCTAAAACGATCGCAATATATTTTAGTTTCCAATCCTACCAT |
| OCTA-126-57 | TTTAATGTCAGGCGCGTCACCGAAAAACTTTTTCAGTATTTC |
| OCTA-126-58 | TTATGCCTCTCCGGTTGGTTTGAAATACCGACTTAGGTAATTGCG |
| OCTA-126-59 | CTGAGAGCCAGTTTTTTTCAGCAAATGAAAAATTATCAGAAACGT |
| OCTA-126-60 | CCAATGTCAGGTTTTTTTAACCTCCGGCCGTGTGGCAGTGT |
| OCTA-126-61 | ACTGGCGAACCACCATTATTTGAATGCTGATGCAAAAATTTCATGTCCTT |
| OCTA-126-62 | TAGATTTATGTTACATCAAGCATTTTATTGCGCTGATAAATA |
| OCTA-126-63 | ATCAAAAGCAGAAGATTAAAAATACCGATTGTAATTCTTCTG |
| OCTA-126-64 | AGAACGCGAGCTGAATAATGGAAGGGAGGCGGTTGATAGC |
| OCTA-126-65 | TTTGGGATTATCACCAATCACGAATGAAAATGGCTCAGTGCCACG |
| OCTA-126-66 | AATCACCAAATTATTTGGTTGATTTTGATTTTGATGACGCAACCGA |
| OCTA-126-67 | CACCATTTTTTTTTTACCATTAGCTAGCGTCAGACTTTTTTTTGTAGCGCGTTACACCA |
| OCTA-126-68 | TATAACTTTTAATGCATTCGATTCCTGTACTGACGGAAT |
| OCTA-126-69 | AGGTGAAATTAGAGGATTGTTTGGATTACCTGCAAATTAGTC |
| OCTA-126-70 | AGGCGTTGACTACCTAACGTCAGTTTATACAGTAGGCAAAGGTTTTTTTAGCGTTG |
| OCTA-126-71 | ACCTAAAATATGTACACGTAAAACAGAAGACTAA |
| OCTA-126-72 | GTCTCGCCGCGAACCAGTATTAACACCGTACTTCTTGAGCCA |
| OCTA-126-73 | TCCTGCGTCCGACCAGATGAGATGGTCAATAAAGATGGGTTA |
| OCTA-126-74 | TTAACAGCATCGCCATAAAACAGAGGTGTTAGAACGCAAGACAA |
| OCTA-126-75 | CACCACCGATTAAGGGGCGGAGCAAAACTTGTCCTAAAAGAA |
| OCTA-126-76 | CGCCACCGAGGAAAGCATGACTTTTCTAACTCCACCAACTCTTTTTTTTTTGAAC |
| OCTA-126-77 | CGGTCATACATATATAGTCCTGTCGGATATTACGTAGAAG |
| OCTA-126-78 | GAAACATACAGTACAGTCAATAGTGAATTGCAGATAAAACCAAATTTGAGAGGCT |
| OCTA-126-79 | AAGAAGAGATTGCTCCTTGAAAACATAGAGCAGTACATA |
| OCTA-126-80 | TTAATTACATAGAATTATTCATTTCACAAAATTAATTAA |
| OCTA-126-81 | TTGCTCCGGTCAGGCCCGGAATAGGTGTGGGTTGAATTTAGA |
| OCTA-126-82 | TTTGCGGATGTTTTTTTGCTTAGAGCTTAATGTTTTAACAGAACC |
| OCTA-126-83 | GTGCGACCATAACCCGAGGCATAGTAAGCGATAGCTAACGGA |
| OCTA-126-84 | TCGATTATGAAACAATTTTAATGGAAACAACACTATAATCTA |
| OCTA-126-85 | TCCAACATTTTGATTTAGAGCCGTCAATAGATAGCCCTTTAA |
| OCTA-126-86 | ACCGGATACTCAGGAGGTTTGGCGGATCAAACAATTTTACTCGTATATCT |
| OCTA-126-87 | TGCTTCTGTAAGATTAGAGAGTTTAACAATTTCGAATAACCT |
| OCTA-126-88 | CGCCACTTTTTTTTCCTCAGAACCGAGACGT |
| OCTA-126-89 | GCCAAAATGAGAAGCTTTTACATCGGGACCAGAGTCAATCAG |
| OCTA-126-90 | AATCAATATATTAGAATTTGAATACCAAGTTAATTACCGAAAACAAAA |
| OCTA-126-91 | TTTTCCCTGTGAGTATTTGAATTACCTTACATCAATGAGCAA |
| OCTA-126-92 | CGCTATCGCGCAGAGATACATTTGAGGTATAAGTATAATCGT |
| OCTA-126-93 | AAATCATTTTTTTTTAGGTCTGAGAAAATAAGAATAATTTTTTTTTACACCG |
| OCTA-126-94 | AAGACGCGGAATTACTCGTTTACCAGACTGTCGGGTGTTTCT |

| OCTA-126-95  | AGTATTATAATAGAAAGAGGTCATT |
| OCTA-126-96  | TTCGCCTTGGTATGGGAAGCCCGATGCGGAAACAATTAGATT |
| OCTA-126-97  | AACCCTCAAAGAAATGCCTATTTCGGAACTTGAGTGGAACCC |
| OCTA-126-98  | CTTGCTGGGAATTATCGATAGCAGTTAATGCCCCCCCACCAG |
| OCTA-126-99  | TTAGTTTTTTTGAGCACTAACAACGACTTTAAAGTGCCGTCGAGAATCACCGAGCAAAC |
| OCTA-126-100 | AAAGGAAATTAATTAGAGGCTGAGACTCATACAGGAGCCGGC |
| OCTA-126-101 | TTCTTTGATTATTTTTTTGTAATAACATCACTTCAGGGCGCATTTT |
| OCTA-126-102 | TTTAGAGCGAGAACATGGCTTTTGATGCTCAAGATTTGCCC |
| OCTA-126-103 | TATAATCAGCCCCCAACGGGGTCAGTGCCCTATTATCATTTT |
| OCTA-126-104 | CACCCAATCAAGTTTGCCTTAAGGCCGGATGATGGTTTATCAATATAATCCTCCAGCAA |
| OCTA-126-105 | GAACGTGCAGGAACAGGTTATCTAAAATTAAATCCGAAGGAT |
| OCTA-126-106 | TAAAGGGAGTGAGGATCTGGTCAGTTGGAACATTATTCTGAA |
| OCTA-126-107 | CGTATAAACAGCGGTGCCGTAAAGCAAGTCTGTAATATCA |
| OCTA-126-108 | TCAGTAGGAAACCATCATCATATTCCTGATCTAAAGTAGCAATAC |
| OCTA-126-109 | GTAAGAGCGGGGTTTTGTTTTTTTCTCAGTACCAAGTAC |
| OCTA-126-110 | TGGTAATAGTATTATTAAAAGTTTGAGTCAAATCAGTGTTTT |
| OCTA-126-111 | CTGGCATGGAACCGAGTGTTGTTCCAGTAATCTCCATGGAAA |
| OCTA-126-112 | CGGAATTTTTTTTAAGTTTATTTAGGGAAGGTAATTTTTTTTATATTGACGGAGTAG |
| OCTA-126-113 | GGTGGCAAGCCCCCGAAAAACCGTCTATGCCTGACGCCAGC |
| OCTA-126-114 | CACCAATCGACAGAAATCAAGTTTTTTGAACCGTTGCATCAC |
| OCTA-126-115 | TAGGATTCGTCATAAGGAAGGGATTTCGAAAGGAGGCCGATTTTTTTTAAAGGGAT |
| OCTA-126-116 | ACATGAAAAGTTTTGATTTAGAGCTTGACTGAGAAACAGTTG |
| OCTA-126-117 | AAGGAGCAACCTCACCATCACGCAAATTGGGTCGAACCGTAA |
| OCTA-126-118 | GAACGTTTTGAGGAGGTACGCCAGAATCCGGGGAAAGTGTAC |
| OCTA-126-119 | GCGGAACAATCAATCCACCGAGTAAAAGCTAAATCAACAGTGCC |
| OCTA-126-120 | ACTATGGCTTAATGTGAGGCAGGTCAGAAAAGGAAACTAGCA |
| OCTA-126-121 | GGAATATAGTTTTCATTTGGGGCAAGAGAACGTTGAA |
| OCTA-126-122 | TTAGAATAGTGTAGGGATCCTCATTAAACTTTCAAAGCCCCA |
| OCTA-126-123 | CATCAATTCTATTTTTTTCTAATAGTAGTAGAGAGGTCTGCTTTC |
| OCTA-126-124 | TCCCAATTGAAGATTGTTTTGCTAAACAAGCCAGAACTAGGGC |
| OCTA-126-125 | CATTTCGCATATGTAGAAAGGAACAACTCGATTGGCCACACC |
| OCTA-126-126 | AAAACAGCTGCGAAGGAGCTAAACAGGAGCGGGCGTGGAAAG |
| OCTA-126-127 | TGTCAATCAAATGGGACGAGCACGTATAGTAACCACCTTGAT |
| OCTA-126-128 | TAATAATAGGAGGTCGCCGCTACCCTGTATCACATCCGCCAG |
| OCTA-126-129 | AATCTCCAGAGCCGATTCTGCTGACGCATCAGTGAGAAAGGTGG |
| OCTA-126-130 | CAAAAGGGCCGCCAAGCCTTTTTTTTGCCACATGAAGCGTCAGAAC |
| OCTA-126-131 | TTAATTTTTTTTTGTATCGGTTTAATAACATAAATTTTTTTAACAGGGAAGCACAAAG |
| OCTA-126-132 | AGCGGAGAACGAATCGGTCACGCTGCGCACGTGCTTTAGATA |
| OCTA-126-133 | CATTGACTTTTTCATCGATGAACGGTAAAGCTATAGGCGCGT |
| OCTA-126-134 | CACCACCAAAAAAAAGTCTGGAGCAAACGCGAGCTGCGAGGAAGC |
| OCTA-126-135 | CGCAGTCGTATGGGGTATAAGCATTTTAAATTGTATGTAACATTTTTTTGTTGAT |
| OCTA-126-136 | ATTCACATGAGAATACCCCGGTTGATAATTGACCATTCCTCG |
| OCTA-126-137 | GCTGGCACAGAGCGCGAGTAGATTTAGTTCAGAAACAGTTTC |
| OCTA-126-138 | CGCCGCGTTGCTTTTCAATAACCTGTTTTCGTAAATTGCGAA |
| OCTA-126-139 | TATAAATATAGGCCACAAACTACAACGCGGAACCCCAAAGCG |
| OCTA-126-140 | TCATTCCAAACGTTTTTGTCGTCTTTCCCCACCCTTTCGAGCTTTTTCGAACCAG |
| OCTA-126-141 | GTTACATTCTTTACGGCTACAGAGGCTTCGGAACGCCCTCAT |
| OCTA-126-142 | ATCATCATGATTTTTTTACAATAAAACTGTCTGCTTTAACGTAAT |
| OCTA-126-143 | CTGTAGCCCGAAAGCACCACCCTCATTTCCTCATATTAAATT |
| OCTA-126-144 | TGTTTTAGGGGGGGGTCAAAAATAGCAAGCCCAATACTGTAGCTCATTTT |
| OCTA-126-145 | TAGCCTTAACTGGCCGCTCGCCGCA |
| OCTA-126-146 | GATTTGACCGCTGGTAATATCCAGAACAGTTTCGCACATAAA |
| OCTA-126-147 | GATGCTTTTGCCGCGGCATCTTCCAGGATTGGAACCCAGAGC |
| OCTA-126-148 | AATCAGGTGCACAATATTTCAACGCAAGGATAAAACTCTGAT |
| OCTA-126-149 | ACCATATAAAGACTTTTCACTCAGCAATGCCTGATTTTGTAGGTACCTG |
| OCTA-126-150 | GATTGCAGGGGGAAATAGCCCGAGACCTAATCAGCATTCCAC |
| OCTA-126-151 | AACACTGAGTACCCTGACTATTATAGTCAAAATAATCCCT |
| OCTA-126-152 | AAGTTTTTTTTTTCCATTAAACGGTAGCCG |
| OCTA-126-153 | GCCACCCCTAAAGTAATATTTGTTAAAGGTGTCTTCTGGTG |
| OCTA-126-154 | ACCAGTGAAATCGGCAAATTTTTAGAAAGGGTAGCATTCGTC |
| OCTA-126-155 | TAGTAATTTTTTTTATGAATTTTCTTCTGAATTTACCTTTTTTTTGTTCCA |
| OCTA-126-156 | TAACGATTCAGAGCACTTCAAATATCGCTGCTGAAGGAAGTT |
| OCTA-126-157 | AGACAGCTCAGGGAGATTAAGAGGAAGCTCAACA |
| OCTA-126-158 | ATATTTTAAGCCTTGATAAAAATAT |
| OCTA-126-159 | TTTGTTAGCAAATATGCAACTAAAGTACATTCGCGGTTAGCG |
| OCTA-126-160 | TTAACCACAAAAGAAGCCACGTTGTGTCTCAGAAGATGTACCGT |
| OCTA-126-161 | AAATCATCGGAGAGAAACAGCTTGATACTTGTTTAAAAATTA |
| OCTA-126-162 | TAATTTTTTTACTTTTGCGGGAGAAATGCAGCGAAAGACAGCATTGAGGACAATCAAA |
| OCTA-126-163 | TTATGACAAGATTCCAGGGAG |
| OCTA-126-164 | AAAGCCTCATCAACCATCGCCCACGCATACAGCCATCTGTAC |
| OCTA-126-165 | TATCTACACCGTGCACGTCAAAAATGAAATTTCTTGGTAGCT |
| OCTA-126-166 | GATAAATACATAATCTTTCCAGAGCCTAGGCTTGAAAAGGG |
| OCTA-126-167 | TGCAGCATGCCCTCTATTATTTATCCCACAACAACATGATAT |
| OCTA-126-168 | TTATCACTCCCGTAAGGCAAAGAATTAGTTCTAGCTGCGCCG |
| OCTA-126-169 | AAAACATTCTCGCGTAAAGCTAAATCGGGGCCGGAATATTCG |
| OCTA-126-170 | CGATTTTCGATAGTTGATAAATTCATGCACAGGCATCGTAGT |
| OCTA-126-171 | CAGAGAGATCAGCTTACAAAGGTTTGGTCATTGCCTGAGAAGGCTC |
| OCTA-126-172 | CGCTACCGCTTTTGCGGTTTTTTTGATCGTCACCTGAGG |
| OCTA-126-173 | AAAATAAAACCAATGACAGTCAAATCACCAGAGCAGTATCAT |
| OCTA-126-174 | ACAATGAATCCAAAGTGGAAACGATACTCTGGGGCAAGCAAT |
| OCTA-126-175 | GAGGTGAAATAGCAGAGAAATGACTGATACGACGGATCCAAT |
| OCTA-126-176 | TTAAAGGACGAGCGCAATCCTCGTTTCGGGGATTTGGCTGGTTTTTTTTTATTGCT |
| OCTA-126-177 | GTCGCTGAATTTGCATTAAACCCAATATTCTGGAGAAAAACA |
| OCTA-126-178 | ATTTTTGCGTTAACGGAGTCAGGCAATGATCTTCTTTTTTTTGAGATCG |
| OCTA-126-179 | TGAGAAATTGTACCCCGGTGAGCGTGGGATAGATGCAGTTAC |
| OCTA-126-180 | TCAACCGCAAAATTCAGATGGTAAGCCCGAGTACGTAAGAAA |
| OCTA-126-181 | TCGCGTCAAGACGAAATAAGAGCAAGAATCAGAGAGTAGATG |
| OCTA-126-182 | ATCGGTGGGGGTGTATTAAACCAAGTTATTTCTCCGGTG |
| OCTA-126-183 | CAGGGCGGTTGCTGCCGAAGCCCTTTTTAATGAACACATGAAGG |
| OCTA-126-184 | TTCAACACACAGGAAGCGAACCTCCCGAGGCGCATTTATCAG |
| OCTA-126-185 | GAGTTTATAGGAGTCCGTGAAGACGTGTGCAAGAGTAAT |
| OCTA-126-186 | GACCACTGGCGGGTAAGATTAGTTGCTATGACCAATCAGGCA |
| OCTA-126-187 | GTTCGCCATCCTTTTTTTTGGGAAGACTCCTGCGATCCGTGAATAA |
| OCTA-126-188 | ACTCAACGAAGCGGCTAAGGGAACCGAACTTTTGCATCTGAAC |
| OCTA-126-189 | GGGAGTAAGAGGTTCGGTGTACAGACCACTTGCGGATGCACT |
| OCTA-126-190 | TCGTTCGATATTCATTACCCAATCAGAAGATTGGGTTTCAATGTTGTGCAGAATCGTAG |
| OCTA-126-191 | CTGCGTGAGCAGCGGAGCTCGAATCGGCTATTTTACCCAGCT |
| OCTA-126-192 | TAAACAGAGCGTATCGAAGCGGAGAACAATGATAAGAGGTTT |
| OCTA-126-193 | CTGACCTGCGTTTTGAATGGAATAATGAGTATCAATCTAAGA |
| OCTA-126-194 | CTTGACACCGGTATTGAGTTAGAGTCTGACGACATGTTTCTTGTT |
| OCTA-126-195 | AACGTATTTTTTTTACAAAGCTGCTTAGGCA |
| OCTA-126-196 | AGAGGACTTAAATCTTTGTTTTATGGAGTCAACGACTCGGAT |
| OCTA-126-197 | ACGCGAGTCATCAAGTTGCCAGGAGGATGCAACGAAGAGCCA |
| OCTA-126-198 | GGCTTATAGAACCGAGGGTTGTCGGACTGAAACCAACATTGCAAG |
| OCTA-126-199 | ACAATTTTCAATCAAGAGCAGGCTTTTGACGACTACAGCAAATTTTTTTATCAAA |
| OCTA-126-200 | TGAAGCCAGATGAACGAAGTGAGCGAAAGTGTTATCTGCTAC |
| OCTA-126-201 | TCGCTACGGATGATTGAGTATTACGAAGTTAACTCCTTTGAA |
| OCTA-126-202 | TCCGAGTGAGTTATTGCTAAACTGGAAACTGGAACAGGCTGG |
| OCTA-126-203 | GACTTGACGCCCGATTAAATTCCAAAGAGGGGATTATTA |
| OCTA-126-204 | GCTTGAAGGGGATTTGCTCCATGTTACTGTAAAATAACAGTTCATTTCGAGAATG |
| OCTA-126-205 | GAACGAGCGATTGTATTTGTATCATCGCAAAGAGGGAATCGT |
| OCTA-126-206 | CTTGCTCTTTAGACTCATCTTTGACCCCCAGCGACGTAATAG |
| OCTA-126-207 | ATGGGTATAAATTTTTTTGGGCTCGCGATAAGACGATAACATAAC |
| OCTA-126-208 | GTAATACATTCATTACCAACCTAAAACGCTGATAAATTGGTA |
| OCTA-126-209 | TAAAATGGAGGCCGAGCAATACATCAAAATGAACGGCGAAAC |
| OCTA-126-210 | TTTGCACATCAGTTGAGATTAGAAAAAGCTGAGTTTTTTCAGATCACGCATCAGAACTG |
| OCTA-126-211 | CATAAATAAGGGGTCGCTAACAGTAGGGTCCATGAATTGTGT |
| OCTA-126-212 | CAGGTAGACGAACTTGTTGAATAAATCGTGGCAGAGCTGATTTAT |
| OCTA-126-213 | ATACCATTTTTTTTCATTCAACTAATTATCA |
| OCTA-126-214 | GCCACTACGCGACCTGACGCAGACCTTTAAACTGCATAAACA |
| OCTA-126-215 | ATACACTAACGGAGGCTGGAACTGCTGGCGCCGCGGAAACGT |
| OCTA-126-216 | ATTGCTGATATTTTTTTCCGTTTAGCTGAAAGCAAAATATAGAA |
| OCTA-126-217 | TTTTGGGTCTAATACGGGAGAATTAACGAAAAGTCAATTAC |
| OCTA-126-218 | CGAAAATGTAGTGTCAGATATAATTGAGCGCTAATAACAATGATCCGGGT |
| OCTA-126-219 | AAAGTACAAAACACTGGATAGCGTCCAATCAACGGACCAGGA |
| OCTA-126-220 | CCAAGCGGAAAGAAAGGACGGCGGCTTAACGGAACAATTATA |
| OCTA-126-221 | TAATAAAAAGATTAAAGAAGTTTTGCCCATGGATGCATTACGCT |
| OCTA-126-222 | GAACGATTTTTTTGGCGCAGACGGTATCCTGAATCTTTTTTTTTTACCAA |
| OCTA-126-223 | CGAAATCCGAAGGCGAATCCCCCTCAAATGCTTACGTGTTGA |
| OCTA-126-224 | ACACCATGATATTGGTTATGAGCCATATTACTGCGCGAAAGA |
| OCTA-126-225 | TCTTGTCCTGTTTCAAGATAACGCTTGTGGCGCAGTCATCGAGAGTTAAG |
| OCTA-126-226 | TTTATAATCTGCGCAGGTTCTCTGAGCTTCTAAATAAGCAGA |
| OCTA-126-227 | AAGGTATAGATGATCAGCGATGCCAGAGATGTCTTAATAGCA |
| OCTA-126-228 | CAAGAATTGAACAGACATCACGAAGGATCGGGTCTGAATT |
| OCTA-126-229 | CCGTTTTACCGCACCGGTCGGACTGAACTTTGTAATTTACATAAAC |
| OCTA-126-230 | GAATCATAGAACGGTCGTGCATTTTCAGCTTGGTCAAAGGATTTTTTTGAGTGAG |
| OCTA-126-231 | AAAGTCATCTTACCAGTGGTGCTTTTGCTCAACCGCCTTG |

**Supplementary Table 8 | DNA sequences of cube DNA origami (Scaffold 10004)**

| Name | Sequences |
|---|---|
| CUBE-126-1 | ATTAAACCCTCTCTCGAGGCATAGTAAGGATAAATTAGTCCT |
| CUBE-126-2 | TACGGGGTTGATCGGGTTTTGTTTTACGCCAGGCACCGGA |
| CUBE-126-3 | CGACATGAGTATCCCACATTCAACTAATTACTAGGCGCACG |
| CUBE-126-4 | TGAGTGTTGTGTGCAAAAGACGACGATAAAAATCTTCTTTCGTGA |
| CUBE-126-5 | CAATCCTTACGGTGCTATCATAACCCTCGGTTTGACCACTGATTTGAGCCGATCC |
| CUBE-126-6 | GCACTTCCAGGAGAAAAAAGC |
| CUBE-126-7 | CAGCCTCAGGGGGTTTTAGTTAATTTCACCAAAAAAATGAC |
| CUBE-126-8 | GTCCCGCACTCGCGGTTCGCCAAATACCGACCGTGTAGCAACAGAAACGA |
| CUBE-126-9 | AGTCAGGAATAGCAAACAAGTCTGGAAAGAGTGATGAATGAA |
| CUBE-126-10 | TAAGAACAATAACAAGATGAACGGCTGGCCTGTTGGCCTTTA |
| CUBE-126-11 | AGGTAGAACAGCCAGTGATTTCTCTTTTGATAACCAGTTTCTTTTTTTGTTGTTC |
| CUBE-126-12 | GTGTCAGTGTCCTTGTCGGACTTGTGCACTCATTGGACGGGA |
| CUBE-126-13 | GAGAGTGGCGGCAGGACTCCTGTTATCATTCTCACTAAGAAA |
| CUBE-126-14 | GCTGAATCAATCACTTTGATGAC |
| CUBE-126-15 | ATAAAGTGCAAGGCAAATCAACGTAACAAATCACTGGATGC |
| CUBE-126-16 | GATTATGCAAAAATTCATCAAGAGTAAGAGCCGCAACGGGA |
| CUBE-126-17 | TGATGCGCTGACATAAAGCCCTGACGAGAAAGTCATAGTCGGGCA |
| CUBE-126-18 | ACAGAGATTTAGCTCTCATTCAGTGAATTTTGCCAATAAATGGGCTCGCTGATTC |
| CUBE-126-19 | TATTTTTTGCTCGAAACCGCCTCCCTCATCTTGACGAGTTAG |
| CUBE-126-20 | AGCATTGCGACAACATCGGCATTTTCGCACCAGACTTACTG |
| CUBE-126-21 | AGGTGAGAACCCTTCTATGGGTTCTTTTCATAATCAAAAGCTGGAAACGA |
| CUBE-126-22 | AACGTCTGAATGGCGGCAGATTCACCAGTATCAATAAGAACC |
| CUBE-126-23 | ATCAGGTCCAGGTATCCTGCGCCGGTTGGTATTATACGAGTA |
| CUBE-126-24 | TGATTTATGACCTGCTGGCCA |
| CUBE-126-25 | AGAGCCACATTACCAGTTTATAAATGAGTCACACGCAGACAA |
| CUBE-126-26 | AGAACCGGCTGACCGATTGGGCGTTTCAATGTTGTCAATCGTCTTTTTGAAATG |
| CUBE-126-27 | CCACCCTCTTTTCACCACCGTAGCATTTTTTGACAGCCCACAGACCAGTTTTTAGGCTG |
| CUBE-126-28 | ATTAGCGAAGGCTTCATTGCTGATACCGTAAAATATTGT |
| CUBE-126-29 | GGATATTCCACCGGGGCCGCGATTAAATCGTGGCAACCAGTA |
| CUBE-126-30 | GTAAATTGCGTTTTTCTATCGATTTTTTGGGAAGCCCGATCCCCTTTTTGGGAAAAC |
| CUBE-126-31 | AGTCTGATTACATTTATTAGTCTTTAATCATATTCCACCCTC |
| CUBE-126-32 | TTTCTTTGCAGTGTTTAGAAGAATATCCGATAATGCCCCCTT |
| CUBE-126-33 | CATACATGGACATTAAAGCGTAAGAATATCCAACACGGAACC |
| CUBE-126-34 | TGGACTCCTAATGAATAGTAAAATGTTTCATAAATGAATACA |
| CUBE-126-35 | ACGCTCATGGCATGCAGCTGACCAACTTTGAAAAGTTTTCTGTAT |
| CUBE-126-36 | CACCCAATTTATGTGTCAATCATAAGGGACATTTGTATC |
| CUBE-126-37 | CGTAAAGATTATACGTCGAAATCCGCGACCTGCAATTTGACC |
| CUBE-126-38 | AGAGCTTGGCGAAAATTCATT |
| CUBE-126-39 | CTGGTATGCTAAATAGCGTAACGATCTAAGAGGACGGTGTC |
| CUBE-126-40 | ATTGCACAAGTTTTTAGTACAACGGAGGTTAGTAAACCAGCC |
| CUBE-126-41 | AGGGAGCTGGAGATCATCGCTAT |
| CUBE-126-42 | TGCTTGTTTTTTCTTGGAACAAGAGTCCGTGCGAGTAGCGAG |
| CUBE-126-43 | GTCGGGTGGATTTATACATAA |
| CUBE-126-44 | CCCAGCGCACTAAACAGGGCGATGGCCCATATAGAAGCCGGA |
| CUBE-126-45 | CTAAAACGGAGCCCAAAGGGCGAAAAACATTCCATATAGCGTCCTCCATG |

| CUBE-126-46 | GGGGTTTATATCCATACATTTTGACGCTGCAGATCCAGATGA |
|---|---|
| CUBE-126-47 | GAATCGTAGACTGGGAAGGTTTATAAGTCAACGTCCCGATTT |
| CUBE-126-48 | GAATCCCGGGGGTAAGACAAATCTTTTCTGCAATACTATTAATTTTTAGAACG |
| CUBE-126-49 | TTTCCAGAACCGAAACATCACGAAGGTGATACAGGAAAA |
| CUBE-126-50 | ATCGCCTGATGCAGACGAGATGAAGGTATAGACTACGTGAGGTGC |
| CUBE-126-51 | ACGAGGCAAATTGTCAAGCGCGAAACAATGGGGTCGAACCAT |
| CUBE-126-52 | TTACTTGTCGGCATACAAATCGTCTATTCGGAACCCTAAAGACTCATCTACTGCG |
| CUBE-126-53 | ACGGTGTTCATAGTCAACTTTCAATTTTTCAGCGAAACTATTTTTCGGCCTTG |
| CUBE-126-54 | TTGTCTCAAATACCGAACAATATTACCGTGAATTTTGTCGTC |
| CUBE-126-55 | CGAACGAATGTAACATCTGGTGCTGTAGAGGTCATGCCCACG |
| CUBE-126-56 | AACAACTAATCCTTAAAGGGATTTTAAGTTGCGAAGTATA |
| CUBE-126-57 | CTAAAAATTTGAGAGGCGGATAAGTGCATAGGTGAGCTTGA |
| CUBE-126-58 | TATATTTAAGATGAATTACCTTTTTTAAGACCGGAGGATC |
| CUBE-126-59 | AAACAGGAACCATCTTTTGCGGA |
| CUBE-126-60 | GCCAGAACTTAAACTATCACC |
| CUBE-126-61 | CACGTATTTGCAGGATTAGAGAGTACCTAATAATTACAT |
| CUBE-126-62 | TACCGATGACAGGAAGGAGCACT |
| CUBE-126-63 | GTCACTTTCAGCAGCGCGCTACATTTTGGGCGCGTACTATGTTTTGCGAGCAAAC |
| CUBE-126-64 | CATAACCTCAGAGCTAGTTTGACCATTAGGTGTCTTGTTTTA |
| CUBE-126-65 | TGGCTTAGAGATCCGACAATGACAACAGGCCGAAATTCTG |
| CUBE-126-66 | GCCCGGACGTCGAGTCAATAGATAATACTATCTTTACGGTAC |
| CUBE-126-67 | GTACTCACAGTACCGATTTAGAATTTTAGACTTTAATTGAGGATTTTAGGTTAT |
| CUBE-126-68 | GGAGGTTTTTTTCGCCACCCGCAG |
| CUBE-126-69 | TCCAACACATTTGATGAAACAAACATCAATAACCTTGACGAG |
| CUBE-126-70 | TGATAAGCTCAACAGGAAGTTTCATTCCGTAGATTGGGAGCT |
| CUBE-126-71 | TGCTGAAGTTGATTCAGATTAGAGCCGAGGGTTGATCTTAAT |
| CUBE-126-72 | TTAACAATTTGGTCAGGGAGTTAAAGGCCGCGTTGCTTGTTTAGC |
| CUBE-126-73 | AATATGCTTAATTGTCGGTCGCTGAGGCAACGTGCTTCGCAAATGGTCAAGAAAA |
| CUBE-126-74 | CAAAATCTAAAGTACGATACATTTTCCTCGTTAGAAAATATATCTCCTTT |
| CUBE-126-75 | TAAATAAGGAATTAGTACAAAACGAGGTCAATGGGACATAAC |
| CUBE-126-76 | ATCATAATGCAGATTTCAGGATGCAGGTCATTACGGATAAAT |
| CUBE-126-77 | CTGTTTAAGGAATATGCATATGATTTTGACGCTGGCCCGTTAGTTTTTTAACTAT |
| CUBE-126-78 | GTATCATATTTTTATACAAAATTTGCCTTTTCAAAATAAAGATTCATTTTTGAGATTT |
| CUBE-126-79 | ATTTAATGTTTACCATTATTATCACGAGCGAGATAGGGT |
| CUBE-126-80 | ATGATATTAACATTTGCCCCCTGCCTATTCAGTGCAATTTTT |
| CUBE-126-81 | AAAGGGCGGAATTAAAAGGGCGACATTCGGAAATATTTAAA |
| CUBE-126-82 | TGGTCAGTTGTTAAATCAGTATTAAGAGGCTTTTGATGATCGGCA |
| CUBE-126-83 | TCAGGTCCGTTAATAAGGTAAATATTGACAACCGAACCACCA |
| CUBE-126-84 | AAAGCATTATAAACGTCATACATGGCTGAGACTCTTCGACA |
| CUBE-126-85 | CCCTCACATGCCGGACAGCTCAAAGTTTTAACGGGGTTCGGAATTTAAAA |
| CUBE-126-86 | TTGTAAAATTGCCTGGCCGGAGACAGTCCAAAGAATTGAGGGAGATAAAC |
| CUBE-126-87 | AAATCCCTCACCTTAACAGTTGAAAGGACAAACAACTCAAGA |
| CUBE-126-88 | GTTAAATGAGGGTATTCTAGCTGATAAATATTAATCCTATTA |
| CUBE-126-89 | CAGTGCCCGTGGATTTTGTTAAAATTGTCTACACATCAAC |
| CUBE-126-90 | TAAAGGTCAAAGACATCATCATATTTTGATTATCAGTAGGTAATTTTTAGATTCA |
| CUBE-126-91 | GAATTATCTTTTCACCGACCAGATAGCTTTTAAAGTTACATTCATATGTTTTCCAGCGC |
| CUBE-126-92 | AGTGTACACATGAACTTTGCCCGAACGTTTATCAATATC |
| CUBE-126-93 | AGTTAAATCATTTTGCGGAAAAATCACAAGGCTA |
| CUBE-126-94 | GAAGGATCAGTAAGTCAAAAGAATTTTCCGAGACCTAGCAAATGTTTTTAAAAATCT |
| CUBE-126-95 | TTCTGAATGGTAATTTTTTTAACCAATAATCAAA |
| CUBE-126-96 | GAAGGAGTGAGAAGAGAGTCTGGAGCAAGCAAATTATTCAT |
| CUBE-126-97 | ACTCGTAGCAAATCGCTGAACCTCAAATGGCCGAAATACAGG |
| CUBE-126-98 | GTTTGAGTCAACCGGCTATTTTGAGAGCGCGTTACTTGAGTAA |
| CUBE-126-99 | CTTTATTTTAGAACATACATAAGGTGGTCCTTATTTGATGC |
| CUBE-126-100 | CCAAAAATAAAGATTTATTTTCATCGTTCAGATATATGGTA |
| CUBE-126-101 | TATTTTATAATCCTACGGAATAAGTTTACAATAATTTGTAAC |
| CUBE-126-102 | GATATTGTCCTGATCAATAGCAAGCAAAAGGAATCCACGTAA |
| CUBE-126-103 | TAACACGAGCATTAAACCGAGGAAACGTTTTGTCGCAATTC |
| CUBE-126-104 | AGAAACCGTTTTTCGGCATGATTAAGACCAACATAGAATAAT |
| CUBE-126-105 | TTGATAACGCTAACGACCCTGTAATACTTTATTTGATTACCGCGCGTAGA |
| CUBE-126-106 | ACTGGCACATCGATTGCCTGAGTAATGTGATGATGACAATCA |
| CUBE-126-107 | TCGATGAAGCAATAAAGGATAAAAATTTTACTTCTTAAAAGA |
| CUBE-126-108 | TGTTAGCAAACCATGAATAAATTGCAACCAGGAGAGAAGC |
| CUBE-126-109 | CTTATCCAGCCGTTAATTGCGTATTTTTCAGGTTTGCTAAATCTTTTTGGTTGTA |
| CUBE-126-110 | GGTATTCTTTTTCGCGAGGCGACA |
| CUBE-126-111 | TACCCAAAGACACCGATTGTTTGGATTATTAGAACACGGGAA |
| CUBE-126-112 | GCCAAAAGAATAAAGGGAAACGCCTGGTGCTACGGATAGATG |
| CUBE-126-113 | AAAGATTTTTGCCAGACCTCAAATTTTTAAACAGTT |
| CUBE-126-114 | AGGCTTTAAATATAGCGGAGCCTATTTAAAAACGGCTGCTGACGTTTTTCACCGGTG |
| CUBE-126-115 | AAATACCTACCATATCAAAATTTGCGGGAACGAG |
| CUBE-126-116 | ATAGAAACAGAAGGACGCTGACTTTTTGGGACGGCTTTTCCATTTTTTGAATTGG |
| CUBE-126-117 | AACGCAAAAGAACTTAATCAGAATTGGTTGGATGACCTCATA |
| CUBE-126-118 | AACAGAAACATTATAGTAGGGAAACTGCTCAAAAATAGAAGG |
| CUBE-126-119 | ATCAATAAATGCAATGTGCTGGAACTGCTAATTGGAACGGAA |
| CUBE-126-120 | GGAAGGGTCAACGCCATCAAACGCCGCGGTTTCATTACGCAGTA |
| CUBE-126-121 | AACATCCTGATTGCTTTACGAGCATGTAGATAAGTTACCAGT |
| CUBE-126-122 | AGCTGAAATTATTCGAGTGAATAACCTTCCCTTAGTTAACCG |
| CUBE-126-123 | ATTAAGCAACAGTACTTATCATTCCAAGACATGTTGGTTGGA |
| CUBE-126-124 | AAAACCGTTTAGCCAATCCTT |
| CUBE-126-125 | GGTCTCCGATAGTGACGACAATAAACAAACGGGTGATGAAT |
| CUBE-126-126 | CCCTCCCGCCAGTGGTTTATCAACAATAGAAACCAATAACGG |
| CUBE-126-127 | GCGCATGGCTCTGACATCAATTCTACTATACAAAATAAATCGTCATCCCA |
| CUBE-126-128 | CTCAAGAGCGGTATCCTCAGAGCATAAAAACGTCAATTAAAC |
| CUBE-126-129 | GGCTGCTGTATCGTCATACAGGCAAGGCAGAAACAATCAATA |
| CUBE-126-130 | GAAAAATAATGCGACTAACTCCTCTACGGGGAGTAGCGTT |
| CUBE-126-131 | TAATTTTGCTTCTGTCGCGCAGAGGCGAAAGGTGGAAACGAA |
| CUBE-126-132 | GAAAACATATATGTATTTCAATTTTTGAGCAAAAGTCATTTGGTTTTTGGCGCG |
| CUBE-126-133 | TGCAGAAGTCTTTCCCTTTTACATCGGGAAAGAATTGGTAAG |
| CUBE-126-134 | TCCTAATTTGAATACCAAGTATAGTAGTCAGGCAATCTCTTACTTCAATATTAAT |
| CUBE-126-135 | CAAGTACTCCAGACTACCGGATAATTTGCAGCGGTCGGAGCCGGTTTTTTGAGCGTG |
| CUBE-126-136 | ATCGGCTCGCGCCTGTGCTTTTGCATGTGGCCAGATAGCAAA |
| CUBE-126-137 | ATACAGTAATAAAGCATTGCAGCACTGGCTTTCCGCAGCTAA |
| CUBE-126-138 | ATTCGCCAATAAATAGTTATCTACACGAAATCAATCCTGAACAA |
| CUBE-126-139 | GAATTTAGATGGTCCCATTAGCAAGGCCAAGAGCAGCATCTT |
| CUBE-126-140 | CATGTCAATCTGGGCAATGCATGACGCTTTTGCGCTATCT |
| CUBE-126-141 | TGAACGAAGCCCCATTTGGGAATTAGAGCCCTTTTTGAATA |
| CUBE-126-142 | CCTGATGGAAACATGCAGCACCGTAATCATCAGAGGTTCAAA |
| CUBE-126-143 | GCAGACCGGCTTTGTTAAGAA |

| CUBE-126-144 | ATCTGGAAACTGGTCAATTGAGCGCTAATAGTAGCGCCAGAGT |
| CUBE-126-145 | ATTAACTACGCAGATTAAGCCCAATAATGGAAACGGATGTTA |
| CUBE-126-146 | TGATACCTCCAGTTCCTGCCACATGAAGGTCAGATGACCTAA |
| CUBE-126-147 | TACTTGCCCAATATTTTTATCTGAACTCATCTTTAAAGGCGT |
| CUBE-126-148 | AATCGAAACTGGGGTAAAACTAG |
| CUBE-126-149 | ATCACCACTTATCAGTTACTCACCACTGCGATGCGACAGAAT |
| CUBE-126-150 | CCCGACACTCAGGCTCCGACCATCAAGCTGCCAATTCACCAA |
| CUBE-126-151 | ATGAAATACCATTAAGACTAAACATATGTACCCCGAAATCAC |
| CUBE-126-152 | TACCGAAGCCAGCAGTTGATAATCAGAAGTAATCGATTTGAC |
| CUBE-126-153 | AAGTAAGTTGAGCCAAAAACAGGTTTATTGTATAAACAAGATTTTTGAATCGA |
| CUBE-126-154 | CCACAAGATCGATAGGCAAAGGTAGCGTATTTTATGAGCGAA |
| CUBE-126-155 | CAGTAGCAGCAATATGAGTTGAAGGATCCAGAGCAGCTGACG |
| CUBE-126-156 | TTAGCTTTTACTGTAGCGGGCTTGAGTTTTTAATTT |
| CUBE-126-157 | CAAGTTTAGAGGGTGATCCCCGGGTTTGCTCGCCAAGTTGCCTTTTTAGGAGG |
| CUBE-126-158 | TGAAACCAATTGAGCCGTTCCGTGGCAATCGAAGTCCGTACT |
| CUBE-126-159 | TGTTTCTATGCATGGTAAACAGAGAGGTAGCAAAAGATAAC |
| CUBE-126-160 | CAGATGATGCCTCTACTGCGTGAAGCGGAGATCACAGAAACA |
| CUBE-126-161 | CAGAAGAAAAATAGGAGGTTGAGGCAGAGCCCAATCACGTT |
| CUBE-126-162 | CCCTAAAACAGTACCGCCACCAGAACCCCAGTACAATAGAA |
| CUBE-126-163 | CTGCAACCACGTTGACGAATGGATCCTCCCGCCACTTTAATT |
| CUBE-126-164 | AACATCAACTAAAGACTGAGTTTCGTCAACCACCATGTCTGC |
| CUBE-126-165 | TTTTATAATTTATCACGCCACCCTCAGAAATTAAAGGGGGGG |
| CUBE-126-166 | ACGCAAATCCAAAATTTCAGGGATAGCAGTCAGACACATTGC |
| CUBE-126-167 | AGGAACACTTGCCTCCATTAAAAATACCATAAAACGAGCCGC |
| CUBE-126-168 | GTATCGGTCAGTGACGCTGAGAGCCAGCGCAGGGGCCAGAAT |
| CUBE-126-169 | GAAAATCTTAACCGGAGGTGAGGCGGTCTGATGTTGATTGGC |
| CUBE-126-170 | CCATGTAATTGACATATCATCATGAACAGAACGAATAGTAAT |
| CUBE-126-171 | AACGCCTCTCAGAGATACAAGGGTTTTTATGAGCGCGCGAATTTTTCTGATAG |
| CUBE-126-172 | GCCACCATTCACAATGTCTCAAAATCTCAGTATTAGTCCATC |
| CUBE-126-173 | CGCCAGCCCGTAACGAATTGCGAATAATCTTTGATCCACCAG |
| CUBE-126-174 | TCTCTTTTTTACCGTTCTAGGATTAGTTTTTTTGCT |
| CUBE-126-175 | GGAAAGCTCAGAACGCTTGCTTTCTTTGTGAATTTTCCTGAGTTTTTAAGTGT |
| CUBE-126-176 | CTTGATACCCTCATAAAAAGGCTCCAAAAGAGTCTACACCGC |
| CUBE-126-177 | TTACATAAACATCGGAGTAGAAGAACTCGAGTGAGAAACTAC |
| CUBE-126-178 | GGAAAGCAGTGCCAGGCCACCGAGTAAAAGGAGCCCCTCAGA |
| CUBE-126-179 | ACAAGATTAAAACATTGTAGCAATACTTAATTTTTTAGGAAC |
| CUBE-126-180 | TGTAATATGATGAAATCATTGTGAATTCGCATTATTGTGAG |
| CUBE-126-181 | AAGAGTTTCTAGATCGGAAACGAACTAACGGATCCAAACGGATTC |
| CUBE-126-182 | TAACGGTACGCTTGGAAGAAAATCTACTTGTTTAATAAGCTTTTGCCAAGCACT |
| CUBE-126-183 | CCGCAGTCCCGTTCACTACCTTTTTAACCAAAGCGCTAAAGA |
| CUBE-126-184 | CCAACTCTTTTTGCGCGTAACCACCAGTAGCAAAGGAAGC |
| CUBE-126-185 | TTTTCGCGCAGTCGAAGAGTCAATAGTACTTCAAGCATCGG |
| CUBE-126-186 | AGCGGAACCCTGTTTATGTAAATGCTGAAATCAAAATACGTA |
| CUBE-126-187 | AATGCGCAAAGACAATATCGC |
| CUBE-126-188 | GGCGAACGTACGAAGCGAGAATGACCATATGCAAATCGGCCCT |
| CUBE-126-189 | GGCGCTAATGAGGAATTATAGTCAGAAGCTCCGGCCAGGAAA |
| CUBE-126-190 | AACGAGGCACCCGCTCTGAGCTA |
| CUBE-126-191 | ATGCCACTGGCGAGTGTATCACATATTCTTTGCCGCCAATCG |
| CUBE-126-192 | CTTTTTCGGGCGCTCCGAGCGTAGCGAGCATCTTCTTAGGTT |
| CUBE-126-193 | TCAAAAACTGAGAGGTAAGCCATGAACCGAGGTAATCAAAAT |
| CUBE-126-194 | CCGAAAGGAATTTACTGGCTTGGAGGAGAAGGTTCCGCGCTT |
| CUBE-126-195 | GTTTTAAACGCTGAACCAAAACTTTTCCTTTCAGCCTTGCATTTTTGGGCGGT |
| CUBE-126-196 | TTCGAGCTTTTTGCGAACCATGGAAACAGTTTTAAATCAATAGCGATATTTTGATTAAG |
| CUBE-126-197 | TCTTTACATAACTAAAGTATCTTCCTGGTCAGTGAAGGAGCG |
| CUBE-126-198 | CATAGGTGATTAAGCGGCTACAGAGGCTTCACGCTCCGCTCG |
| CUBE-126-199 | AAGAACGCGTTTTACTTTTTCTGCA |
| CUBE-126-200 | CAAGACACAGAAAAGCACCAACCTTTTACGAAAGAGACGGGGTTTTTAAAGCC |
| CUBE-126-201 | GGGTTATCCTGACTAGTTTCCATTAAACAAGCGAAGCGAGGA |
| CUBE-126-202 | CTCACTTTATATCCAAAGGAAGGGAAGAGGGTAAAAATCAGG |
| CUBE-126-203 | TCTCCGCCGAACGAGGCAAGTGTAGCGGTTGAGGAGATTGCA |
| CUBE-126-204 | ACATCAATGTCAGGTCGCCATATTTAACTTGCACCCGGAATC |
| CUBE-126-205 | ATGGAATGACACCCCAGTATAAAGCCAACGAGCGTGACGAG |
| CUBE-126-206 | GGCTGGCCATACAGATTTTCGAGCCAGTTGCGGGACTGCCTC |
| CUBE-126-207 | AGCAACGTAATAGGTCTTACCAACGCTAACGCTCACCATAAC |
| CUBE-126-208 | AACAGCAATTTCTCCCGAACCTCCCGACTAATAAGAACGGGGG |
| CUBE-126-209 | GGTGTTAGATACCAGATTAGTTGCTATTAACGCCATACCCGG |
| CUBE-126-210 | GGGAAATAAGTCCGCCATTCAACAGAGTAACGCGGACAGTAG |
| CUBE-126-211 | GGTGAGTAATCAAATTGCTGATAAATCTGGACTGAGAATATA |
| CUBE-126-212 | GCAGACCTCTCGGATACGACCACTGGATAACTGCCACATGTA |
| CUBE-126-213 | AATTTTATTGAGAACGTGGAATGAGACATATCGAAACTGGAA |
| CUBE-126-214 | AGAGCCTATTCTTAGGTAAACCGTTTGGCAGGAACGAGTCATTTTTCAGGAGA |
| CUBE-126-215 | AAGCCTTCAGAGGCTCCAGCTTGGAGCGGATGAGCTTACGAA |
| CUBE-126-216 | GAGCGTAATGAAGTCTCGCTCAGGCGTTCGCGTTTATACC |
| CUBE-126-217 | GGTCGTTATCCGATTAAAAACAGGGAAGACCTTATGTCTGTA |
| CUBE-126-218 | GCCATCACTCATGTATTATTTATCCCAAACAACAATCAAAG |
| CUBE-126-219 | GCACTGGGTTGATGCGAAATGCACGTCAAAAATGAAACGTTGGTGAAAAT |
| CUBE-126-220 | GGCTTAATCCTGAATTGTATTGATGTTGTTGCTAAGCGGAGA |
| CUBE-126-221 | AAAGGTTTTTAATTCTGCGCACTCATTTTACAAGCA |
| CUBE-126-222 | AAGTACCGTTTTAGTTCATTACAGTTTCGGCTTTTGTGTTGATTTTTGCTTGA |
| CUBE-126-223 | ATTTAGGAAATCAAGGATCTTGCCATCCGTGAGTATCGAATC |
| CUBE-126-224 | AGCGGAATGGAGAGTGAAGACGGAAACCTTATTTTTCTTTCC |
| CUBE-126-225 | GTTCGTGTGGTTTAACTCAACGAGCAGCTATGGAAGGTTTTG |
| CUBE-126-226 | AACTGAGCGACTGCTGGGAGTAAGCGTAGACGAGTCAGCTAC |
| CUBE-126-227 | TGTGGCGCAGGGTTTTAACAGCGATCGCTGCCAGAGCGATTT |
| CUBE-126-228 | AGTCGTCCTGGGAATTAATCGAACAAGACATTCGCTTATTAC |
| CUBE-126-229 | CAGAGAGTGGCTCACGTCTTCACAGCGAGTATTTCGACGCCT |
| CUBE-126-230 | GAATTAACAACTTTCCGTACTCATTTTCGGGTTGACATTCGATTTTTTCCTGTT |
| CUBE-126-231 | CTGAACACTTTTACAAAGTCGCCT |
| CUBE-126-232 | CGATTTTGTTAATATTTTGTAAAGATATTGTGACCTGG |

# Supplementary Table 9 | DNA sequences of 48helix DNA origami (Scaffold 7249)

| Name | Sequences |
| --- | --- |
| 48h-pore-1 | GATCACGTTGGTGTTTTGAGGGGACGAAAGCTATTCAAGAGA |
| 48h-pore-2 | CCAACGTCAAAGCAATTTTGGAACAAGATGCCCCA |
| 48h-pore-3 | AAGAAAAGTCTTTCCTTATAGTCCCTTAGACTCTATGCCAAGTTTAAGA |
| 48h-pore-4 | ACGCTCACTGAGAGACTACTTTTTTTTTT |
| 48h-pore-5 | GCTTCTGGTGCCAATCAATAGAAACGACAGTATCGCGGCACC |
| 48h-pore-6 | TGAAATATTACTAGAAAAGAGGGTAACTACGCCAAGTCAAAGGTCAGA |
| 48h-pore-7 | ACCTAAATTTAACTAACTTCGCTAAGAACGCGAGATCTTCTG |
| 48h-pore-8 | TTTTTTTTTTCGTTACCAGATGCAAAGAAGTTTTTTTTTT |
| 48h-pore-9 | TTTTTTTTTAACGCCATCAAAAATAATTCGCGTGCAATCAGCTAGCCCC |
| 48h-pore-10 | TTTTTTTTTTTCATTGAATCCCCCTCAACATAAATTAGTAGCCATTTGG |
| 48h-pore-11 | TCAAATATCAAACCCAGTTGAAAGGATTAATATCCAAAAACG |
| 48h-pore-12 | TTTTTTTTTAGTATAAAGCCATGTAATTTAGGCTTTTTTTTTT |
| 48h-pore-13 | AAAGAATCTAATAGCAAAAATCAGGTTTTTTTTTT |
| 48h-pore-14 | CTAATGAGTGAGCCCAAAAGAATGGAAGCATAAAG |
| 48h-pore-15 | GCCCGGGCGAACTGAACGAACCAGACAATATTTTGAATGGT |
| 48h-pore-16 | TGTGATAAATTTCAAAACTTTTTGGGAA |
| 48h-pore-17 | TTATCCTATTCTGTATCAACAATAGATAAGTCCATAGAAAGG |
| 48h-pore-18 | GGTAAAATACGTGCCCCTAAAGACTTTTAGGCTTG |
| 48h-pore-19 | TAAGAGGTTCGAGCACAGGTCGATAAGA |
| 48h-pore-20 | ATGTGAGTGTAGCGGTCACAACACTAGG |
| 48h-pore-21 | CATGACAAATTTACCGTTCCCCCGAAAGCGCAGTC |
| 48h-pore-22 | CTTTAGCGTCAGACCGAGGCGGCAAGCAAAGCCGT |
| 48h-pore-23 | ACCTCTGTGGAAATGATTATTTATGCCTATTTCGG |
| 48h-pore-24 | GGTAGTGTAAATATCCATATAACTTTTGCGGATGAGCTCA |
| 48h-pore-25 | CGTAGAACCTTATTACGCAGTACTGGCACAATAATTTACCAGAAAGTAA |
| 48h-pore-26 | GAGCCACTCAGAACAAGGAATTAGAGCCCAAGGCCGGAAAGTGCCTTGA |
| 48h-pore-27 | TCAGTACCAGGCGGATAAGATGCCTTGCTGATTAG |
| 48h-pore-28 | TTTGACCGCCCGAATTTCATTGCAACTAAAGTACGAGCATCGGAACGAG |
| 48h-pore-29 | TGACAAGAGAGGCAAAAGAATACACTTTTTTTTTT |
| 48h-pore-30 | TAACACCCAAATGATTAGTAATAAAAGGGACCTGAAAGCGTTTTAAAAG |
| 48h-pore-31 | TCAGCATTGCAAAAAAAGGCTCCAAAAGGAGCTTATTTCAA |
| 48h-pore-32 | CCAGACGTAATAAGTAACAACGCCAACA |
| 48h-pore-33 | TTTTTTTTTGTACCTTTTCTAACGGAACAATTTTTTTTTT |
| 48h-pore-34 | TAACTGAACACCAGCGCATTACACCGGACACGTCACCAATGA |
| 48h-pore-35 | TTTTTTTTTTGATGCAAATCCATATTCAACCGTTTTTTTTTTT |
| 48h-pore-36 | ACCGCCTATCGTCGACCGAGCTCGAACATGCGTAT |
| 48h-pore-37 | CACCGGATCAAAATGACGGGAAACATAAAAATAGCAGCCTTT |
| 48h-pore-38 | AACATTAAAAACCTAGTAAGAGCAACAAAAGGACGTCAGA |
| 48h-pore-39 | TTTTTTTTTAGAGGCATTTTCGAGCCAGACGACAA |
| 48h-pore-40 | ATTACGAAGGAATACCACATTATTCATCGTTAATACAGGACGGTGAATT |
| 48h-pore-41 | TTTTTTTTTAAGAGGACAGGCTGACCTTTTTTTTTT |
| 48h-pore-42 | AACCAGTTTGCCTAACCATTACCATTAGAGCAAAACAGAGCCCACCCTC |
| 48h-pore-43 | CTAAGTATTAAGAGGCTGAAGACAAAAGAAATATTGAGCCACGCCACCC |
| 48h-pore-44 | TCCTTTTAGGATTAGAGAGTACCTTTTTTTTTT |
| 48h-pore-45 | AATCATTTGGGAAGAAGGAACAACTAAAGGAACA |
| 48h-pore-46 | TTTTTTTTTTCTTTACCCTGACTATTATA |
| 48h-pore-47 | CAGAATTTTTAAATAATATCTAGTTGGCGCGCAATTCATCAA |
| 48h-pore-48 | TTTTTTTTTTTTAATATTTTGTTATAGAAGGCTTTTTTTTTTT |
| 48h-pore-49 | AGGCCGCGTCGCTGTCATGAGGAAGTTATGCAACGGCTACAG |
| 48h-pore-50 | TCCTGTTGCTGGTTGTCCACTATTAAGGACAGATAGGGTTGA |
| 48h-pore-51 | CTTTCACGTTGAAAATCTCACGAGCACGGGAGCTACCAAGCC |
| 48h-pore-52 | TTTTTTTTTTCATTATTACAGGTAGAAAGCAACTAATGCAGTTTTTTTTTT |
| 48h-pore-53 | AACCCTACACAAGCGGTCCACTGATGGTGGTTCGA |
| 48h-pore-54 | TATCTTACAAGAAACAATGAATAAGCCCCCGGGTCTATTAA |
| 48h-pore-55 | TTAGAATGGGATTTTACGTTGCGCTCACTGCCCCG |
| 48h-pore-56 | ATTCGCCAGCTTTCGCCTCAGGAAGATCGCTTTTTTTTTT |
| 48h-pore-57 | AATAAATCATACAGAAATCGGTTGTACTATAGCGTAAACAGT |
| 48h-pore-58 | TTTTTTTTTTCTGAATATAATGTTACTTAGCTTTTTTTTTT |
| 48h-pore-59 | TATAACGTACTATGGTTGCTTGCGCCGCAACTACATCAATAT |
| 48h-pore-60 | CATTAAAACGGGGTACGAGTGTACTGGTAATCCCTTATAAAGGCGAAAA |
| 48h-pore-61 | AGTCAATTAGCTTAGATTAAGACTTTTTTTTT |
| 48h-pore-62 | TTGACCGATTCTCCGTGGGAACATTTTTTTTTT |
| 48h-pore-63 | ATCAGAGATCCCAATCGGCGAAAAACCGGATTGCCGAGCCAGGGTGGTT |
| 48h-pore-64 | TCACATTAATTGGACACCCTCGAGGTTTATGTGTACAATTGA |
| 48h-pore-65 | GTGTTGTTCCAGGAAAACAGGTCATCAG |
| 48h-pore-66 | GAGCGTCAATCGTCAGTCACACGACCTTGCGGAACAAAGAATGAGTAAC |
| 48h-pore-67 | CCCGGAATAGGTGTATCACAAACTGCCGTCGAGAGAAATCCTGGTAAAG |
| 48h-pore-68 | TTTTTTTTTTATGTAGAAACCAGTACCGCACTCATTTTTTTTT |
| 48h-pore-69 | TGAATTTCTTAACCGCCTTTAATTGTATAACAAACTTAATGG |
| 48h-pore-70 | TTTTTTTTTTTTAATTGCAGTTTGACCATTTTTTTTTTT |
| 48h-pore-71 | CAAAGTCGTCAAAAAATAAGAAACGATTATTATTTAGTGAAC |
| 48h-pore-72 | TTTTTTTTTTAACAATTTCATTTGAATTCAGGCGC |
| 48h-pore-73 | ATTGTTTGGATTAGAACTCAATATTACCAAATCAACTCAATC |
| 48h-pore-74 | GTTTTAAAAGCCCGAAAGACTTCTTTTTTTTTT |
| 48h-pore-75 | TTAATTTGGTAATTGAGCGTCGCCTGAA |
| 48h-pore-76 | TTTTTTTTTTAACGGCGGAAGCATGTCAATCTTTTTTTTTT |
| 48h-pore-77 | TTTTTTTTTTACTGGCTCATTATACCAGTAAACGAAACATCGGTGAATAT |
| 48h-pore-78 | CATAGGTACAGTAGTCATATGCGTTATA |
| 48h-pore-79 | TTTTTTTTTTGCGCAACTGTTCAAATATTTTTTTTTTT |
| 48h-pore-80 | ATTCTGCCCTGTTTAGACAAACAATTCGACAACAT |
| 48h-pore-81 | TGTAAAGCCTGGGCTTCGTACTCCACACAACATACGAGCCGTGCTATAG |
| 48h-pore-82 | AACAAAGAACGGAATAGCTGCAAGGCGAACGACGTTGTAACA |
| 48h-pore-83 | CACCGACTTCATTAGGAAGGTGGCGACAGGTTTACGTCACAAAGACACC |
| 48h-pore-84 | GACGGGCAGAGAGTTGCAGAGTGAATAATCGTCACAACCCAT |
| 48h-pore-85 | TTTTTTTTTTGTCATTGCCTGAGAGTCTGTAAAACT |
| 48h-pore-86 | GGAGGTTTGCACCCAAACCAAATCAATAATAGCGAAGTGAAT |
| 48h-pore-87 | CCGTATACGCAGTAAGCGTCATACATCGTGGCCAATTGGCAG |
| 48h-pore-88 | TATAAGGAATAGATTAGAGTCTTAGGAGATCAAAATTATTTCGAATACT |
| 48h-pore-89 | GCACACCAGCAGTAGATAGAACCCTTCTGACATTCCTGAGAGTTGCTGA |
| 48h-pore-90 | CAATATGAATCGCAAGACAATGCGGTTATATAACTATCATAACCGACCG |
| 48h-pore-91 | ATTATCAAATAGAAGTATTAGACTTTCTCGAACCATGCATCA |
| 48h-pore-92 | TACCAAGAGCCGATATATTCGTTTTGCGGGATCAG |
| 48h-pore-93 | ATCAGAAGTAATCGGAGCAAATTTGAGAATTAATGAAGCGCCGGGCGAT |
| 48h-pore-94 | ATTCACCTGAAATGACCTACACAACAGGAGAACAAACTATCGCACTTGC |
| 48h-pore-95 | TTCATAAACCGCCTCCCAGGAAGATTGTAATTTTTCCCAATATTTTAGC |
| 48h-pore-96 | TTTTTTTTTTCATCAAGACAAATCAACGTATTTTTTTTTT |
| 48h-pore-97 | AAAAGATGTCAGAAGAATGACATGCTTTCCAATACGTAAAATTAGCGAG |
| 48h-pore-98 | ATTTTTGAAGATGTTGCTTTGAATACCAATAACGAATCTACAGTTGAG |
| 48h-pore-99 | AGGCTTTGAGGACCCGTGGCGGTGCTAT |
| 48h-pore-100 | TTGATACGCCCACGCATAAACACCTTTTATCAAGAAGCAAAA |
| 48h-pore-101 | TTTTTTTTTTCTAGCTGATAAGATCTAC |
| 48h-pore-102 | ACGCAAACACCGAGTAAAAGATTTTATAGGTACGCGATTAAACAGAGCG |
| 48h-pore-103 | TTTTTTTTTAAATATCGCGCTGAAAAGGTGTTTTTTTTTT |
| 48h-pore-104 | GTGAGAATAGAAAAGATTCGCATTTTCAATTTTTAAACCAAA |
| 48h-pore-105 | TTTTTTTTTTATACATAACGCCCTATCATAACCCTTTTTTTTTT |
| 48h-pore-106 | TTTTATTGGGTATTAGCTACAATTTTTTACTCCCGACTTGCG |
| 48h-pore-107 | TTTTTTTTTTCATCGGCATTTTCGGTCATCATTTTCGCATTAATAAGCA |
| 48h-pore-108 | TTTTTTTTTTAGTAAATTGGGCTTGAGAACCTTAT |
| 48h-pore-109 | TTTTTTTTTGCATCAATTCTATAGCAAAATTAATTTTTTTTT |
| 48h-pore-110 | TCTGAATAATGGAAGGGTTAAGTTTAGACTTCCTG |
| 48h-pore-111 | TTTTTTTTTAGAAGCCTTACAGTAACATTTTTTTTT |
| 48h-pore-112 | TTTTTTTTTAGATTAGTTGCGTCTTTCTTTTTTTTT |
| 48h-pore-113 | TTTTTTTTTTATCCGGTATTCTCTTAAATCATTTTTTTTTT |
| 48h-pore-114 | TTTTTTTTTTTCAGCTAATGCAGAACGCGTAATTTACGAGCTTTTTTTT |
| 48h-pore-115 | TTTTTTTTTTACTCCAGCCATTCAGGCTTTTTTTTTTT |
| 48h-pore-116 | TTTTTTTTTTAGATACATTTCGCAAATGGGGCGCGA |
| 48h-pore-117 | GTAACGAAGTTTTAAATACCGATAGCCCTAAAACAGAATCAAAAGAATA |
| 48h-pore-118 | GGTCAAGTTGATTTGGAAGCAAACTCCATTCAAAGATATTTTATTAACA |
| 48h-pore-119 | CCTCATTGAACCGCCAGGAACATCAGTGAGGGTGC |
| 48h-pore-120 | CACCGGAATATGTAAATGCTTTTTTTTTT |
| 48h-pore-121 | GCCTGCAAGGTGAGGAAGAAACGGCGAACGATTTAGAGCTTG |
| 48h-pore-122 | ATAACTTTCAACAGTTTCAGAGAAGTGTGTTTTCTGTATGGG |
| 48h-pore-123 | CTGCCAGAGATGGGCGCATCGTATTTTTTTTT |
| 48h-pore-124 | TTTCTTTTCACCCCCGGGAGAGGCGGTTCTGAGTTCCTTGCT |
| 48h-pore-125 | TCCTGATTATCAGATGATGCAGCCATTGTTTTGACCATCACCCCAGCAG |
| 48h-pore-126 | AGAGCCGACCCTCAGAATTATTCTGAAA |
| 48h-pore-127 | CTGGAAGCGTTATTGTTCGTATTAAATCAGCGAAAGACAGAAGAGTTAA |
| 48h-pore-128 | TAGCTGTTTCCTAGAATTCGTAATCATGGCGCTTT |
| 48h-pore-129 | CGATTGGAAGAAACCATACATAAAGGTGTAGCAAAAAAGGGG |
| 48h-pore-130 | ATGCTGTGCTTAGAGCTTAATTGTTTTTTTTTT |
| 48h-pore-131 | ATTTCAAAAAATCGCGCAGAGGCTTTTTTTTTT |
| 48h-pore-132 | CATATATGTCGTCTTTCCAAGCTAACGATCTAAAGCTGCGGA |
| 48h-pore-133 | AATATTTGTTGATATAGCCAGCTTTCTTTTTTTTTT |
| 48h-pore-134 | TTTTTTTTTTAATAAGAATAAACAAATTCTTACCTTTTTTTTTT |
| 48h-pore-135 | TTTTTTTTTTACCGTGCATAAAGGCTATCAGTTTTTTTTTT |
| 48h-pore-136 | ATTCAAACGCTCACAATTCAGAGGTGTGAAATTGTTACTAAC |
| 48h-pore-137 | TTTTTTTTTGAATTATTCGCGATTTTAAGATTTTTTTTT |
| 48h-pore-138 | GAAGGGTAAAGCCAGAATGGCAACAAATGGTTGATATAAGAAAATTTGC |
| 48h-pore-139 | TTTTTTTTTTGCAATAAAGCTTTTTGCGGGTTTTTTTTTT |
| 48h-pore-140 | TAGTCTTTAATGCGGGAAAGCGCCGAAATCGGCAAAAAGAT |
| 48h-pore-141 | TTTTTTTTTTTTGCCAGAGGGGGTAATATGCGGAATAAAGCTGCAAGGC |
| 48h-pore-142 | CTCCTGTCGTGCCAGCTGCAGCAATAGCGCAGATAGTAACGA |
| 48h-pore-143 | ATAGGTTATGAACGAAGCCCCAAAAACTTCACCAGTCACCGT |
| 48h-pore-144 | TTTTTTTTTTATCAACATTAAATGTGAGCGAGTATATACTGGCCTTCCTG |
| 48h-pore-145 | TAGCGACAGAATCAAGAGAATGAATCATTACCGCGGTATCTT |
| 48h-pore-146 | GGAACCGGGTGTACCGCGAAATTTGACCCCAGCGAAACGAAAACAGC |
| 48h-pore-147 | AAAGTACCGACAAGCCGAAGCCCAATCGCCATATTAGAATAT |
| 48h-pore-148 | CCAGTCGGGAAACGGTACCGTCCACCAC |
| 48h-pore-149 | AAGGAGCGGGCGGGAGGAAGGGCGGTCAAATAAGAATACGTG |
| 48h-pore-150 | ACGGAATCATATAACCTTGATGAAAGGCCGGAGACGCTGGCG |
| 48h-pore-151 | AAACAGTAACCATCCGATAGTAAGGCACCAACCTAATTTCCATTAAACG |
| 48h-pore-152 | TTTTTTTTTTATATGTACCCCGAAATTGTAAACGTTTTTTTTTT |
| 48h-pore-153 | TATTTCAGTAATACCAGAGCATCGTCATAAATATTTTTTTTTT |
| 48h-pore-154 | TGGGCGTATATTAATGAATCGGCCAACGCGCGGCGTGCTTTC |
| 48h-pore-155 | AAAACAGACAGTGCCAGGCTTTTGATGATACAGCA |
| 48h-pore-156 | TCAATAAGAACGAGTCAATCACCGCGACATCGCCTGATAAAT |
| 48h-pore-157 | TAAAGTAGCAGGTCGAATCCTTGAAAACATCGGCTATAATATAAATAAA |
| 48h-pore-158 | TTTTTTTTTTAAAACACTCATCCAAAGTACAACGTTTTTTTTTT |
| 48h-pore-159 | GCGCTGGAAGGGAGCGTAAAGCACTAAATTTTGGCAAGTGA |
| 48h-pore-160 | CGGCCAGGAGGATCAATAATAAACCCACAAGAATTCTCTAAT |
| 48h-pore-161 | TCTGTAAGGCCCTGAACAGCTTCTATCAATCAAGTTCAGAACGTGGACT |
| 48h-pore-162 | GGAAGGTAGTAATACCGTTGTAGCAATAGTCCATCCCACCCT |
| 48h-pore-163 | GTAAAACAGAAATAAAGAACAAAAGAACCTACCATCACTAACGAACCCT |
| 48h-pore-164 | GAGAAACAAGTTACTTACCTGAAACAAAGAATAAGGGATATT |
| 48h-pore-165 | CAATAGGCAGTACATACAGGGACACCCGCCGCGCTACGCTGC |
| 48h-pore-166 | GTTTAACAGCATTCCAAGAACTTCATCGTAGGAAT |
| 48h-pore-167 | TTTTTTTTTTCGAGAACAAGCAATCAGATAAAATTTTAACCAATAGGTTTTTTTTTT |
| 48h-pore-168 | AACCTCGGAAATTATTGAGCCATTTGCAGTTAATGCCCCCCACAAGTGC |
| 48h-pore-169 | AATATCTTCTAAAGGCGAATTATCATCA |
| 48h-pore-170 | CAGCCATTTATCCTGAATCTTACCAACGAGTTACACCCATCCCCTGTTT |
| 48h-pore-171 | TTTTTTTTTGAGATTTGTATCCTGCTCC |
| 48h-pore-172 | TTTTTTTTTTACAAAGCTGCTCCACCAGAACGAGTTTTTTTTTT |
| 48h-pore-173 | AAGGAAACCGCCTGTTTAGTAGGCTTAATTGAGTTCTTGCAT |
| 48h-pore-174 | CATAGTTAGCGTGAATTTCCACAGACAGAATTGCGAATAATA |
| 48h-pore-175 | AGAGCCGGAGGGAGAAGGTGAATACAACCCGTCGGTAATGGG |
| 48h-pore-176 | TTTTTTTTTCAGAGCCTAATTTGCCCTAACGAGCTATTT |
| 48h-pore-177 | TTTTTTTTTTATTTTAGTTAATAAGGCGTTATTTTTTTTTT |
| 48h-pore-178 | CTGAGTATTTGATTTATCTAAGCAATGCCTGAGTAAGTACCG |
| 48h-pore-179 | TGACAACACATAAAACGCCTGTAGCAGACGGTTTA |
| 48h-pore-180 | CATTACCGTAATCTATAGGCTGATGAACAACTGACCAGACGGTAGATTT |
| 48h-pore-181 | TGGTTTAGCCCTGACGAGAAAATTCAGTATTAATTACATTTTTTTTTTT |
| 48h-pore-182 | GGAGGCCCAGAATCTTCCGCCACCCTCATTCAGGGATAGCCT |
| 48h-pore-183 | GATTAGCGGGGTGTTTATTTTCAGCGCCGACAGGACACCACC |
| 48h-pore-184 | ATCGCCACCAGAACGGTTGAGGCAGGGTTTCATATTTCAACC |
| 48h-pore-185 | GGGTTTTCCCAGTCTTAAGTTGGAAACGTGATTAAGATGGTT |
| 48h-pore-186 | TATTCAGCATTTGAGGATTAAGGCCGTCAATAGAT |
| 48h-pore-187 | TTTTTTTTTTGCTGAGAAGTAAACAACATGTTTTTTTTTTT |
| 48h-pore-188 | TCCAATAAAAACGAGCAAAGCGGAACCACCAGAAG |
| 48h-pore-189 | TTGAAGCAAGAACGTGTAGCGCGTTTTTTTTTTTT |
| 48h-pore-190 | AGGCTTTCGACGATTGACCCTACGCAAGGATAAAAGGTTTAA |
| 48h-pore-191 | GAACCAGAGTTTGCCTTATTAGCTCGATAGCAGCACCGTAGC |
| 48h-pore-192 | CGGTGCGGGCCTTAGCGGAAACCAGGCACCGGAGATCACCAT |
| 48h-pore-193 | GCGTAACGGTCGAGGTAATGCCACTACGTGCGCCGTATTGCTTTCGAGG |
| 48h-pore-194 | TTTTTTTTTCGGAACGAGGCGCAACTTTGATTTTTTTTTT |
| 48h-pore-195 | ATTTGCTAAACTTGGATTGCGTGACGTTAGTAAATGAATCTCTTGCAC |
| 48h-pore-196 | GATGTTCAGGGTGAATTCACACAGCATTAAGACTCCTCAAGA |
| 48h-pore-197 | TTTTTTTTTCTTTTTAACCTCCGGCTTAGGTTGCAGCAAAAT |
| 48h-pore-198 | ACATGTTTTCGAAATTAGTCACCCTCAGCCTTCCCA |

**Supplementary Table 10 |    DNA sequences of 54helix DNA origami (Scaffold 8064)**

| Name | Sequences |
| --- | --- |
| 54helix-1 | AAAACCGGATTGCCCGGCCAGGGTGGTTTTTCTTTTCACCCTCAGGAGA |
| 54helix-2 | AGGCGCATGAGGAATGGTAGCAACGGCTACAGAGGCTTTGAAGAGCGAA |
| 54helix-3 | GTGGACTCCAACGTCAAAGTAAGTTTGG |
| 54helix-4 | TAATAAGCCCACAAGAGCTTATCCGGTAAAGCCTT |
| 54helix-5 | CGGGAGAATTAACTGCGCTAAAACGCGAAATAGCA |
| 54helix-6 | TACAGACAGGGAACAGAGGACTAAAGACAAATACG |
| 54helix-7 | GCTCCATGTTACTTAGCCGGGGACTGAT |
| 54helix-8 | GACAGGACAGGAGGAAAGTGAGACGGGCAGAGAGT |
| 54helix-9 | CTTATTACTAAGAACTGGCATGAATTGAGTGGCGA |
| 54helix-10 | AGAAGTTGTAGCAAGCCGTTTGCCTCCTACACCGGGCTATTAGCGGAAA |
| 54helix-11 | TCGGCCTGTCCTGAGTAGAAGCCGATTAAAGAACG |
| 54helix-12 | GTATGTTCCCGGGTTAAGGCGAGAGGGGACGACGA |
| 54helix-13 | CATCGTAACCAAGTTTATCAT |
| 54helix-14 | AGACAGCTAAATGCATTTTAGATGATATGAATCATATGTACCAAGAATA |
| 54helix-15 | GGCGGTTGAACAACAAATACCGCCTTCCCGCGTAACCGTGCATCCTGTT |
| 54helix-16 | GCCAACGCGCGGATTCATGGAAACCTGTCGTGCCAGCTGCGG |
| 54helix-17 | TCGAGAACAAGAGAAACGCAGGCTGTCTTTCCACCGCACTCA |
| 54helix-18 | TCACCCTCAGCAAAACATTATTAGGGAGTTAAAGGCCGCTGC |
| 54helix-19 | TGGCAACAGTTTATGTAGCACGCGTGCCGAGGATCAGCAAACAAGACTC |
| 54helix-20 | AACAGGACTCAATCAACAACAGTTTCAGATTGCGATCCAGAACAAACTA |
| 54helix-21 | GGAAATAAGAATAGTGTATGGATTCGCGTTAAATCTAATGGGGAATCAT |
| 54helix-22 | TTGCACAATAACTATAGTGAATCACCATAAATGCAGAATCGAATATTAA |
| 54helix-23 | CGCTCACTGCCCTGTCACAATAAGACAATAGTGCTGCGGCCAGGGGTTT |
| 54helix-24 | AAAATAACAATCAATAATCAACGCGTCCTCACGGTCGGGCCT |
| 54helix-25 | ACCGATATATTCCTGAAATGGGACCAGTTGGTCTTTCCAGACCTAAACA |
| 54helix-26 | AGCGCCACAATAGACGGAATAATATAAAAAATACATACATA |
| 54helix-27 | ATAACGCTACCACAAGATTCACGGAACAAATCTAC |
| 54helix-28 | GTCACACATTATTTTTTGACGAAAACGCATTACCGCCAGCCAGGTAATA |
| 54helix-29 | GTGAGGGGCGACATACGGAAGCATAAAGTGTAAAGCCTGGGACGCCCTC |
| 54helix-30 | CATGTTCAGCATAAGTCCTGAATGGCGGGCC |
| 54helix-31 | AATGATAAAAGGGACCTTTCTTAAACAGCTTGATACCGATTATTTCTGT |
| 54helix-32 | CAATAAACGGAGCAGCGCAAATAAGAGAATACCAGACGACGA |
| 54helix-33 | ATCACACAACATACGAGCCGGACACGCCTGTAGCATTCCACA |
| 54helix-34 | AACTTGCTTTCGAGGTGAATGTTTATATGTGAGTGAATAACC |
| 54helix-35 | AACAAGATGCCCCAACCGCCTTGATAAAACCGAGC |
| 54helix-36 | AAATTGTCTAAAACCCATTAACATCTTCTTTTTTC |
| 54helix-37 | AATTCAGTTGTGTGAAATTGTAACTAACTCACATTAATTGGA |
| 54helix-38 | GTCACTGAGGCAAATTATAAACAGTTAACCATTTGGAATTAT |
| 54helix-39 | CCAGTGTGTTTTAACAACGCCACAATAGTAATGCA |
| 54helix-40 | CGTCGCTTCATATAATGACCCTGTAATAGATAGCGAAAGAA |
| 54helix-41 | ATCAGATTCGGAGCCTTTAATAACAACAACCATCGCCCACTGAGAAGAG |
| 54helix-42 | AATGAATTTAGCGTTTAAATTTTGATTATACTTCTATTAGTCGCACAGA |
| 54helix-43 | TTCCTTCAATTCGTAATCATGTTGCTTTCCAGTCG |
| 54helix-44 | CTGCCCAAGTTGGGAAGGGCGTGGCGAATTAAATACGGAATA |
| 54helix-45 | AAATCATTTAGGTTAATCCAATTGGGAACTGGCTC |
| 54helix-46 | TTTATCAGACGCTGCAGATACGAGCAACACAGTTGCGCCGAC |
| 54helix-47 | CAAAAAATAGAAAATCTCCAAGTGGTCGCTGAGGC |
| 54helix-48 | ACTTTAAATAGGAACGCCATCAGCTTTCAATTTAATAACATC |
| 54helix-49 | GATTTTGGTTAGTATTCACCAGCCAACAGAGGTGCCTAATGA |
| 54helix-50 | TCGAAAGTGCGTATTGGGCCAAAATTAATGAATCG |
| 54helix-51 | TTACCGATTATTTTAGCAAATCAGATAT |
| 54helix-52 | ACGTTTGATCGGAACGAGGCTTTTTTGCGGGATCG |
| 54helix-53 | ATAATAATGACCTAATCAGAAAAGCCCC |
| 54helix-54 | GACAGTGTTGTTTCGTCACCACCTATTATTAGCTTTCCGGCACGCTCAC |
| 54helix-55 | CACCGTCAAATTATGTGCACTCTGTGTCCCGCTTCTGCCTAT |
| 54helix-56 | CGATCGTGGAATCGTTAACGGTGGGCGGGCCAGCGGATCAAATTTCGAG |
| 54helix-57 | GTTTTGCATAGCGAAATTTTCAGATTAA |
| 54helix-58 | TTGCTGCTCTAATGGAAACAGCTACCATTTATTGTAAACGTTTCGGTTT |
| 54helix-59 | CAATATTAAAGCGTTAAAGTTTTGTCTAAATATTT |
| 54helix-60 | CCGTAACACTGATATTAAAAATAGGATAGCAAGCC |
| 54helix-61 | GTTCAGCAAAACCATTAGCGCTGGTAATGGGTACACTGGTGT |
| 54helix-62 | ATACTGCGGAATCGTAGACTGCTTTTGCAATTTTTAGTAATGCGGAGAC |
| 54helix-63 | AATTACCTTTTTCACAGTTCAGAAAATTAATTACA |
| 54helix-64 | CAATACACCAGTAGCCAATGACTTGAGTTAATAAGACGACGT |
| 54helix-65 | CACCATTTTAGAGCCAGCAAAACTTGAGTGCCCCCTGGTGCCGCTGCGC |
| 54helix-66 | TCAAATGAAATCAAGTTGTAC |
| 54helix-67 | TTTAACCCTAAAACCAGAAGAATATAATATTATCACGGCGGA |
| 54helix-68 | ATCGCCAATGCGCGAACTGATAATGGCTGATAATTGTTAAATTTTTAA |
| 54helix-69 | TCCAGCAAGTAAGCGACACCCTCAGAGCCACCACCCCCTTTTACCCTAA |
| 54helix-70 | GTTTAACGTCAAAAATAAAAACACCCAGTGCGGGAATTACTACAAATTC |
| 54helix-71 | ATCGCGAAAAGAGTGAAACAACTAATAGATTAGAGGAGGCCAAACGAAA |
| 54helix-72 | CTGCGAAGTACATCAGGCGGCCAGTGCCGATAACCTCACCGAGCGTCCG |
| 54helix-73 | GAGCCCCATCGATATTTGCCTCATGGCTATTTACCCCAGAAT |
| 54helix-74 | TATAGTCGAGGAAGAAGGCAATTAACATATCAATT |
| 54helix-75 | TGAAAAACAGAGGGTGCCACCAGAAGGTCATTTTAATTTTATTAAATC |
| 54helix-76 | CAGAAAACTCCTCAAGAGAAGGGTATTAAGAGGCTATAGGTC |
| 54helix-77 | TGAGCGCAGAAATTGCGTAAGGCCGTAAAACAGAATGAACGG |
| 54helix-78 | CGCCACCCTCAGAGCGTCAGATAGCCCCCTTATTACTCAGGA |
| 54helix-79 | CAGCTTATGCAGGCGCTTTTAAGTGATGCCGGCAAACGCGAGATAATCA |
| 54helix-80 | TTTAATTCTCCAACCTTTTGATAATTGCATATGCA |
| 54helix-81 | GGCGAATTATTCTGAGAGCCACTGAACCTCAAAAGTTACAAA |
| 54helix-82 | GGTTTCTCATAGGTGTATCACTAAGGATTAGGATTAGCGGGT |
| 54helix-83 | GGCATTTTCGGTCACTGTAGCAATCAAGGCAGCACAACGTCA |
| 54helix-84 | GTGCCATCCCACTCAAGCAACCGCAAGATGCCGTTAAGGGTAGCCGCAC |
| 54helix-85 | AGACCGGAAGCAAACGAGCTTAGATTAAAGAAGCATGACCATCTTTAAA |
| 54helix-86 | ATCGCCTGTGCTTTGAATACAAGATTTTCAGGTTAACGGTTCATTTT |
| 54helix-87 | TAAGAGGAGTAGCAAGAATTAGCAAAAAATCAAAACAAAGCGAAGCAGA |
| 54helix-88 | AATCAATAACATTAAGCGGAATTATCGACACCGCCATGAAAATAGTAC |
| 54helix-89 | ATCTGGTAAAGCATCACCTTGGCAGCAATGCAACATGAGGCGCCACCAG |
| 54helix-90 | TATAGCCCGGAAGTCGAGAGGAGTATTATAGATAA |
| 54helix-91 | GTCTGGTCAGCAATGGAACGTGGTTGAGCCGCCAC |
| 54helix-92 | GATTCGCCTGATCGTCGGGAGCAAATGGTAGTTTG |
| 54helix-93 | TCCCTCAGAGCCACGATTGACCACCGGAACCAGAGTCTTTTC |
| 54helix-94 | ACTAATGTTTAGCAGATGGCTTAGAGCT |
| 54helix-95 | AGGTTATCTAAAAACAATTGTTGGCAAATCAACAG |
| 54helix-96 | ATAAGTGAATCCCTTATAAATGCGAAAATCTGCCAGTTTGGG |
| 54helix-97 | ACAAACACAAAGTTTAAGAAAATAGCTATCTTACCAGAAACAATAAGAT |
| 54helix-98 | TTGGGGCAGTGAATATCAACGCAAGAACCGGATATCGCATAGGACCACT |
| 54helix-99 | TACCTTTCCAGCGATTATAGTGAGGCAACCGGTTGATAATCA |
| 54helix-100 | CAGTACCATCTTGGTGTAGATGGGCGCATAGGAGGTTTTGCT |
| 54helix-101 | TTTTTTCATCCTCAGTTACAAATAAGAAACGATAGCTCAGAGGCAGGTC |
| 54helix-102 | TCTCTGATTTGATGATACATTGCGACAGGCGTTTTCAGCAAC |
| 54helix-103 | AGCCTCATGGAGCAGCTATCAAGCTATT |
| 54helix-104 | GCAATAACAAAAACATTATCAACGAGAAAAGCGGATTAAAAGAAGATGA |
| 54helix-105 | ATTCCTGCCTGATTGTTTGCTGAACGAAGTCAGTATTCCGCCACCCTCA |
| 54helix-106 | GAAAGGGTTCGGAAGTACAAACTACACGATGGAACCCATGTA |
| 54helix-107 | TCAGTGCAAACCACCCTCATTTTCAGCCGAAACAT |
| 54helix-108 | CCCGTGTGGCCGGACGTAATCAGCGCACTCAATCCTGTTGCC |
| 54helix-109 | GACATAACAGCAGTCATCAGACCCCCTGCATCAGA |
| 54helix-110 | TTGCAAATTAGAACTACATAAATCAATGAGCAATTTCATTTG |
| 54helix-111 | GGAAGGGTTCATCATACAAACATCAAGAAAACAAAAAATTAT |
| 54helix-112 | ATTCGCAAACGATCAAGAATATAGAACCCTTCTGACATTCTG |
| 54helix-113 | AGGCCACCCTCAGAACCGCGGATTTAGAGTTGATACCGGCGG |
| 54helix-114 | GACAGGAGCCGGACAGAGCACGTCTCGTGGGAACG |
| 54helix-115 | CCAGTACGGTGTCTAATTCTGCGAACGATACCCAAAACATCT |
| 54helix-116 | TTTAGGAGCACTTACATTTCGAAACAATCGAACAG |
| 54helix-117 | TACATTTAGATCAAAAGAATAGCCCGGTCCGTCAAGACTTTACAATATC |
| 54helix-118 | CAGAACCAATCCAAAATAAACCAACGCTCGAAACGTGAGAGA |
| 54helix-119 | ACCATTACTCCAAGCGCGAAACAAAGATGTAGATTTCAATAA |
| 54helix-120 | AACTGATCCCAGTCTTTTAACCGTCATATTAACCGCCACCCT |
| 54helix-121 | CATTCAGGGAAACCCGCGCCTTCATTAAGAAGGTAAATATTGTCAACCG |
| 54helix-122 | CATTCGCCGACGGCCTGGAGTGTACTGGAACAGTG |
| 54helix-123 | ATGCCTGAGAACCCATTAATTGAGGCTTTCATAACCCTCGTTATAGTAA |
| 54helix-124 | CCAATAAGAACAAAGATGATGAAACCACGCATTTCAATTACC |
| 54helix-125 | ATCGGTGCATACCGGAATGCGGTTACCATTGAGGGAAA |
| 54helix-126 | CCGTGGTAGAATCGCCATATTGTAGTATAAAGCCATAGAAACTATCCCA |
| 54helix-127 | TTTGCCATAGCCTTGATATTCGGAAAGCTTAAGTTTGTAAAA |
| 54helix-128 | TTGACCCTACTATATTTTCATCTACTAATACAAAGAACAAGA |
| 54helix-129 | GCAAATTCCGAGTAAGTGTTTTTATAATTACGCCAGTCAAGC |
| 54helix-130 | TCGGCAACCCGACAACTCGTAAAAGTTTTAACAACTTGACCGAGCTCAT |
| 54helix-131 | CAGTATGGACCGTGGGCCCTGAACAGCTTCTATCAAGCTAAAACAGAAC |
| 54helix-132 | TAGACTCAATCATAGGTTTTGTTCTAAGTATCAGAGAGATAAAGTTAGA |
| 54helix-133 | TTCTAGCACTTTTTGTTTAATGATTTTACGAACTG |
| 54helix-134 | AAAAAATTTAATTTACGGGTATTTTTCAGACGGTCAATCATACAGACCT |
| 54helix-135 | CAATCGGAACGAGCTATTTTGCAGGGAAGCGCACAGATTTTT |
| 54helix-136 | GCCAGGGATTAAAGGTTCCAGCGTAGCGTTTGCCACCACCACCGGAAAG |
| 54helix-137 | GGGTAACGGATGTGGAAGGAAACCGAAGAGCCGAAAATAAATTACCGCC |
| 54helix-138 | GCATGTCATCTATCAGATGAATATACAGTAAGTGTAAAACTA |
| 54helix-139 | ATTCTCCCGTTATTGCGGAACCATATCAAACCCTCTTGAAAGGAATTCG |
| 54helix-140 | CCGTCGGACATTAAGTTGTAGCAATAAACCATCACCTTTGCCAAGAGGA |
| 54helix-141 | ACGTTTTCGCACTCCAGCCTGTCTCGGCCTCAGGATAGCTGT |
| 54helix-142 | AAGCTTTGAGACGCAGAAAGCCTTTTCT |
| 54helix-143 | TAATCGGGCAAATATTTAATTCACAGGAAGATTGTAAGGCTC |
| 54helix-144 | CTTGCCCGGCTTGAGCTGGCTAACGGTG |
| 54helix-145 | GCTGGTTGTCCACTATTAAGGCAAGATAGGGTTGAGTGTTGT |
| 54helix-146 | TAGTTGCGTCTTTCCAGAGCACTGCAAGCGCCAGC |
| 54helix-147 | ACGAAGGCACTAAAACACTGGTTTGAGATCAACCGAGTCAAA |
| 54helix-148 | CACCAACGTCGAAATCCGCGGTCTACAACGGAGATTTGTATC |
| 54helix-149 | GGTCCACTGATGGTGGTTCCCATGTGAGTGTAGCCAAAAATA |
| 54helix-150 | CCCAATCGCGGAAACGCAATAATGAAATAGCCCAA |
| 54helix-151 | AAATCAAAGAATAACACAGTTTGTTCTTGACACCAAAATTCATACAAGA |
| 54helix-152 | TAATGTGCAAATATTGATGCAGGGTTATGGTAGAATTCAACTAAGCATA |
| 54helix-153 | ACTTGTTGACTTCTTTGATTAGAATCCTGATTTTA |
| 54helix-154 | TGCAGAATGGTTTGTAAAGGACGGAGTGCCTACATACATTGGCACGTTG |

**Supplementary Table 11 |** DNA sequences of 2-layer DNA origami (Scaffold 8064)

| Name | Sequences |
| --- | --- |
| 2layer-1 | GACGGTCTGTTACTTTAGTAAATGAAACTTGTCGTCAGTTTG |
| 2layer-2 | GACCAGGCGCATCGCTGACCAACTTTGATAGCCCGATCCAAT |
| 2layer-3 | AACGGCGACCCGTCTAGCAAAATTAAAAACAGGCAATAAATC |
| 2layer-4 | CGACAGTATCGGGTAGGTCACGTTGGTGGGTGGTTGCGGTCC |
| 2layer-5 | TTTTTTTCATCTGCCACTTTCCGGCTTTTTT |
| 2layer-6 | TTATCAGAGCGGAAGGCGCTAGTAACCACCCTGTC |
| 2layer-7 | ATCAAAATTATTTATTATACTTCTGAATAAAACCGCGTTTTT |
| 2layer-8 | ACATAAATGTTAGCGCCCAATAGCAGTTCGTAGGACGTCAA |
| 2layer-9 | AATCAATAGAAAAAAAACGCAAAGACACAGTCCACAACGATC |
| 2layer-10 | TTTTTTTCATCAATATAATCCTGATTGAAAGCCCGAAAGG |
| 2layer-11 | CACTACGATTAAACAGGAGCCTTTAAAAGGAAAAAAAGGCTC |
| 2layer-12 | AAAATCTACGTTACGAAACTGGCTCATTCATTTTTTTAATTG |
| 2layer-13 | TATTTAAAAAACAGAATGCAATGCCTCCTGACCCTCATATAT |
| 2layer-14 | TGAAGATTTCGGCGAAACGTAATCCCCGACGCGTG |
| 2layer-15 | ATAATGTTTAGAACGGAGCTTTCAAGGGTTTCAAGGCGGGGC |
| 2layer-16 | TTTTTTTGAATTTGTGCAGAAACAGTTTTTTT |
| 2layer-17 | CAAACCCCATCACCTCACGCAGTAATAACATCTTA |
| 2layer-18 | TTTCATTTGAATCTAAAAACATCAAGAATCGAGCCACATGTA |
| 2layer-19 | AGAAACATGAGTTATTTATCCTGAATAGCATGCACCCAGCTA |
| 2layer-20 | CAGAATCAAGTTATTCTGAAACCATCGACAGTACCGGATTAG |
| 2layer-21 | TTTTTTTTGGTCAGTTGGCATCTAAAATATCTTTTTTT |
| 2layer-22 | AGGAAGTTTTGAGGGCTTGCTTTCGAACTGTTGTATCGGTTT |
| 2layer-23 | AGAAAGATTCATTTATAATAAAACGAACAACTCCATTCAAAG |
| 2layer-24 | GAAAAGCTCATATGAAGATTCAAAAGGAGAGAGTAATGTGTA |
| 2layer-25 | CGCCATGGTGGATAGCTCTCATTTCACGCGCCTGT |
| 2layer-26 | TTTTTTTCGGATCAAAATAAAAAAATTTTTTT |
| 2layer-27 | TGAGAGCAACACCGACTTGCCGGTAATATCACCGG |
| 2layer-28 | ATGTGAGTGAATGTACTACCTTTTTTAATAATTGAAAAGCCA |
| 2layer-29 | CCCACAATTGAGCGAGCGTCTTTCCATGACCTTACCAACGCT |
| 2layer-30 | CATTTTCGGTCAGATATGCCTTTAGCGTAGTATTAGGAACCT |
| 2layer-31 | TTTTTTTAAAAATCTAAAGTCAATCAATATCTTTTTTT |
| 2layer-32 | GCTACAGCAGCATCAGCTTGATACCGTTTAGGTGAATTTCTT |
| 2layer-33 | AGATACATAACGGTACCAGTTGAGATTTAAGACTTTCAAAAA |
| 2layer-34 | TAGCATGAATCGATGACAGTCAAATCTTAGGGTGAGAAAGGC |
| 2layer-35 | TAGTGTCGTGCAGTTGGGCGGGCATCAGCATCAGA |
| 2layer-36 | TTTTTTTCCCGTAAAATTGCCGTTTTTTTT |
| 2layer-37 | AAGATAAAAAATACGAACAATTCATGGAAATTTAA |
| 2layer-38 | TTCCCTTAGAATCACAAACCTTGCTTCTTCATATGATTACTA |
| 2layer-39 | TCAGAGGAATTAACTTACAAAATAAAATCCGAGCCTAATTTG |
| 2layer-40 | CAAAATCACCGGTGCATAGCCCCCTTATGTATAAAGTCAGTG |
| 2layer-41 | TTTTTTTGCGGTCAGTATTCAGCAGCAAATGTTTTTTT |
| 2layer-42 | GCAGCGACGCTTTTTGACAACAACCAAGTTATAGTTGCGCCG |
| 2layer-43 | ATCATAACCCTCTCGGCCAAAAGGAATTACTATTAAATCAAA |
| 2layer-44 | GCAAACATCAGGTCTCAACCGTTCTATCTCACCATCAATATG |
| 2layer-45 | TCGCTGAAAATGAAGGGTAAATACACTGGGCTGGT |
| 2layer-46 | TCCGATGCTGAAAAGCCCATCGACCTTAAATAGAGACGAGAG |
| 2layer-47 | TTTTTTTCCGGCAAACCATAACGGATTTTTTT |
| 2layer-48 | GCGAACTTTGAATGCCTACATACATTGG |
| 2layer-49 | GAGAAGAGTCAACTACCCTTGAAAACATGCGTTAAAATACC |
| 2layer-50 | TTAGACGAATAACACCCAATCCAAATTTTGCAGCCATATTAT |
| 2layer-51 | CGCCACCCTCAGGGCCAACCAGAGCCACCAGGAGTGCGTCAT |
| 2layer-52 | TTTTTTTACATCGCCATTAAACAGAGGTGAGTTTTTTT |
| 2layer-53 | GAGTTAAGGATTCGGTCGCTGAGAGAAGTTTTGCCTAGCGTC |
| 2layer-54 | GGCTTTTGCAAAGCCGGTTTACCAGACGTTAAACAATTCATT |
| 2layer-55 | TACAAAGTAAGGGTAGCTATTTTGTCTGGTCAGCACCAGCTT |
| 2layer-56 | AGCGTGGTGCTGTGGAGGCAGCCTCCGGTGCGGTACGGGGTC |
| 2layer-57 | TTTTTTTACGTGCCGGGCCAACGGC |
| 2layer-58 | AGCGTAAAGAGATATTCACCAGTCACACGACCAGTAA |
| 2layer-59 | CCTTTTTAACCTTGGTTAGTGAATTTATTTTCATCACTTTTT |
| 2layer-60 | TTTACAGTACGTCAAAAATGAAACACCAGAGCCGCGACGATT |
| 2layer-61 | CACCAGAACCACATAGAACCGCCACCCTAGCGCAGAATCCTC |
| 2layer-62 | TTTTTTTCAGACAATATTTGATAGCCCTAAATTTTTTT |
| 2layer-63 | GACCTGCCCTGATATTGCTAAACAACGCCATTTTCTGTATGG |
| 2layer-64 | GAACCGGATATTGAAAAGGCTGGCTGACCATCAATTTTGGGG |
| 2layer-65 | AGCGAGTCCAGCTTGCATAAAGCTAAATTGGCAATAAAGCCT |
| 2layer-66 | CGCCATTCCCCTCAGGAAGATTCACCGCTGGTTTT |
| 2layer-67 | TTTTTTTACCGCTTCTGGGAAGGGCTTTTTTT |
| 2layer-68 | CATTTTGATTTTAAACCCGCCTGGTTGCTTAGAAA |
| 2layer-69 | TTCAGGTTTAACTACATGCACGTAAAACCAAGTACATCATTC |
| 2layer-70 | TATTACGAAGAACTAAGGCTTATCCGGAGGAGCAAATCAGAT |
| 2layer-71 | GACATTCAACCGAAAGATTCATATGGTTGCATTCCACCAGTA |
| 2layer-72 | CAATACTGCGGACCGATATAGGTTGAGGCCTCGCCCACGCAT |
| 2layer-73 | ACGGCTGGAGGTGCCGGAGAAATGTTTACCGCTGATAAATTA |
| 2layer-74 | GTGGTGCCATCCGCGCTTTCGCACTTTTTTT |
| 2layer-75 | ACAAAGAGAACCCTTCCCGGCTTAGGTTAATCCAATCGCAAG |
| 2layer-76 | GGCCTTGATATTTGTTTAATGTAAATGCATAAGAAACGATTT |
| 2layer-77 | ATTCTGGCCAACGAATACGTGGCATTTTTTT |
| 2layer-78 | TTTTTTTTCAATCCGCCGGAAAGGTTTCTTTTTTTTTT |
| 2layer-79 | AAATCGTCATAAATGTTCAGAAAACGAAGCGGGATTTGCAGG |
| 2layer-80 | GAATCCCGACTGGAAGAGGGGGTAATAGGCAAAATAGCGAGA |
| 2layer-81 | ATGTCCAGCATCAGTGAGCCGGGTCAATATTGCCTAGAGATC |
| 2layer-82 | ATTGCAGCACGCAAGCAACCGCAAGAATACTTGTAACATCCTGCGG |
| 2layer-83 | CAGAACGCGAGAAATTCTGACCTAAATAGCTATTAACCTGAA |
| 2layer-84 | TTTTTTTCTGAAATGGATTATTTTTTGACGAAAA |
| 2layer-85 | CAAATATTGATGCAGGGTTATATAACTAAGTCTGAGAGACTA |
| 2layer-86 | TTCACAAACAAATATCTCTGAATTTAATTAAAAACAGCAGCC |
| 2layer-87 | ATTAAAGCAGGTCACGCCAGCATTGACAAGCCTCAGAGCCGC |
| 2layer-88 | TTTTTTTGCTCGTCATAAATTACCTGCAGCCTTTTTTT |
| 2layer-89 | ACAGAATGACCATATAGTCAGAAGCAACGGAACGAACCCTCA |
| 2layer-90 | AATCAGGAAATGCTACGATAAAACCTAAGAAGAGCAACACT |
| 2layer-91 | ATCTGTTGCCCTGCGTGTGTTCAGCAGAGAACGGTGTCTGGA |
| 2layer-92 | AATGGGTGCGCGGTCCAGAGCGAACGTC |
| 2layer-93 | CGCATTACCGCCTTGCTTGAGTAGTTGATTAAATTAACAGTA |
| 2layer-94 | TGGTTTGATAAGAATAAACCACGAACGATTAATGC |
| 2layer-95 | TTTTTTTGCAACAGGACTCAATCGTTTTTTT |
| 2layer-96 | GACCGTGTAGTTAACAAATCATAGGAGGGGATTAAGACGCT |
| 2layer-97 | TTCCGTTCCAGTAAGTACTGGTAATAAGTGAACACAAGCGCA |
| 2layer-98 | ACATGGCAATGGAACAGAGCCACCACGCAACTCCCTCAGAGC |
| 2layer-99 | TTTTTTTAGCGGTGCCGGTGTGGTGCTGCGGTTTTTTT |
| 2layer-100 | AAAAGCGGATTGCACAAATATCGCGTCAACTAAAGAGCAACG |
| 2layer-101 | GATTAAGTACCCTGACGAGGCATAGTAGTCTTCAACTAATGC |
| 2layer-102 | CGAATCGTTAACGGACGATCCAGCGCTATACCCCGGTAAAAC |
| 2layer-103 | TGCCGGGCATCCCTGTTAAACGTTTTTTCGTCTCG |
| 2layer-104 | AATCATACGTTATACAAATTCCCTGCAACCAGCAG |
| 2layer-105 | TTTTTTTAACTATCGGCCAGCCATTTTTTTT |
| 2layer-106 | GAAAAAGAAATAAGAGCGATAGCTTAAGGTTATTAATTAATT |
| 2layer-107 | CCAGTTTTAACGGGCAGTTAATGCCCCGCTAATATAACAAAG |
| 2layer-108 | CCTTGAGGATGATACACCGGAACCGCGAAGTCTTTTCATAAT |
| 2layer-109 | TTTTTTTCCAGAATGCGGCTTCGCGTCCGTGTTTTTTT |
| 2layer-110 | ATTTTAATTCGAGCACAGGTCAGGATAAGGGTAAATTTCATG |
| 2layer-111 | CGAACCAAGCCCGAAGGAATACCACAAACCATTATTACAGGT |
| 2layer-112 | GGAGTGTCACTGCGGTCATACCGGGGTAGAAGATTATAATCA |
| 2layer-113 | GCACTCTGCCCCCTTTGTGTAGCACAGGCGGCCTT |
| 2layer-114 | CCAGTATGAATCGCCATATCATTGCTGAGCCACGC |
| 2layer-115 | TTTTTTTAATACTTCTAAGAACTCATTTTTTT |
| 2layer-116 | ACGCTCATTTAGTAGTAAATCGTCGCAAGAATAAATCAATAT |
| 2layer-117 | AACCTGCCTATTTCAGAGGCTGAGACATAGCCCAAGAGATAA |
| 2layer-118 | ATTATTCAGTGCCCTAGCGTTTGCCAGCTTCGTTTTCATCGG |
| 2layer-119 | TTTTTTTACCACCAGAAGGATGATGGCAATTTTTTTTT |
| 2layer-120 | TTTTTTTAGCCTCCTCACATTCCTGTGTGAATTTTTTT |
| 2layer-121 | CATAGAGAGTACCTGCGGATGGCTTAGTCTAAAACGTAATGC |
| 2layer-122 | CTCCTTTGGAAGCATAACGGAACAACCAATACGTTGGGAAGA |
| 2layer-123 | TTGTTCTGCCAGCGGTACCGAGCTCTATAATATTAAGCAAA |
| 2layer-124 | CCTGTTCGGGCCGTCGGAAAATTCTGCTCATTTGC |
| 2layer-125 | AAAGTTTTTAACGGTACCCGATTAATCAGAGCGTATAACGTA |
| 2layer-126 | AACGCCAAGTAATAAGAGACTCAGTTGACAAATAT |
| 2layer-127 | TTTTTTTAGGCCACCGCGTTGTAGCTTTTTTT |
| 2layer-128 | ATTTAGGTAGGGCTTGGAAACAGTACATATTACATTTAACAA |
| 2layer-129 | CATCCTCAAGAGAAAGGCGGATAAGTATATAGCTAAAGAGCA |
| 2layer-130 | GATTAGCACATGAACAGACTGTAGCGCCAAAATCAGTAGCGA |
| 2layer-131 | TTTTTTTATTGTTATCCGCTGCCTAATGAGTTTTTTT |
| 2layer-132 | ACGAGCTTAATTGCATATGCAACTAAAATCATCTTGAGGCAA |
| 2layer-133 | AATGCTGAAGAGGTATACCAGTCAGGTGAATGAATTACCTTA |
| 2layer-134 | GAGAATTCGTAATCAACATACGAGCCGGAATCAGCTAAAATT |
| 2layer-135 | TAGCTGTGTTGAGGCAGCGCCGACTTTCTCCGTGG |
| 2layer-136 | AAGTACCGACAATAAACAAAAATAGATAATTGAG |
| 2layer-137 | TTTTTTTAGACAGGATAATCAGTGTTTTTTT |
| 2layer-138 | GGTAAAGGGCATTTAACAAAATTAATGATACTGAGCAAAAGA |
| 2layer-139 | AGGCCGTCGAGAGGCACCGTACTCAGGCAGCAGATACCGAAG |
| 2layer-140 | TAAGTATTTTGCTTAGCAGCACCGTGGCATAGCAAGGCCGG |
| 2layer-141 | TTTTTTTGAGCTAACTCACCAGCTGCATTAATTTTTTT |
| 2layer-142 | TAAGTACGGTGTCTAATTCTGCGAACGCACAAAGTCCCCAGC |
| 2layer-143 | TCATTCCTCAACATCTTTAATCATTGTTGAGAGTAGTAAATT |
| 2layer-144 | GCGGAAGCATAAAGGCTCACTGCCCGAAAAAATAATTTTTAA |
| 2layer-145 | CCTGGGGTCACAATGTGCCAATAACCTCACCGGAA |
| 2layer-146 | TCAGCTAAAGTCCTGAACATGTACAAACGCCGTCA |
| 2layer-147 | TTTTTTTCTCGTTAGAAAGGGATTTTTTTTTT |
| 2layer-148 | ACGCGCCTCTGTCCCATTTCAATTACAGGAATACCAAGTTAC |
| 2layer-149 | TTGAGGTTTAGTACCCACCCTCAGAGCGAGGAAACGAACAAA |
| 2layer-150 | CTCAGAACGGAATACACCATTACCATCTCCCCATTTGGGAAT |
| 2layer-151 | TTTTTTTTGAATCGGCCAAACCAGTGAGACGTTTTTTT |
| 2layer-152 | CAGAGTAGATTTAGATAACCTGTTTATTAATTGTGCGGAGAT |
| 2layer-153 | ATTAGATAACAGTTAAACACCAGAACACGTGCTGCTCATTCA |
| 2layer-154 | AACTTTCCAGTCGGCGGTTTGCGTATGATCATCAACGTCTGG |
| 2layer-155 | GTCGTGCATTAATTTGTGCTGTCCCAGTCACGACG |
| 2layer-156 | CTAGCGCTTAGCTGCGCGGGCGCTAAGAAAGGGCGAACGTGGCGAGATTTTTTT |
| 2layer-157 | AATAATAATCAATAATCGGACAAGTTTGCGACAAC |
| 2layer-158 | TTTTTTTACAGGGCGCGTGCTTTCTTTTTTT |
| 2layer-159 | CTAATTTTATCAACTGATTGCTTTGAAGCCACCTTTTACATC |
| 2layer-160 | CGCCACCACCCTCATGTACCGTAACAAGGGCATGATAATAAC |
| 2layer-161 | GGATAGCCACCCTCCACCGACTTGAGAAATGACGGAAATTAT |
| 2layer-162 | TTTTTTGGCAACAGCTGATTTGCCCCAGCATTTTTTT |
| 2layer-163 | GAGCTATATTTTCATCTACTAATAGTCGTAGCCGGAATCCGC |
| 2layer-164 | CGCGAGCTTCGCAACAACGTAACAAAAGAATAATCTTGACAA |
| 2layer-165 | CATGGGCGCCAGGGCTGGCCCTGAGAATGGATTCTAAATGTG |
| 2layer-166 | TCTTTTCCGCGCGGGCTGCGCTTCGCTATTACGCC |
| 2layer-167 | TTTCCTTCGCACTCATCGACGTTATCATACATTAT |
| 2layer-168 | TTTTTTTAGCGGTCACATGCGCCGCTTTTTTT |
| 2layer-169 | CAAGAACGCATGTATACAGTAACAGTAACAATTGCGTAGATT |
| 2layer-170 | ATCTGAGTTTCGTCACAGACAGCCCTCGCAAACGTGACTCCT |
| 2layer-171 | CAAACTACCAATAGAAGGTAAATATTCGTCAGACAAAAGGGC |
| 2layer-172 | TTGTATCGCGCGAAGGAGTGAGAATAGCAATTTCAACAGTTT |
| 2layer-173 | GTGAATAAGGCTAATCCATTACCCAAATATGGTCATTTGACC |
| 2layer-174 | CCTTCCTGCCATCAACATTATGACCCATACATCGGTTGTACC |
| 2layer-175 | AGCTGCGCATTCGCCATTCAGGGAGAGGGAAACCT |
| 2layer-176 | CTCAACTGTTGGTGCCGCAGCCAGGTTTGAGGCGCATCGTAACCGTGTTTTTTT |
| 2layer-177 | TTTTTTTGATCGGTGCATTAAGTTGTTTTTTT |
| 2layer-178 | TCGTATTTAGACTTACGAGCACGGGAGCTACATGT |
| 2layer-179 | GGGAGAAACAATTTAGGTCAGATGAATAGAAACCATCCCATC |
| 2layer-180 | GGAATACGAAACCGAGGCGTTTTAGCTTACGTATTCTAAGAA |
| 2layer-181 | TCATTAAAGGTGAATTATTGAGGGAGGGGAACCCATTTTCAG |
| 2layer-182 | GAACAAGCACGGAATAAGTTCATAAGGGGTTGTTCCTTTCCA |
| 2layer-183 | AAAAGAAAAGAGGACAGATTTGACCGTAATCCCTTAGGCAAA |
| 2layer-184 | TTTTTTTGGCGAAAATCCTGTTTGATTAGATGGGGACGA |
| 2layer-185 | AGGGCGAAATGGAAGGGTTTGGCAACATGACTCCAAATCATT |
| 2layer-186 | GAAGTAGCATTAACAGATAGGGTTGAGTAACCGAAAGGCGCA |
| 2layer-187 | AAATCATAAGGTGGCTTCATCAAGAGCAGAGAACGGTGTACA |
| 2layer-188 | GAGAGTTGCAGCAACCGAAATCGGCAAAATGGGATGGGAACA |
| 2layer-189 | ACGCTGGTTGCCCTCGCACTCGAAACCAGGCAAAG |
| 2layer-190 | AGGAACAAGCAAGCTCTATCATGACGGGGTTTGGATTCCTGA |
| 2layer-191 | TTTTTTTAAGGAAGGGGGCAAGTGTTTTTTTT |
| 2layer-192 | ATTTTCAATTAAACAGAAATAAAGAAGTGGAGAACCTACCAT |
| 2layer-193 | ACCATAGTTAGCGTTATTAAAGAACGTGATAAAAGAATACAT |
| 2layer-194 | TAAAGTTGCCTGTATACCAGCGCCAACAAATTATTTTGTCAC |
| 2layer-195 | TTTTTTTCGAACGTTATTACGGAACAAAGAATTTTTTT |
| 2layer-196 | GATTATAAAAACACGGAATTGCGAATCCCCGAAAGGAACAAC |
| 2layer-197 | GGGCTTGAGATGACACTGCCCTGACGAGGATTCCCGGAAGTT |
| 2layer-198 | CCAATAGTTTGTTAGAGAAGCCTTTAGCATTGTAATACTTTT |
| 2layer-199 | TTGTAATTTGCGAAAGGGGGAGCGTTGCTGTAAAG |
| 2layer-200 | TTTTTTTGGTAACGCCGAGGTGGAGTTTTTTT |
| 2layer-201 | ATAGATAAACAACTCAGGAGGGCCAGAATCATATA |
| 2layer-202 | AAAATCGCGCAGGAAAAACGGATTCGCCAATAGATATGCAGA |
| 2layer-203 | GTTACCAAAAAGTAGGGAGGTTTTGAATTGGAACCTCCCGAC |
| 2layer-204 | TAGAGCCAGCAACCGCAATTATCACCGTAGAACCGCGCCACC |
| 2layer-205 | TTTTTTTATTTAGAAGTATAAATCCTTTGCCTTTTTTT |
| 2layer-206 | AAGAATACACCAACTGAAAATCTCCAGGAGAATAATTTTTTC |
| 2layer-207 | TGCGATTTTAAGAATGGTTTAATTTCAAGTTTTAATGAATAT |
| 2layer-208 | CGCATTATAAACGTAAAATTTTTAGAATTATTTCAACGCAAG |
| 2layer-209 | ACAATGTTCAAACGACGGCCATCCACACATGGTCA |
| 2layer-210 | TTTTTTTCCGCCACGGCCAGTCCCGTTTTTTT |
| 2layer-211 | GAAGGTTAAATCAAGAGAAGTGAGTCTGTCTTAAC |
| 2layer-212 | AGATGATGAAACGGTAAGGCGAATTATTAGACGACGACAAAA |
| 2layer-213 | CCCTTTTAATAGCATAGTTGCTATTTGATAAGCCTTAAATCA |
| 2layer-214 | AAACGTCACCAATTAGAATCACCAGTAGGGTGTATGTTGATA |
| 2layer-215 | TTTTTTTTTAGGAGCACTATACATTTGAGGTTTTTTT |

Supplementary Table 12 |   DNA sequences of 24helix DNA origami (Scaffold 8064)

| Name | Sequences |
| --- | --- |
| 24helix-1 | ATCACCGAGCGACACATCGATTTAGCGTTTGCCTTAAGGTCA |
| 24helix-2 | TAGACGGCAATAGCATAAGAGGAAACGCAATAACCTTCCAGA |
| 24helix-3 | ATCCCATTTCATCGCGAGAACCGAGGCGTTTTAGAACCGGTA |
| 24helix-4 | CTTAGGTTTTGAAATCATCTTTGTTTAGTATCACTAGATTAC |
| 24helix-5 | TTAGAACACGGATTACCTTTTCAAAAGAAGATGCAGAATTTC |
| 24helix-6 | GAAGATAATCAATACTGAACCTAACAACTAATACCTAATATC |
| 24helix-7 | AGCGTCAATTCTGACCCCCTGCCAGGCGGATAAGATAGGGGT |
| 24helix-8 | TAGAGCCATTAGAGCGGAAGCATTCGAGGATAGCG |
| 24helix-9 | AAGTCAGTTTAAATAAAAATTGAAGCCTCTGTTTA |
| 24helix-10 | AATCAATGGAACAAGTAACAATTTCATCCATGTCA |
| 24helix-11 | CTGATGCTCACGGATTTCTCCGTCCCGGGCCTCTT |
| 24helix-12 | TAAAACATTCTTTGCCTGCGGGCGGTATCACATCC |
| 24helix-13 | AGAGGTGTGCCGGGTTACCGGTACACTGGTGTGTTGTGGTTTACGATCC |
| 24helix-14 | TTTCTGTAAACACTACGAAAGGCCACTAACAACCA |
| 24helix-15 | TGTACTGCGGAACAAATCTACTACCAGTCTTCATC |
| 24helix-16 | AAGGAACCGCCTCCATAATCAGTGAGGCAAAGACATATT |
| 24helix-17 | TGATTACGCAGTATCAAATTAACCGTTGAAACGATAGTT |
| 24helix-18 | CCAAGATTAGTTGCATCACTTGCCTGAGACAATAAATAT |
| 24helix-19 | AAAACAGTAGGGCTGGTAATATCCAGAATAGATTAATTA |
| 24helix-20 | CCAATTTCATTTGAAAAAACGCTCATGGTATTCCTTTTT |
| 24helix-21 | CAGTATTAGACTTTCTGAAATGGATTATGAATGGCTTCT |
| 24helix-22 | GAACTATGGTTGCTAGAATCAGAGCGGGAGTTTCGTTCA |
| 24helix-23 | CCTGTATCACCGTATTTAGACAGGAACGGCCTTGAGCCG |
| 24helix-24 | GTCTGTCCGCATATGGTTTACGATTGAGGGATAGCCGAACAA |
| 24helix-25 | TGATTAGAATTATCCCAATCCAATGAAAATGATATAGAAGGC |
| 24helix-26 | CTATCGGAGTAATTCTGTCCAGCAGAACGCTAAACACCGGAA |
| 24helix-27 | GCCATTGTGTGAAAACATAGCATAGTGAATGCAGAGGCGAAT |
| 24helix-28 | TGACGCTGCAGGAGCGGAATTATTCATCAAGAGGA |
| 24helix-29 | CCGATTAAGCCCATGTACCGTACGCCTGTAGAAGGATTAGGA |
| 24helix-30 | TGAGAAGCTGAGGCAGGTCAGTCCTCATTATTTTCATCGGCA |
| 24helix-31 | AGTGTTGTTGCTCCGGTCAGGAGCAAAATCACCGAATTAACTTAATTGC |
| 24helix-32 | AAAATCCGTAATGTCATATATAGGGTAATTGAGTAATTTACGAGTCAAA |
| 24helix-33 | CAGCAGGACCGTAACTCCGTGAATCGGCTGTCTGGTTATATACCGTGCA |
| 24helix-34 | GCCCTGAAGAAACAATAGCTCAAATCCAATCGCACCATATCAAGTTGGG |
| 24helix-35 | CACCAGTCATCCCTTAAAGGTGAAATAAAGAAAAC |
| 24helix-36 | GATGGCCCCCAGCGTACACTAATGGGATTTTGCCATGGCTTTTATCATC |
| 24helix-37 | GTGGACTGGTAGAAGAACTAAGTAATAAGTTTTACCGACTTGTAATGCA |
| 24helix-38 | CCTTTAATTCCAGTAAGCCCGGGATTGCTGACTATTATAGTCCCTACTG |
| 24helix-39 | TGCCTGACTTATAAACATTAATAAAGCAAATTAAGCAATAAAGATTTT |
| 24helix-40 | CGGATTGCGAAAATATAATTCTTAACCAAATTTTTGTTAAATTGATGTA |
| 24helix-41 | GAGACGCGAGAGTTATCGGCGGGAACGGAGCTTTCAGAGGTGTTTTACG |
| 24helix-42 | TCATAAAGAGACGGCGCTTTCCCAGCATGCAACCAGCTTACGATACGGA |
| 24helix-43 | TTTGACCCACTACGAAGTTTCGAGGACTGGGTAGC |
| 24helix-44 | TATTACACCAACGTACCTTATATTTCAAAGTAAATTGGGCTTCCTAATC |
| 24helix-45 | TGAATCATTTTGATAAGAGGTCCGAGATAACTAAA |
| 24helix-46 | TCACCCTGTAGGTAAAGATTCGTTCCGAATTAATG |
| 24helix-47 | TCTGCTTTGGGATAGGTCACGACGCTGGGGAAGA |
| 24helix-48 | CGGTTGGGCGGATCAAACTTACCCTTCAAGCCGCACAGG |
| 24helix-49 | GCCTGAAATTATACCAAGCGCCCGTCTAGTTACTT |
| 24helix-50 | GATACACAGATTCATCAGTTGCACTATTAGCAACA |
| 24helix-51 | GGAATGAATGAAACGAATCAAGTTTGCGGAAATCA |
| 24helix-52 | CTCAACGCCCAATATATCTTACCGAAAATAACAAG |
| 24helix-53 | TTCAATGCACTCATTAGGAATCATTAGATAAGTAA |
| 24helix-54 | ACGACGAGTTAATTTACCGACCGTGTTGAGAGAGA |
| 24helix-55 | ATAAACATAACAGTCGCCTGATTGCTGATTATAAG |
| 24helix-56 | TAATCACGTCAGATGAAGAACAAAGTTTGAGTGCCCGAGGTGCTG |
| 24helix-57 | TGCCGCCCACCTTGTCTGGTCAGTTGTAACGAACCCAGTATT |
| 24helix-58 | GAAATGGGTTAATGAACATGAAAGTACAGTCTCCT |
| 24helix-59 | AAATAAGAGGTTGGAACAAGAGTCAGATTTAGGAATTA |
| 24helix-60 | TTCTACCAAAATCAAAAGAATAGCCATTTTGCGGAAA |
| 24helix-61 | ATTAATCAAAACCTGTTTGATGGTGAAAAGGGTGAGAGG |
| 24helix-62 | TTTATAATCAGGAACAACGTGGCGAGAAATCATAGTCA |
| 24helix-63 | TGTGGAAACAGCAGCAAGCGGTCCTTGGTGTAGATGAA |
| 24helix-64 | CGGTTGCAGGGCAACAGCTGATTGAATTTCTGCTCACG |
| 24helix-65 | ACGGTGAATTCAAAGGGCGAAAAAGAAACAAAGTACAT |
| 24helix-66 | CTATCATGCTACGAGGCATAGAATAGTACAAAAGAAGAATGA |
| 24helix-67 | GTACGTGAACATGTTTTAAATATTTCGCTAGATTTCCAATAA |
| 24helix-68 | CCGGACCAACCGTTCTAGCTGGAACGGTGTCTGGATTGTAAA |
| 24helix-69 | TCGCACCCGAGACAGTATCGGCTGGGAAGTTCGCCACCAGTCA |
| 24helix-70 | CCAGTAAAATCCCGTAAGCTGGCAACGCGGTTGCCAAC |
| 24helix-71 | AGCGCGGTTGTGCCCCTGCATTCTGCCCCGTTTTGTTTCCT |
| 24helix-72 | AGCCGTTTTCCGCGACCTGCTCAGGCGCGAGGACATAAGGCT |
| 24helix-73 | CCTCGTTTACCAGATGGTTTAGCGATTTTAAGAAGAAAGAAC |
| 24helix-74 | TGGAAGTTTCATAAGCAAAGCAAAGACTTCAAAGCAGGGTTG |
| 24helix-75 | TTTTGCCTCAGAGCTGACCCTGTAATAAAATCGGC |
| 24helix-76 | TCCGGGCTCATTTTGCGTCTGGCCTTCATTTGCCC |
| 24helix-77 | TTAGTGATGAAGGCCGCCACGAAACGTACAGCGAACCGCCTG |
| 24helix-78 | CACTGCGCGCCTTGGAGGTGTGCACTCAATCCGAGTTCTTTT |
| 24helix-79 | AGGCGCAGACGGAGAGGCTTTCATTAAACGGGTATTCAGGGC |
| 24helix-80 | ATAACAGTTGATTCCCAATCATAAATCGGTGTGTC |
| 24helix-81 | TCTACAAAGGCTATCAGGTTAATAGGAACGGAGGGTAGCTAT |
| 24helix-82 | GGTTTATCAGCTGAAAAGACTTTAGGCTAAAAATTACTT |
| 24helix-83 | CTTCTGGTGCCGGAAACCACAATAACCTCATCCAGCCAGCTT |
| 24helix-84 | AGTTAAACGATGCTGATTGACCAGCGGGGTGGCCT |
| 24helix-85 | AACGGCTGGCACGCATTCGGAACTGCTTTCGAGGTCGCCGACCACA |
| 24helix-86 | CATAAGGGAACCGAACTGAGTCTTTAATCAGAACG |
| 24helix-87 | CGATAAAACCAAAATAGCCCATCAAAAAGATAAC |
| 24helix-88 | CGGAAACTCACAATGACACCATGGCAACCCCCTCA |
| 24helix-89 | AGAATTGAAATAAATCCAGAGTCTTACCATCAATT |
| 24helix-90 | TAAACCAGTACCGACGAGCCACATGTAAAAACAGG |
| 24helix-91 | TCAAATAAATTTTCCTTCTGTCAATATAAGGCGAT |
| 24helix-92 | AGTGAGAATAGCAAAGAGCCACGCCACCTGCAGGG |
| 24helix-93 | TGCCCGTCCAGCATACCAGAACAGAGCCAATCAAC |
| 24helix-94 | CGTCTTTACGAGAGGCTTTTGAAATGTTAGATCTTTTCATAA |
| 24helix-95 | CAGAATTAGTCTGCGAACGAGAAATGGTTGTAACG |
| 24helix-96 | CCTTCGCATCATTGCCTGAGAAATCGTAGAGCGAA |
| 24helix-97 | CCCCAGTGCGGCAAAGCGCCAGGCGATCCGTATGC |
| 24helix-98 | ACCCATCCCCGTTCCGGCAAGCCTCCGATATGAA |
| 24helix-99 | GTCGCGTGCACTCTGTGGTGCTGCGGCCATCAAAGCCTGGAGTGT |
| 24helix-100 | TTAGAACGACCAACTTTGAAAATAGGCTCAGTGCCGTCGAGA |
| 24helix-101 | TCCAAATACGCAAACAATAGAAAATTAGCAAGAAAAATAATGCTGTAG |
| 24helix-102 | GCTATAGGCGTCTTCAGCCATATTATCGAAGCAAGCCATCAATATGATA |
| 24helix-103 | ATCATCGGCATTTTCAAAAGGTAAAGCCCTGACCTGACAGTTTGAGGGG |
| 24helix-104 | CGCTACGAACCTTGCCTTAGAATCCTAGACATCGGAGGTGTACATCGAC |
| 24helix-105 | TCATAACTTTAAAAGAAACCACCAGAACTCAAATATGTGCAGCCAGCGG |
| 24helix-106 | ACACGAAGGCCATGAGGTGAACCATCACCCAGCACTAAATCGGTC |
| 24helix-107 | CTGGCCCGCTTGCCTAATATTGGGCGCCAGGCAGCAAATCGTTCA |
| 24helix-108 | TCGCCTGCACCCTCGCCCAATAGGAATACCTATTTAGATAAATTGTGTC |
| 24helix-109 | AAGAGAGAGCCGCCTGACAGGAGGTTTAAGCAGCAAAATAACGCCAAAA |
| 24helix-110 | AATGCTTTCAAAAATCAGGGAATCGTCATAAATATAGTTTGAAGTCCAT |
| 24helix-111 | CTACTAAAGGCAAGGCAAATTTGGGGCGCGAGCAAGCAAACAAGAGAGA |
| 24helix-112 | AAGATTGATTTTGTTAAAACCGGTTGATAATCATCTTCAGGCCACACCG |
| 24helix-113 | TAAGTTGGTAAAACGACGGAGCTGGCGAAAGGGAACCGTTTTGAGGTAA |
| 24helix-114 | GTCTGGTCCGTCGGTGGTGGTGCCGGACTTGTACTCACG |
| 24helix-115 | CGAATTCATTGTTAATGAATGCGGCGGGAGCACGCACGATTA |
| 24helix-116 | AGTTAAACGAAAGACAGCAAACCGATATATTCGAAGATGAACACTCAAT |
| 24helix-117 | GTAACAAACGAGAAACACCGACAAGAACCGGATCGAGTTTTGGAGACGA |
| 24helix-118 | CCATAAATAAACAGAGAGCCACCACCGGCGGAATATAGACTGCTTC |
| 24helix-119 | ATCATACTAGTAGTTTAAGACTCCTTAACCTAATTCAATAACTTAT |
| 24helix-120 | CGTTAATTATAAGCAAGCCTTAAATCAAGTAATAAAAACTAGAAC |
| 24helix-121 | TCAGCAGGGCCGCTTACAGGGCGCGTACCCACCCTAATG |
| 24helix-122 | CGACGTTGGTAACGAAGCCAACGCTCATAAATCGTGGTGCGGAATT |
| 24helix-123 | GGCAGCACAGCAGCATTACATTTAACTTTAACATTGCCAGAGGAGC |
| 24helix-124 | GTGTGAAGTAATCAGAGGATTTAGAACGAGATAGAGTGC |
| 24helix-125 | TGCCCTGAGCTGCTGCCCGGAATAGGCTCCACCACGGCTGACCAGG |
| 24helix-126 | TTAGCATGGAAGGTAAATATTTTCAACCCAGCGCCCACCGAGATACATA |
| 24helix-127 | ATAACGTCAGCCTTTACAGAGCGTCAAAAAATAAGTAGCAATTTTATCC |
| 24helix-128 | GAACGATCTGTTTATCAACAAAGCTAATGACGACGTAGAAGAAACAACG |
| 24helix-129 | GAAAAGAATCAAAATCATAGGAGAGTCAGATAGCTCAATATTAGTACAT |
| 24helix-130 | GTTTAAATAATCCTGATTGTTGATGGCAATCATCAAAATACCATTAAAT |
| 24helix-131 | GTTTCGAATTCCACAGACAGCAACTACAAACACTGAGCTAAAACCCTCA |
| 24helix-132 | TGAGTGTGCCAGAATGGAAAGAAATAAAACGATTGGTACGCCACCGCCA |
| 24helix-133 | AGCAGAGGAAATAGGAAGTAGCACCATTACTATTCATTAAAGTGGGAAT |
| 24helix-134 | AATCAAGTTTTATTCCCGCTAATATCAGAGTAAAAACAGGGACTGAACA |
| 24helix-135 | AAGAAGCTTAATGGTGTTCCTTATCATTCCCCTGAACAAGAAAGAAACC |
| 24helix-136 | ATCGCTTAACAATACTAAGACAAAGAACGCCTACCTTTTTAAGTAAATG |
| 24helix-137 | GAATTTAACCCTCAAATTGCGTAGATTTTCCTTCTGAATAATTTGCACG |
| 24helix-138 | CAAGAGCACCTATTTATAAACAACTTTCAATAGCGTAACGATAATGAAT |
| 24helix-139 | GCGCGAAAATCAGTTCAACGGGGTCAGTGCTGAATTTACCGTACAGGAG |
| 24helix-140 | TTTTCAGGGCGACAGACCTTTAGCGTCAGACTGTA |
| 24helix-141 | AGTTATTTGTTTAAAGGCCCTTTTTAAGAAAAGTA |
| 24helix-142 | TTATCAACATGTTCTACCGCGCCCAATAGCAAGCA |
| 24helix-143 | TCATAACGCTGAGATCGATAAATAAGGCGTTAAAT |
| 24helix-144 | TATTCTTATCAGATTGTTGAATACCAAGTTACAAA |
| 24helix-145 | AGGTTATCTAAATTAGTCTTTATGCAAATCAACAGTTGAAAG |
| 24helix-146 | TTAGCTTCACAAACCGTTAAGAGGCTGAGACTCCT |
| 24helix-147 | TAGCCCCTGTTTTTCTCAGAGCCGCCACTCATCAC |
| 24helix-148 | AGGAAACCATCACGGTTAGCAAACGTAGAGGAACT |
| 24helix-149 | TTCTAAGTAATAACTATTTTGCACCCAGACCGGGA |
| 24helix-150 | TAGAAACCTTGCTTAATTGAGAATCGCAGCTTAC |
| 24helix-151 | AATTACCCAACAGGATTACCTTTTTTAAGAAAAAC |
| 24helix-152 | TTTAGGACAATCGTACAAACAATTCGACTAGATAA |
| 24helix-153 | TTTGCTCAAGGGATCTCAGGAGGTTTAGCCGATAT |
| 24helix-154 | TTGGTCACCAACCAGACAGACCACATTCAACAGCCATTGTAATT |
| 24helix-155 | ACAAGTTAAGCAAGGATGCTGGCTTAGAGCTGAACACCAGCGCAT |
| 24helix-156 | AAAAGTACCGTGAGCGAACAAGGCCGGAGACAGCATGTAAATAAT |
| 24helix-157 | ATTTATTTTAGATAGACAAGGCGCATCGTAAACTATATCCTCCGG |
| 24helix-158 | GCGTATACAGCTGTTGCCTTTTGCCGCCAGCAAATTATGGAAGGG |
| 24helix-159 | GGGATAGAAAACCTAAACAAACCCTAAAGGGGTTAGTACTAAAGT |
| 24helix-160 | GACAAAGCATAGTCGGGGCAACGGCATCAGAAGGCGGTACCAGCA |
| 24helix-161 | CCGATAAACAGAAGAAAACAACGGAGATTTGTGATGATTCCAGTA |
| 24helix-162 | GAATACCCAAAAAATAGATACATTATCGCGTTTTAAAACTCCCAAAGGC |
| 24helix-163 | CCTCCCGACTTGGAAATCGATATACTTTTGCGGGATTTAGAACCCCACA |
| 24helix-164 | CCAGTAACTTTCCAAGTGCTGGACCAGTACACCGGAAGGGAAGAA |
| 24helix-165 | GTTATACAAATTAGCAACTGTCTCCTGTAGCCAGCCCCGTCGAAGGTAT |
| 24helix-166 | ACAAACATCAAGATTCTCGTCAACCATGTTTACCAGTGGTGAGACTTTT |
| 24helix-167 | GAGCCGTCAATAATCGGGGGTTCCCGGGCGCGGTTCTGG |
| 24helix-168 | GGGTTGCTACAGACCCAAAATACGTAATAGGCAAACGAGCGG |
| 24helix-169 | TCAAAAGAGGGGGTTAACTGGCTCATTAGTTAATACCAACAG |
| 24helix-170 | GGCATGAAGCATTAACATTCATTGAATCATATAAAAAAATACTAAAAGA |
| 24helix-171 | GGTTTTGAAATATTTATGAAAAGGTGGCAACGCTACTACAATACTTCTT |
| 24helix-172 | CAGTATACCAGGGTTTGAAAAGCCCCAATTTAGGCCATATTTACTCAAA |
| 24helix-173 | AAAATTAAACCGCAAGGGATGTGCTGCATGTGAGTTGGAAACACCGCCA |
| 24helix-174 | TACATTTTGGTCATAGGAACGTCAGCGTACGTTATAACTCGTTACATTT |
| 24helix-175 | AAGTATACATTCAGTGGTCGCTGAGGCTCTCAGAATACCGCCCAGGAGG |
| 24helix-176 | CGGAACCTTCAGAAAAATTCATTACCCAACCACCCCCTCAGAAGAATCC |

**Supplementary Table 13 | DNA sequences of *p6mm* DNA lattice**

| Name | Sequences |
| --- | --- |
| XY9-IN1 | GGGCTCCTGACACTACAGTCAGATGTAGCCATAT |
| XY9-IN2 | CACTTTTCATTGGGAGAGCTATGCGTTTCGAACG |
| XY9-IN3 | GTTCCGACTATGCAACTACGTCACTGTGCACATA |
| XY9-IN4 | CCAGTCAGGGTGTGGCTAAATTATAGGCCAATCC |
| XY9-IN5 | CTGATCGGACGAAGCCTCAGAATCCTAAACACAC |
| XY9-IN6 | ACTCAACTTTTCTTTAGCGAGAGGGTCCCTCTCA |
| XY9-IN7 | TTTAATGAAAAGGTCCCGCCTTGTGCGAGATGTA |
| XY9-IN8 | AGCGTGCACTAAGAGTGTCATCAGACGTTCTTTC |
| XY9-IN9 | GACCCGGCACGGTTTTGAGTTACAACTCTCATAA |
| XY9-IN10 | GTTTCTGTCAACTTTATATCGGCCGGATACCGGG |
| XY9-IN11 | AAGGTCGTTCAATATTAGCGGATCGGTGCGAAAA |
| XY9-IN12 | CTGGCACCCATTGGAGAAACGTATCTGTTCATGT |
| XY9-IN13 | GGCCATTCTAATGTAGGTGAATGGAGCATCCGTC |
| XY9-IN14 | TTCGTGCCGTTCCTAGTTTAGTGCTCCTAGTGAT |
| XY9-IN15 | GTTACTAAGTTCAGACCCATAGGCTCAGCTGGGT |
| XY9-IN16 | CTCCCTACCTCTCCTGGCTCTATGAAGTCAGTCT |
| XY9-IN17 | GGCAGTAAGGGGAAAACGTAGTACACTCGAGAGT |
| XY9-IN18 | AAGGCCTTGACTAATCGAGTTGTTGGATGCATCG |
| XY9-IN19 | GGTATTAACCGCATTAACGTGAATCCGACCCGTT |
| XY9-IN20 | TAGCCAAACGTAGAAGGGTCTCATGTATTCGCCA |
| XY9-IN21 | GATCCCTGAGCCTTAATCGCGTTAATAGGGACCA |
| XY9-IN22 | ATCAAATGGCGACGTCATAGGGCCTTGATTATGC |
| XY9-IN23 | ATGTTGCTCTACCGGTATGTCGTACGATGTCCTC |
| XY9-IN24 | TTCTGAAGTGTTTCAAGTGGTGCCGTACAAACGT |
| XY9-IN25 | GTCGGATTATTACCCACACGGGACATCAAACAGC |
| XY9-IN26 | CCGGCTTTGCATTATGTGGACGTGGAACAGATTA |
| XY9-IN27 | GGGATGTTGACTGTACGATGACTCTTCCCGACTC |
| XY9-IN28 | GTAGCAGCGTTACCATGGATTCGCGTATTGAATTAA |
| XY9-IN29 | TATGGACTCTGCGAAAGTTAGTGTCGTCTATCAACG |
| XY9-IN30 | CCTTCTTCCTCGTATCCGTGTAAAGACACATCAGTT |
| XY9-IN31 | GTCCATCTACTGTCCGGGTTTCTGTACTCTTTGTAC |
| XY9-IN32 | CAGTTAGAATTGTTTATGCGTGGAGGATGGCGTAAG |
| XY9-IN33 | TCTTATGAATGACAGGAGGCGAGGACTTTTGATTGT |
| XY9-IN34 | CATACATCGAACATCTTTGTACGACCTGCAAACTCC |
| XY9-IN35 | GCTTATTGGAAAGACTGGTTCTCACCAATTTCGGTC |
| XY9-IN36 | TGTGCTATTCAGTCGTTTGTCGCGTACCTATAACTT |
| XY9-IN37 | TCGCGCTCAATCTTACATCATATTGGACTCTGGATC |
| XY9-IN38 | AACCTCCGGATCAAATGTCCAAACTACATACGGTAC |
| XY9-IN39 | TCGCATGTTTGCCTTATACCCGAATAACAGTACTGT |
| XY9-IN40 | TGGTACCTCTCCCGAATGATCACACACTACAAGTAA |
| XY9-IN41 | GGAACCAACCAGATACGTTAGTATGGGGAAACGAAA |
| XY9-IN42 | TGAAGTTAGGAAGTCGCGATCACTAATCTTGAAGGC |
| XY9-IN43 | CCATTACGTAGCCAATGTGTACAGCAAACTCCAATG |
| XY9-IN44 | TTCTGCAAGTGTTCAGTTAACAAACGCGCATCTTTA |
| XY9-IN45 | CAGGATCAAGAGACATTAGCATTGCATACTCCCGAA |
| XY9-IN46 | CGATTTCGATTATGTCGGATCGGTAAGGTTTACGGT |
| XY9-IN47 | CCAAGCTGTGATAGGCTTGATTTGAATTAGCCGAAC |
| XY9-IN48 | GTACAGTAGTATGTCGTGCCCTCGCGTATCTACTAA |
| XY9-IN49 | CTATCCCTCACGGGACCATATTGTGTAGGGTATCTC |
| XY9-IN50 | GGGACAGCGATACAATAAAGCGGAGGTGTTCTATAC |
| XY9-IN51 | CCTAGCAAGTCTACCGTAAGTGTCATTCACACAGTG |
| XY9-IN52 | GCAGTCCATCTCCGAGGGTCGGATAATTTTATTCTG |
| XY9-IN53 | GTGGCTATATGCCCGTAAAGGACTATGACAACATCC |
| XY9-IN54 | GCAGGTAAATCTAGTCAAGCATCGCTTTACTCAGCT |
| XY9-IN55 | ATGTTATTCGTCGTGGGACGCTAAAACACTCTATGG |
| XY9-IN56 | GCACGAAGCAGCTGACTATTTGACGATCTTCAATTG |
| XY9-IN57 | GTCTTAGGAACGGAGGTCTCAGGAACTGGTCATTAA |
| XY9-IN58 | GACCGAGAGTCAATGGAGACTAATGTTTACGCGTTT |
| XY9-IN59 | GTCTCAATCGCTTTAGAGTCCGAATCCGTTACGATC |
| XY9-IN60 | TTCAGTACATCTCTTCGCGAACTTATCGCGCTAATC |
| XY9-IN61 | CGACCTCGTATTTCTCAAATCGGTTGACTTCCTTGT |
| XY9-IN62 | TCCCAATTGAAATCTAATGGACAAACTGTAGCGGCA |
| XY9-IN63 | AGTTAATTCGCGACCCGGTTAAAACGATTGACAAGT |
| XY9-OUT1 | TTTTTTTTCCATAGAGTGATTGAGACTTTTTTTT |
| XY9-OUT2 | TGGTGAGAATTTTTTTTTTTTTTTCTCCTGTCA |
| XY9-OUT3 | GATTCACGGGATGTTGTCTACTGTACATTAAGGC |
| XY9-OUT4 | GGATTCGGATGGTCCCTATAATCCGACTTGAGAAAT |
| XY9-OUT5 | TCCATTCAGCCTTCAAGGTTGGTTCCACTAGGAA |
| XY9-OUT6 | TCGTTTAAGCTGTTTGAGTTTGGCTATAGTCAGCT |
| XY9-OUT7 | TGTACTACTTCGGGAGTCTAACTTCAGGTCTGAA |
| XY9-OUT8 | CGGCACCAGTATAGAACGAGGGATAGTGACGTCG |
| XY9-OUT9 | AGAGTCATCACTGTGTGTATAGCCACACATAATG |
| XY9-OUT10 | CACAAGGCCTTACGCCAGGAAGAAGGAGTTGCAT |
| XY9-OUT11 | TACGCGAATTTTTTTTTTTTTTTTAAACGACTG |
| XY9-OUT12 | TTTTTTTTGCCGCTACGAATTAACTTTTTTTTT |
| XY9-OUT13 | AGCACTAAACAGTACTGACGTAATGGCCAGGAGA |
| XY9-OUT14 | ATGCAATGCTGAGAGGGAGTGCACGCTAACTGAACA |
| XY9-OUT15 | TCATAGAGTAAAGATGCGAGGTACCACCTACATT |
| XY9-OUT16 | GTCTTTACATTTTTTTTTTTTTTTTCCAGTCTTT |
| XY9-OUT17 | TTGCTGTACCGTTCGAAATGCCGGGTCTAATGTCTC |
| XY9-OUT18 | GTTTTAGCGGAGGACATCCAACATCCCCCGGGTCGC |
| XY9-OUT19 | TTTTTTTTAAACGCGTATGTACTGAATTTTTTTT |
| XY9-OUT20 | CAACAACTCATTGGAGTCCGGAGGTTTAATATTG |
| XY9-OUT21 | TTAACGCGCAGAATAAACGCTGTCCCTACCGGTA |
| XY9-OUT22 | AGTCCTCGCTTTTTTTTTTTTTTTTCGGATACGA |
| XY9-OUT23 | AGTTCCTGAAACGGGTCGCAAAGCCGGATTAGATTT |
| XY9-OUT24 | TTTTTTTTGATCGTAACTCCTAAGACTTTTTTTT |
| XY9-OUT25 | ACCTCCGCTGACGGATGCCTTAGTAACTACGGTAGA |
| XY9-OUT26 | TTTTTTTTACTTGTCAAACGAGGTCGTTTTTTTT |
| XY9-OUT27 | TCCTCCACGTTTTTTTTTTTTTTTTAAAGATGTT |
| XY9-OUT28 | TATAATTTGGAGTTTGCCCAATAAGCCACTCTTA |
| XY9-OUT29 | AGTTTGTCCGAGTCGGGAACTTCAGAAGCGAAGAGA |
| XY9-OUT30 | TTATTCGGGTACATCTCGTCCGATCAGACGTATCTG |
| XY9-OUT31 | CCCTCTCGGACCGAAATAATAGCACACAAAACCG |

| XY9-OUT32 | TTGTAACTTTAATTCAATTCTAACTGAGGCTTCG |
| XY9-OUT33 | TACACAATAAGACTGACTCTTACTGCCTTACGGGCA |
| XY9-OUT34 | TCTGATGAAACTGATGTCGCTGCTACTGTAGTGT |
| XY9-OUT35 | CATCTGACCGTTGATAGTAGATGGACAGCCACAC |
| XY9-OUT36 | GATACGTTGATCCAGAGCTTGCAGAAGTTTTCCC |
| XY9-OUT37 | GGATTCTGGTACAAAGATTCATAAGACTAAAGAA |
| XY9-OUT38 | ATTCAAATCCGATGCATCGGTAGGGAGCCCTCGGAG |
| XY9-OUT39 | CGATCCGCTTACTTGTATGAGCGCGAATAAAGTT |
| XY9-OUT40 | AGGCCCTAACCGTAAACACAGCTTGGCCTTCTAC |
| XY9-OUT41 | AGTGACGTACAATCAAAGAGTCCATATCTCCCAA |
| XY9-OUT42 | TTTTTTTTTAATGACCGAATAACATTTTTTTTT |
| XY9-OUT43 | TTTTTTTCAATTGAAGCTCTCGGTCTTTTTTTT |
| XY9-OUT44 | AGCCTATGGTACCGTATAACATGCGATCTCCAAT |
| XY9-OUT45 | CATGAGACTTAGTAGATCTTGCTAGGCTTGAAAC |
| XY9-OUT46 | CCCATACTATTATGAGAGCAGGAGCCCATGTAAGAT |
| XY9-OUT47 | AATGACACTACTCTCGAGCAAGGCCTTTTGACTAGA |
| XY9-OUT48 | AAGCGATGCCCCGGTATCCGGCACGAATTATTGTAT |
| XY9-OUT49 | CCACGTCCGAGATACCCATGGACTGCGTGGGTAA |
| XY9-OUT50 | AGGTCGTACTTTTTTTTTTTTTTTTCCCGGACAG |
| XY9-OUT51 | ACGACACTATTTTTTTTTTTTTTTTCCATGGTAA |
| XY9-OUT52 | TCCAATATGGAAAGAACGAGTCGGAACTATAAGGCA |
| XY9-OUT53 | CGCATAGCAAGTTATAGCGATGTATGGGGACCTT |
| XY9-OUT54 | TTTTTTTTACAAGGAAGCAATTGGGATTTTTTTT |
| XY9-OUT55 | GTACAGAAATTTTTTTTTTTTTTTTCATAAACAA |
| XY9-OUT56 | GTACGCGACTTTTTTTTTTTTTTTTTACTTTCGCA |
| XY9-OUT57 | GTGTGTGATATATGGCTATGAAAAGTGACATTTGAT |
| XY9-OUT58 | TTTTTTTTGATTAGCGCGCTTCGTGCTTTTTTTT |
| XY9-OUT59 | CATAGTCCTACATGAACAGACAGAAACCCGACATAA |
| XY9-OUT60 | ATTAGTGATGTGTGTTTACCTGACTGGCATTCGGGA |
| XY9-OUT61 | TGTCCCGTGTTCGGCTATTTACCTGCCGTACAGT |
| XY9-OUT62 | CTTACCGATTTTTCGCACAGAATGGCCTGGTCCCGT |
| XY9-OUT63 | ATCGTCAAAGCATAATCAGTTAATACCTCCCACGAC |
| XY9-OUT64 | GTAGTTTGGTATGTGCACAAGTTGAGTCGCGACTTC |
| XY9-OUT65 | ACGCGAGGGACCCAGCTGAACGACCTTAAGCCTATC |
| XY9-OUT66 | GCGTTTGTTGGATTGGCCTTCATTAAAACATTGGCT |
| XY9-OUT67 | CGGCCGATTTTCGTTTCTTGATCCTGCGATTAGT |
| XY9-OUT68 | GATAAGTTCTGGCGAATATCAGGGATCGACCTCCGT |
| XY9-OUT69 | ATTATCCGAATCACTAGGGGGTGCCAGCACGACATA |
| XY9-OUT70 | AACATTAGTACGTTTGTAGAGCAACATCTCTAAAGC |
| XY9-OUT71 | GTACGACAAGCTGAGTATCGAAATCGTTAATGCG |
| XY9-OUT72 | TCAACCGATTAATCTGTTCCATTTGATCTCCATTGA |